\documentclass[aps,rmp,showpacs,showkeys,preprint,eqsecnum,floatfix,longbibliography]{revtex4-1}

\usepackage{amssymb}
\usepackage{amsmath}
\usepackage{graphicx}
\usepackage{amsbsy}
\usepackage{color}
\usepackage{bm}
\usepackage{bbm}
\usepackage{slashed}

\usepackage{hyperref}

\newcommand{\complexn}{\mathbb{C}}
\newcommand{\realn}{\mathbb{R}}

\newcommand{\bq}{\begin{eqnarray}}
\newcommand{\eq}{\end{eqnarray}}
\newcommand{\bqn}{\begin{eqnarray*}}
\newcommand{\eqn}{\end{eqnarray*}}

\renewcommand{\r}{\mathbf{r}}
\newcommand{\rr}{\mathbf{r}}
\newcommand{\sss}{\mathbf{s}}
\newcommand{\gggg}{\mathbf{g}}
\newcommand{\qq}{\mathbf{q}}
\newcommand{\pp}{\mathbf{p}}
\newcommand{\zz}{\mathbf{z}}

\newcommand{\crr}{{\cal R}}
\newcommand{\css}{{\cal S}}
\newcommand{\caa}{{\cal A}}
\newcommand{\cdd}{{\cal D}}
\newcommand{\cll}{{\cal L}}
\newcommand{\cvv}{{\cal V}}

\newcommand{\Tr}{\mathop{\mathrm{tr}}}
\renewcommand{\det}{\mathop{\mathrm{det}}}
\newcommand{\sgn}{\text{sgn}}
\newcommand{\cotan}{\text{cotan}}

\newcommand{\qqq}[2]{{q_{(#1)}}^{#2}}
\newcommand{\ppp}[2]{{p_{(#1)}}_{#2}}
\newcommand{\om}[1]{\tilde{\omega}^{\hat{#1}}}
\newcommand{\omud}[2]{{\tilde{\omega}\mbox{}^{\hat{#1}}}_{\hat{#2}}}
\newcommand{\omdd}[2]{\tilde{\omega}_{\hat{#1}\hat{#2}}}
\newcommand{\omdu}[2]{{\tilde{\omega}\mbox{}_{\hat{#1}}}^{\hat{#2}}}
\newcommand{\ruddd}[4]{{R^{\hat{#1}}}_{\hat{#2}\hat{#3}\hat{#4}}}
\newcommand{\ruudd}[4]{{R^{\hat{#1}\hat{#2}}}_{\hat{#3}\hat{#4}}}
\newcommand{\rc}[2]{{{\cal R}^{\hat{#1}}}_{\hat{#2}}}
\newcommand{\oh}{\frac{1}{2}}
\newcommand{\dt}{\tilde{d}}

\newtheorem{theo}{Theorem}[section]

\begin{document}
\title{Plasma living in a curved surface at some special temperature}

\author{Riccardo Fantoni}
\email{rfantoni@ts.infn.it}
\affiliation{Universit\`a di Trieste, Dipartimento di Fisica, strada
  Costiera 11, 34151 Grignano (Trieste), Italy}

\date{\today}

\begin{abstract}
The simplest statistical mechanics model of a Coulomb plasma in two
spatial dimensions admits an exact analytic solution at some special
temperature in several (curved) surfaces. We present in a unifying
perspective these solutions for the (non-quantum) plasma, made of
point particles carrying an absolute charge $e$, in thermal
equilibrium at a temperature $T=e^2/2k_B$, with $k_B$ Boltzmann's
constant, discussing the importance of having an exact solution, the
role of the curvature of the surface, and the densities of the plasma.  
\end{abstract}


\maketitle
\tableofcontents
\part{Introduction}
\label{part:introduction}

The physics of fluids of particles living in (curved) surfaces is a
well known chapter of surface physics. It arises in situations in
which particles are adsorbed or confined on a substrate with nonzero
curvature, be it the wall of a porous material, or a membrane, a
vesicle, a micelle for example made of ampiphilic surfactant molecules
such as lipids, or a biological membrane, or the surface of a large
solid particle, or an interface in an oil-water emulsion
\cite{Fantoni12c}. On the other hand it often occurs that by
lowering the number of spatial dimensions, the statistical mechanics
problem of a given fluid in the whole space, greatly simplifies, to
the point of becoming, in certain cases, exactly solvable
analytically in the continuum. A relevant feature of such low
dimensional exactly solvable fluids is that they often play an
important role as exact standards and guides to test approximate
solutions and numerical experiments for (higher dimensional) 
fluid's models. In a more general context, the few exact analytical
results have helped form new qualitative insights given by sum rules
and in clarifying the nature of the long distance asymptotic decay of
the truncated two (or more) particle distribution functions
\cite{Martin88,Sausset2011}. 

In the statistical physics of continuous fluids, those where the
particles are allowed to move in a continuous space, one finds
examples of exactly solvable ones especially among the
non-quantum in lower dimensions (one and two). 

Coulomb systems \cite{March84,Henderson05} such as plasmas,
electrolytes, or generally ionic materials are made of charged
particles interacting through the long-range
Coulomb law. They are an important chapter of ionic condensed matter 
(in systems like molten salts, transition metal ions in solution,
molten alkali halides, $\ldots$) or ionic soft matter (in systems like 
natural or synthetic saline environments like aqueous and non aqueous
electrolyte solutions, polyelectrolytes, colloidal suspensions,
$\ldots$). The simplest model of a Coulomb system is the one-component
plasma (OCP), also called {\sl jellium}: an assembly of identical
point charges of charge $e$, embedded in a neutralizing uniform
background of the opposite sign. Here we consider the classical
(i.e. non-quantum) equilibrium statistical mechanics of the
OCP. According to the proof of Sari and Merlini \cite{Sari} which goes
through ``H-stability'' and the ``cheese theorem'', the OCP must have
a well behaved thermodynamic limit. Though this model might seem, at
first sight, oversimplified as to bear little resemblance to molten
salts or liquid metals, it is nevertheless of great value in
clarifying general effects which emerge as a direct consequence of
long-range Coulomb's interaction. This model constitutes the basic
link between the microscopic description and the phenomenology of
ionic condensed and soft matter.

The two-dimensional version (2D OCP) of the OCP has been much
studied. Provided that the Coulomb potential due to a point-charge is 
defined as the solution of the Poisson equation ``in'' a two-dimensional
world, i.e., is a logarithmic function $-\ln r$ of the distance $r$ to
that point-charge, the 2D OCP mimics many generic properties of the
three-dimensional Coulomb systems. In this case the electric field
lines are not allowed to leave the surface as it happens in the
satirical novella of Edwin Abbott Abbott \cite{Abbott}. Of course,
this toy logarithmic model does not describe real charged particles,
such as electrons, confined on a surface, which nevertheless interact
through the three dimensional Coulomb potential $1/r$. One motivation
for studying the 2D OCP is that its equilibrium statistical mechanics
is analytically exactly solvable at one special temperature: both the
thermodynamical quantities and the correlation functions are available. 

The OCP is exactly solvable in one dimension
\cite{Edwards62,Fantoni2016}. In two dimensions, Jancovici and
Alastuey \cite{Ginibre65,Metha67,Jancovici81b,Alastuey81} proved  
that the OCP is exactly solvable analytically at a special value of 
the coupling constant, $\Gamma=\beta e^2=2$ where $\beta=1/k_BT$
with $k_B$ Boltzmann's constant and $T$ the absolute temperature,
on a plane. Since then, a growing interest in two-dimensional plasmas
has lead to study this system on various flat geometries
\cite{Rosinberg84,Jancovici94,Jancovici96} and two-dimensional curved  
surfaces like the cylinder \cite{Choquard81,Choquard83}, the sphere  
\cite{Caillol81,Tellez99,Jancovici00,Salazar2016}, the
the pseudosphere \cite{Jancovici1998,Fantoni03jsp,Jancovici04}, and 
Flamm paraboloid \cite{Fantoni2008}. Among these surfaces only the
last one is of non-constant curvature.

How the properties of a system are affected by the curvature of the
space in which the system lives is a question which arises in general
relativity. This is an incentive for studying simple models. 

The two-component plasma (TCP) is a neutral mixture of point-wise
particles of charge $\pm e$. The equation of state of the TCP
living in a plane is known since the work of Salzberg and Prager
\cite{Salzberg63}. In the plasma the attraction between oppositely
charged particles competes with the thermal motion and makes the
partition function of the finite system diverge when 
$\Gamma = \beta e^2\ge 2$, where $\beta=1/k_BT$ with $k_B$ Boltzmann
constant. The system becomes unstable against the collapse of pairs
of oppositely charged particles, and as a consequence all
thermodynamic quantities diverge, so that the point particle model is
well behaved only for $\Gamma < 2$ \cite{Hauge71} when the Boltzmann
factor for unlike particles is integrable at small separations of the
charges. In this case rescaling the particles coordinates so as 
to stay in the unit disk one easily proves that the grand canonical
partition function is a function of
$\sqrt{\zeta_-\zeta_+}V^{(1-\Gamma/4)}$, where $V$ is the volume
occupied by the plasma and $\zeta_\pm$ the fugacities of 
the two charge species, and as a consequence the equation of state is  
$\beta p=n(1-\Gamma/4)$ where $n=\rho_++\rho_-$ is the total particle
number density. However, if the collapse is avoided by some short
range repulsion (hard cores for instance), the model remains well
defined for lower temperatures. Then, for $\Gamma > 4$ 
the long range Coulomb attraction binds positive and negative
particles in pairs of finite polarizability. Thus, at some critical
value $\Gamma_c \sim 4$ the system undergoes the Kosterlitz-Thouless
transition \cite{Kosterlitz1973} between a high temperature ($\Gamma <
4$) conductive phase and a low temperature ($\Gamma > 4$) dielectric
phase. For $\Gamma\ge 2$ it is necessary to regularize the system of
point charges allowing for a short-range strong repulsion between
unlike charge which may be modeled as hard (impenetrable) disks,
i.e. giving a physical dimension to the particles to prevent the
collapse. The same behavior also occurs in the TCP living in one
dimension \cite{Lenard1961,Fantoni2016}.  

The structure of the TCP living in a plane at the special value
$\Gamma=2$ of the coupling constant is also exactly solvable 
analytically \cite{Gaudin1985,Cornu87}. Through the use of an external
potential it has also been studied in various confined geometries
\cite{Cornu89,Forrester91,Tellez02,Merchan04} and in a gravitational
field \cite{Tellez97,Tellez98}. It has been studied in surfaces of
constant curvature as the sphere \cite{Forrester1992,Forrester1996}
and the pseudosphere \cite{Jancovici1998} and on the Flamm
paraboloid of non-constant curvature \cite{Fantoni2012}. Unlike the
OCP where the properties of the van der Monde determinant allowed the
analytical solution a Cauchy identity is used for the solution of the 
TCP. Unlike in the one-component case where the solution was possible
for the plasma confined in a region of the surface now this is not
possible, anymore, without the use of an external potential. In these
cases the external potential is rather given by $-(\Gamma/e^2)\ln\sqrt
g$ where $g$ is the determinant of the metric tensor of the Riemannian
surface \cite{Fantoni2012b}. On a curved surface, even though the
finite system partition function will still be finite for $\Gamma<2$
since the surface is locally flat, the structure will change respect
to the flat case. 

Purpose of this review is to describe the state of the art for the
studies on the exactly solvable statistical physics models of a plasma
on a (curved) surface. In section \ref{sec:ocp} we will treat the OCP
in the various surfaces and in section \ref{sec:tcp} the TCP in the
various surfaces. Except for the OCP on the plane we will stop at
the solution for the partition function and the densities of the
finite OCP. If the reader wishes he can refer to the original
papers for the resulting expressions in the thermodynamic limit. The
solutions for the TCP do not give the results for the finite system
but only its thermodynamic limit. For the OCP we use the canonical
ensemble for the plane, the cylinder and the sphere, and the grand
canonical ensemble for the pseudosphere and the Flamm paraboloid on
half surface with grounded horizon. For the TCP we only use the grand
canonical ensemble. When appropriate we point out the ensemble
inequivalence which arise for the finite system.

\section{The surface}
\label{sec:surface}

We will generally consider {\sl Riemannian surfaces} $\css$ with a
coordinate frame $\qq=(x^1,x^2)$ and with a metric
\bq
d\sss^2=g_{\mu\nu}(\qq)\,dx^\mu dx^\nu,
\eq
with $g_{\mu\nu}$ the metric tensor. We will denote with $g(\qq)$ the
Jacobian of the transformation to an orthonormal coordinate reference
frame, i.e. the determinant of the metric tensor
$g_{\mu\nu}$. The surface may be embeddable in the three dimensional 
space or not. It is important to introduce a {\sl disk} $\Omega_R$ of
radius $R$ and its boundary $\partial\Omega_R$. The connection
coefficients, the Christoffel symbols, in a coordinate frame are 
\bq
\Gamma_{\mu\beta\gamma}=\frac{1}{2}(g_{\mu\beta,\gamma}+g_{\mu\gamma,\beta}-
g_{\beta\gamma,\mu}),
\eq
where the comma denotes a partial derivative as usual. The Riemann
tensor in a coordinate frame reads
\bq
{R^\alpha}_{\beta\gamma\delta}={\Gamma^\alpha}_{\beta\delta,\gamma}-
{\Gamma^\alpha}_{\beta\gamma,\delta}+{\Gamma^\alpha}_{\mu\gamma}
{\Gamma^\mu}_{\beta\delta}-{\Gamma^\alpha}_{\mu\delta}
{\Gamma^\mu}_{\beta\gamma},
\eq
in a two-dimensional space has only $2^2(2^2-1/12)=1$ independent
component. The scalar curvature is then given by the following indexes
contractions (the trace of the Ricci curvature tensor),
\bq
\crr={R^\mu}_\mu={R^{\mu\nu}}_{\mu\nu},
\eq
and the (intrinsic) Gaussian curvature is $K=\crr/2$. In an embeddable
surface we may define also a (extrinsic) mean curvature
$H=(k_1+k_2)/2$, where the principal curvatures $k_i$, $i=1,2$ are the
eigenvalues of the shape operator or equivalently the second
fundamental form of the surface and $1/k_i$ are the principal radii of
curvature. The Euler characteristic of the disk $\Omega_R$ is given by 
\bq \label{ec}
\chi=\frac{1}{2\pi}\left(\int_{\Omega_R}K\,dS+
\int_{\partial\Omega_R}k\,dl\right),
\eq
where $k$ is the geodesic curvature of the boundary $\partial\Omega_R$.

\section{The Coulomb potential} 
\label{sec:coulomb potential}

The {\sl Coulomb potential} $G(\qq,\qq_0)$ created at $\qq$ by a unit
charge at $\qq_0$ is given by the Green function of the Laplacian
\bq \label{cp}
\Delta G(\qq,\qq_0)=-2\pi\delta^{(2)}(\qq;\qq_0),
\eq
with appropriate boundary conditions. Here $\Delta$ is the
Laplace-Beltrami operator. This equation can often be
solved by using the decomposition of $G$ as a Fourier series. 

\section{The background} 
\label{sec:background}

The Coulomb potential generated by the {\sl background}, with a
constant surface charge density $\rho_b=-en_b$ satisfies the Poisson
equation 
\bq \label{bp}
\Delta v_b=-2\pi\rho_b.
\eq
The Coulomb potential of the background can be obtained by solving
Poisson equation with the appropriate boundary conditions. Also, it
can be obtained from the Green function computed in the previous
section
\bq
v_b(\qq)=\int G(\qq,\qq')\rho_b(\qq')\,dS'.
\eq
This integral can be performed easily by using the Fourier series
decomposition of Green's function $G$.

\section{The total potential energy} 
\label{sec:tpe}

The {\sl total potential energy} of the plasma is then
\bq \nonumber
V_N&=&V_N^{pp}+V_N^{pb}+V_N^0=\frac{e^2}{2}\sum_{i\ne j}G(|\qq_i-\qq_j|)+
e\sum_i\int_{\Omega_R}v_b(|\qq-\qq_i|)\,d\qq+\\ \label{eq:tpe}
&&\frac{1}{2}\iint_{\Omega_R}\rho_bv_b(|\qq-\qq'|)\,d\qq d\qq',
\eq
where the last term $V_N^0$ is the self energy of the background and the
first two terms $V_N^{pp}$ and $V_N^{pb}$ are the interaction potential
energy between the charges at $\qq_i$, $i=1,\ldots,N$ and between the
charges and the background, respectively. 

\section{The densities and distribution functions} 
\label{sec:tcf}

Given either the canonical partition function in a fixed region
$\Omega\in\css$ of a Riemannian surface $\css$, $Z_N(\Gamma)$ with
$\Gamma=\beta e^2$ the coupling constant, or the grand
canonical one $\Xi[\{\lambda_p(\qq)\},\Gamma]$, with $\lambda_p$ some
position dependent fugacities, we can define the {\sl $n$-body density
functions}. Denoting with $\pp=(p,\qq)$ the species $p$ and the
position $\qq$ of a particle of this species, we have, 
\bq \nonumber
\rho^{(n)}(\pp_1,\ldots,\pp_n;N,\Gamma)&=&\rho(\pp_1;N,\Gamma)\cdots
\rho(\pp_n;N,\Gamma)g_{p_1\ldots p_n}(\qq_1,\ldots,\qq_n;N,\Gamma)\\
&=&\left\langle{\sum_{i_1,\ldots,i_n}}^{DP}\delta^{(2)}(\qq_1;\qq_{i_1})
\delta_{p_1,p_{i_1}}\cdots\delta^{(2)}(\qq_n;\qq_{i_n})
\delta_{p_n,p_{i_n}}\right\rangle_{N,\Gamma},
\eq
where $\delta_{p,q}$ is the Kronecker delta, $\delta^{(2)}$ is the 
Dirac delta function on the curved surface such that
$\int\delta^{(2)}(\qq;\qq')\,dS=1$ with $dS=\sqrt{g(\qq)}\,d\qq$ the
elementary surface area on $\css$,
$\langle\ldots\rangle_{N,\Gamma}=\sum_{p_1,\ldots,p_N}\int_\Omega\ldots e^{-\beta
  V_N}\,dS_1\cdots dS_N/Z_N$ is the thermal average in the canonical
ensemble, ${\sum}^{DP}$ denotes the inclusion in the sum only of addends
containing the product of delta functions relative to different
particles, and we omitted the superscript $\mbox{}^{(1)}$ in the one-body
densities. The $g_{p_1,\ldots,p_n}$ are known as the {\sl $n$-body 
  distribution functions}. It is convenient to introduce another set
of correlation functions which decay to zero as two groups of
particles are largely separated \cite{Martin88}, namely the {\sl truncated}
(Ursell) correlation functions, 
\bq
\rho^{(n)T}(\pp_1,\ldots,\pp_n;N,\Gamma)=
\rho^{(n)}(\pp_1,\ldots,\pp_n;N,\Gamma)-\sum\prod_{m<n}
\rho^{(m)T}(\pp_{i_1},\ldots,\pp_{i_m};N,\Gamma),
\eq 
where the sum of products is carried out over all possible partitions
of the set $(1,\ldots,n)$ into subsets of cardinal number $m<n$.

In terms of the grand canonical partition function we will have,
\bq
\rho^{(n)}(\pp_1,\ldots,\pp_n;\{\lambda_p\},\Gamma)=
\prod_{i=1}^n\lambda_{p_i}(\qq_i)\frac{1}{\Xi[\{\lambda_p\},\Gamma]}
\frac{\delta^{(n)}\Xi[\{\lambda_p\},\Gamma]}
{\delta\lambda_{p_1}(\qq_1)\ldots\delta\lambda_{p_n}(\qq_n)},
\eq
and
\bq
\rho^{(n)T}(\pp_1,\ldots,\pp_n;\{\lambda_p\},\Gamma)=
\prod_{i=1}^n\lambda_{p_i}(\qq_i)
\frac{\delta^{(n)}\ln\Xi[\{\lambda_p\},\Gamma]}
{\delta\lambda_{p_1}(\qq_1)\ldots\delta\lambda_{p_n}(\qq_n)}.
\eq
We may also use the notation
$\rho^{(n)}(\pp_1,\ldots,\pp_n;\{\lambda_p\},\Gamma) =
\rho^{(n)}_{p_1\ldots p_n}(\qq_1,\ldots,\qq_n;\{\lambda_p\},\Gamma)$
where for example in the two-component mixture each $p=\pm$ denotes
either a positive or a negative charge. And sometimes we may omit the
dependence from the number of particles, the fugacities, and the
coupling constant. From the structure it is possible to derive the
thermodynamic properties of the plasma (but not the contrary). 

\part{The One-Component Plasma} 
\label{sec:ocp}

An {\sl one-component plasma} is a system of $N$ identical particles
of charge $e$ embedded in a uniform neutralizing background of
opposite charge.  

\section{The plane} 
\label{sec:ocp-plane}
The metric tensor in the Cartesian coordinates $\qq=(x,y)$ of the {\sl
  plane} is, 
\bq
\gggg=\left(
\begin{array}{cc}
1 & 0\\
0 & 1
\end{array}\right),
\eq
and the curvature is clearly zero. We will use polar coordinates
$\qq=(r,\varphi)$ with $r=\sqrt{x^2+y^2}$ and $\varphi=\arctan(y/x)$. 

\subsection{The Coulomb potential} 

The Coulomb interaction potential between a particle at $\qq$ and a
particle at $\qq_0$ a distance $r=|\qq-\qq_0|$ from one another is
\bq
G(\qq,\qq_0)=-\ln(|\qq-\qq_0|/L),
\eq
where $L$ is a length scale. 

\subsection{The background} 

If one assumes the particles to be confined in a disk
$\Omega_R=\{\qq\in\css|\,0\le\varphi\le 2\pi, 0\le 
r\le R\}$ of area $\caa_R=\pi R^2$ the background potential is 
\bq
v_b(r)=en_b\frac{\pi}{2}\left(r^2-R^2+2R^2\ln\frac{R}{L}\right),
\eq
where $r=|\qq|$.

\subsection{The total potential energy} 

The total potential energy of the system is then given by
Eq. (\ref{eq:tpe}). 
Developing all the terms and using $n_b=n=N/\caa_R$ (this is not a
necessary condition since we can imagine a situation where $n_b\ne
n$. In this case the system would not be electrically neutral) we then
find 
\bq
V_N/e^2=-\sum_{i<j}\ln\left(\frac{r_{ij}}{L}\right)+
\frac{n_b\pi}{2}\sum_ir^2_i+n_b^2\pi^2R^4
\left(-\frac{3}{8}+\frac{1}{2}\ln\frac{R}{L}\right),
\eq
where $r_{ij}=|\qq_i-\qq_j|$ and $r_i=|\qq_i|$. This can be rewritten
as follows  
\bq \nonumber
V_N/e^2&=&-\sum_{i<j}\ln\left(\frac{r_{ij}}{R}\right)+
\frac{N}{2}\sum_i\left(\frac{r_i}{R}\right)^2+\\
&&N^2\left(-\frac{3}{8}+\frac{1}{2}\ln\frac{R}{L}\right)-
\frac{N(N-1)}{2}\ln\left(\frac{R}{L}\right).
\eq

We can then introduce the new variables \cite{Jancovici81b}
$\zz_i=\sqrt{N}\qq_i/R$ to find
\bq
V_N/e^2&=&f(\{\zz_i\})+f_c\\
f&=&-\sum_{i<j}\ln z_{ij}+\frac{1}{2}\sum_iz_i^2,\\
f_c&=&\frac{N(N-1)}{4}\ln(n\pi L^2)+
N^2\left(-\frac{3}{8}+\frac{1}{2}\ln\frac{R}{L}\right).
\eq
We can always choose $L=R$ so that in the thermodynamic limit
$\lim_{N\to\infty}f_c/N=-\ln(n\pi L^2)/4$ and the excess Helmholtz
free energy per particle
\bq
a_\text{exc}=F_\text{exc}/N\to-\frac{e^2}{4}\ln(\pi nL^2)+a_0(T),
\eq
with $a_0$ some function of the temperature $T$ alone. Therefore, the 
equation of state has the simple form
\bq
p=(1/\beta-e^2/4)n,
\eq
where $\beta=1/k_BT$ with $k_B$ Boltzmann's constant.

\subsection{Partition function and densities at a special temperature} 
\label{sec:plane-g2}

At the special temperature $T_0=e^2/2k_B$ the partition function can
be found exactly analytically using the properties of the van der Monde
determinant \cite{Jancovici81b,Alastuey81}. Using polar coordinates
$\zz_i=(z_i,\theta_i)$, one obtains at $T_0$ a Boltzmann factor
\bq
e^{-\beta V_N}=A_Ne^{-\sum_iz_i^2}\left|\prod_{i<j}(Z_i-Z_j)\right|^2,
\eq
where $A_N$ is a constant and $Z_i=z_i\exp(i\theta_i)$. This expression
can be integrated upon variables $\zz_i$ ($0\le z_i\le\sqrt{N}$) by
expanding the van der Monde determinant $\prod(Z_i-Z_j)$. One obtains
the partition function
\bq \label{plane-z}
Z_N(2)=\int e^{-\beta V_N}\,d\zz_1\cdots d\zz_N=
A_N\pi^N N!\prod_{j=1}^N\gamma(j,N),
\eq
where
\bq
\gamma(j,N)=\int_0^{\sqrt{N}}e^{-z^2}z^{2(j-1)}2z\,dz=\int_0^Ne^{-t}t^{j-1}\,dt,
\eq
is the incomplete gamma function. Taking the thermodynamic limit of
$-[\ln (Z_N(2)/\caa_R^N)]/N$ $\to\beta a_\text{exc}(2)$ we obtain the
Helmholtz free energy per particle 
\bq \label{plane-a2}
a_\text{exc}(2)=-\frac{e^2}{4}\ln(\pi n L^2)+\frac{e^2}{2}
\left[1-\frac{1}{2}\ln(2\pi)\right].
\eq

One can also obtain the $n$-body distribution functions from the
truncated densities \cite{Martin88} as follows
\bq \label{plane-g}
g(1,\ldots,n;N)=e^{-\sum_{i=1}^n z_i^2}
\det\left[K_N(Z_i\bar{Z}_j)\right]_{i,j=1,\ldots,n},
\eq
where $\bar{Z}$ is the complex conjugate of $Z$ and
\bq
K_N(x)=\sum_{i=1}^N\frac{x^{i-1}}{\gamma(i,N)}.
\eq
In the thermodynamic limit $N\to\infty$, $\gamma(i,N)\to(i-1)!$, and
$K_N(x)\to e^x$. In this limit, one obtains from Eq. (\ref{plane-g})
the following explicit distribution functions \cite{Jancovici81b} 
\bq
g(1)&=&1,\\
g(1,2)&=&1-e^{-\pi n r_{12}^2},\\
g(1,2,3)&=&\ldots.
\eq
This {\sl Gaussian falloff} is in agreement with the general result
according to which, among all possible long-range pair potentials, it
is only in the Coulomb case that a decay of correlations faster than
any inverse power is compatible with the structure of equilibrium
equations like the Born-Green-Yvon hierarchic set (see
Ref. \cite{Martin88} section II.B.3). A somewhat surprising result is
that the correlations does not have the typical exponential falloff
typical of the high-temperature Debye-H\"uckel approximation
\cite{Debye1923}. One easily checks that the distribution functions
obey the perfect screening and other sum rules.

Expansions around $\Gamma = 2$ suggests that the pair correlation
function changes from the exponential form to an oscillating one for a
region with $\Gamma > 2$. This behavior of the pair correlation
function as the coupling is stronger has been observed in Monte Carlo
simulations \cite{Caillol1982}. For sufficient high values of $\Gamma$
(low temperatures) the 2D OCP begins to crystallize and there are
several works where the freezing transition is found. For the case of
the sphere Caillol et al. \cite{Caillol1982} localized the coupling
parameter for melting at $\Gamma\approx 140$. In the limit
$\Gamma\to\infty$ the 2D OCP becomes a Wigner crystal. In particular,
the spatial configuration of the charges which minimizes the energy at
zero temperature for the 2D OCP on a plane is the usual hexagonal 
lattice. Nowadays, the corresponding Wigner crystal of the 2D OCP on
sphere or Thomson problem may be solved numerically
\cite{Fantoni12c}. 

\section{The cylinder} 
\label{sec:ocp-cylinder}

The {\sl cylinder} may be useful to compare an exactly soluble fluid
with the results from its Monte Carlo simulation for example, where
one needs to use periodic boundary conditions. The two dimensional
system studied in the simulation would actually live on a torus but
the cylinder is already a relevant step forward in this direction.

The metric tensor in the cartesian coordinates $\qq=(x,y)$ is,
\bq
\gggg=\left(
\begin{array}{cc}
1 & 0\\
0 & 1
\end{array}\right),
\eq
and again the curvature is zero.

\subsection{The Coulomb potential} 

We now consider \cite{Choquard81,Choquard83} a rectangular disk
$\Omega_{L,W}=\{\qq\in\css|\,-L/2\le x\le L/2,-W/2\le y\le W/2\}$. We
then solve Eq. (\ref{cp}) imposing periodicity in $y$ with period
$W$ expanding $G$ in a Fourier series in $y$ where the coefficients
are functions of $x$ and written as inverse Fourier transforms. The
solution is
\bq \nonumber
G(\qq_1,\qq_2)&=&-\frac{\pi}{W}|x_1-x_2|+\\ \label{cylinder-g}
&&\frac{\sgn(x_1-x_2)}{2}\ln\left\{
1-2e^{-\frac{2\pi}{W}|x_1-x_2|}\cos\frac{2\pi}{W}(y_1-y_2)+
e^{-\frac{4\pi}{W}|x_1-x_2|}\right\},
\eq
where $\sgn(x)=|x|/x$ is the sign of $x$. The term proportional to
$|x_1-x_2|$ comes from the constant term in the Fourier series
solution, while the other terms sum to give the logarithmic part. 

\subsection{The background} 

The potential of the background (\ref{bp}) is then
\bq
v_b(x)=en_b\frac{\pi}{4}(L^2+4x^2),
\eq
since the second term on the right hand side of Eq. (\ref{cylinder-g})
is an odd function of $x_1-x_2$.

\subsection{The total potential energy} 

The total potential energy (\ref{eq:tpe}) for $n_b=n=N/WL$ can then be
written as  
\bq \label{cilinder-tpe}
V_N/e^2=\sum_{i<j}G(\qq_i,\qq_j)+\pi n\sum_ix_i^2+B_N,
\eq
where $B_N$ is a constant irrelevant to the distribution function.

\subsection{Partition function and densities at a special temperature} 
\label{sec:cilinder-g2}

The energy of Eq. (\ref{cilinder-tpe}) can be inserted into the formula
for the canonical partition function $Z_N(\Gamma)$ at $\Gamma=\beta
e^2=2$ to obtain   
\bq \nonumber
Z_N(2)&=&A_N\int_{-L/2}^{L/2}dx_N\int_{-L/2}^{x_N}dx_{N-1}\cdots
\int_{-L/2}^{x_2}dx_1e^{-2\pi n\sum_ix_i^2}\times\\
&&\int_{-W/2}^{W/2}dy_1\cdots\int_{-W/2}^{W/2}dy_1\prod_{i<j}\left(
e^{\frac{2\pi}{W}(x_i+x_j)}\left|e^{-\frac{2\pi}{W}(x_i-iy_i)}-
e^{-\frac{2\pi}{W}(x_j-iy_j)}\right|^2\right),
\eq
Where $A_N$ is a constant. Now we notice that the $y$-dependent part
of the integrand is contained in the square modulus of a van der Monde
determinant. We use the permutation notation to write the expansion of
the determinant and its conjugate as follows
\bq \nonumber
&&\int_{-W/2}^{W/2}dy_1\cdots\int_{-W/2}^{W/2}dy_N\prod_{i<j}
\left|e^{-\frac{2\pi}{W}(x_i-iy_i)}-
e^{-\frac{2\pi}{W}(x_j-iy_j)}\right|^2=\\
&&\sum_{P,Q}\epsilon(P)\epsilon(Q)\prod_{i=1}^N\left(
e^{-\frac{2\pi x_i}{W}[P(i)+Q(i)-2]}\int_{-W/2}^{W/2}dy_i\,
e^{-\frac{2\pi iy_i}{W}[P(i)-Q(i)]}\right),
\eq
where the sums are over the $N!$ permutations, $\epsilon(P)$ denotes
the sign of permutation $P$. Only permutations for which $P(i)=Q(i)$,
$1\le i\le N$ contribute. Recalling that $n=N/WL$ we obtain
\bq \nonumber
Z_N(2)&=&A_NW^N\sum_P\int_{-L/2}^{L/2}dx_N\int_{-L/2}^{x_N}dx_{N-1}\cdots
\int_{-L/2}^{x_2}dx_1\times \\
&&\prod_{i=1}^Ne^{-2\pi n\left\{x_i^2-
2x_i\frac{L}{2}\left[1-2\frac{P(i)-1}{N}\right]\right\}}.
\eq
For permutation $P$, make the substitution $x_i = z_{P(i)}$, $1\le
i\le N$. We then have a sum over ordered integrals over the $z_i$. The
integrand is the same for each permutation and each possible ordering
of the $z_i$ occurs exactly once. Hence, the sum over ordered
integrals may be written as an unrestricted multiple integral over $[-
  L/2, L/2]^N$. Renaming $z_i = x_i$ for $1\le i\le N$ and using the
appropriately defined $B_N$, we obtain 
\bq
Z_N(2)=B_NW^N\prod_{i=1}^N\int_{-L/2}^{L/2}dx_i
e^{-2\pi n\left[x_i-\frac{L}{2}\left(1-2\frac{i-1}{N}\right)\right]^2}
\eq
This equation describes the canonical partition function for an
assembly of $N$ independent harmonic oscillators with mean position
evenly spaced on $[-L/2,L/2]$. Using the correct form of $B_N$ we may
now take the thermodynamic limit of $-[\ln (Z_N(2)/\caa_R^N)]/N$ to
obtain for the excess free energy per particle $\beta
a_\text{exc}(2)=\beta a_\text{exc,plane}(2)+M$ where
$a_\text{exc,plane}(2)$ is expression (\ref{plane-a2}) with the choice
$L=W/2\pi$ and $M=\pi/6 n W^2$ is a Madelung constant for the
potential in the semiperiodic boundary conditions used.

To calculate the one-particle distribution function in the finite
system we simply leave out the integrations over $x_1$ and
$y_1$. Define $x_0 = -L/2, x_{N+ 1} = L/2$, and the ordering of the
$x$ variables with $x_0\le x_2\le x_3\le\ldots\le x_p\le
x_1<x_{p+1}\le\ldots\le x_N\le x_{N+1}$ . There are $(N- 1)!$
orderings, each giving the same contribution to $g(1;N)$. We use the
van der Monde determinant representation of the integrand and carry
out the integrations over $y_2,\ldots,y_N$ giving $P(i) = Q(i)$, $2\le
i\le N$, and so $P(1) = Q(1)$ by default. Collect all the integrals
with $P(1)= q$ and change variables with $x_i=z_{P(i)}$, $2\le i\le
N$; $P(i)\neq q$ and $x_1=z_q$. This generates ordered integrals
with respect to $(N- l)$ of the $z_i$, all possible orderings
occurring exactly once. An unrestricted integral over 
\bq
\{z_1,\ldots,z_{q-1},z_{q+1},\ldots,z_N\}\in[-L/2,L/2]^{N-1},
\eq 
results. The final form for the one-particle distribution function is
then 
\bq
g(1;N)&=&\frac{1}{Wn}\sum_{q=1}^Ne^{-2\pi n\left[x_1-\frac{L}{2}
\left(1-2\frac{q-1}{N}\right)\right]^2}/I(q,L,N),\\
I(i,L,N)&=&\int_{-L/2}^{L/2}dx\,e^{-2\pi n\left[x_1-\frac{L}{2}
\left(1-2\frac{i-1}{N}\right)\right]^2}.
\eq
The higher orders distribution functions are determined in
Ref. \cite{Choquard83}. 

\section{The sphere} 
\label{sec:ocp-sphere}

The metric tensor in the polar coordinates $\qq=(\theta,\varphi)$ is
now, 
\bq
\gggg=\left(
\begin{array}{cc}
a^2 & 0\\
0 & a^2\sin^2\theta
\end{array}\right),
\eq
where $a$ is the radius of the sphere. The sphere is embeddable in the
three dimensional Euclidean space. The intrinsic Gaussian curvature of the
sphere is a constant $K=1/a^2$ and the surface area of the sphere is
$\caa_\css=4\pi a^2$. So the sphere is the surface of constant positive
curvature by Liebmann's theorem. Also by Minding's theorem we know
that surfaces with the same constant curvature are locally isometric. 

\subsection{The Coulomb potential} 

The Coulomb interaction between a particle at $\rr_i$ and a particle at
$\rr_j$ is
\bq
G(\rr_i,\rr_j)&=&-\ln(r_{ij}/L),\\ \label{sphere-rij}
r_{ij}&=&2a\sin(\theta_{ij}/2),\\
\varphi_{ij}&=&\arccos(\rr_i\cdot\rr_j/a^2),
\eq 
where $\rr_k$ is the three-dimensional vector from the center of the
sphere to particle $k$ on the sphere surface and $r_{ij}$ is the
length of the chord joining $\rr_i$ and $\rr_j$. 

\subsection{The background} 

The background potential is then a constant 
\bq
v_b=en_b2\pi a^2\left(-1+\ln\frac{4a^2}{L^2}\right).
\eq

\subsection{The total potential energy} 

The total potential energy of the system (\ref{eq:tpe}) is then
\bq
V_N/e^2=-\frac{1}{2}\sum_{i<j}\ln\left[\frac{2a^2}{L^2}
(1-\cos\theta_{ij})\right]-\frac{N^2}{4}
\left(1-\ln\frac{4a^2}{L^2}\right).
\eq

\subsection{Partition function and densities at a special temperature} 
\label{sec:sphere-g2}
At $\Gamma=\beta e^2=2$ the excess canonical partition function is
\bq \label{sphere-z}
Z_N(2)=e^{N^2/2}\left(\frac{L}{2a}\right)^N\int\prod_{i=1}^Nd\qq_i
\prod_{j<k}\left(\frac{1-\cos\theta_{jk}}{2}\right),
\eq
where denoting with $g=\det[g_{\mu\nu}]$ we have
$d\qq=dS=\sqrt{g}\,dq^1\,dq^2=a^2\sin\theta\,d\theta\,d\varphi$. Introducing
the Cayley-Klein parameters defined by
\bq
\alpha_i&=&\cos\frac{\theta_i}{2}e^{i\varphi_i/2},\\
\beta_i&=&-i\sin\frac{\theta_i}{2}e^{-i\varphi_i/2},
\eq
we can write
\bq
1-\cos\theta_{ij}=2|\alpha_i\beta_j-\alpha_j\beta_i|^2.
\eq
The integrand of Eq. (\ref{sphere-z}) takes the form
\bq
\prod_{i<j}\left(\frac{1-\cos\theta_{jk}}{2}\right)=
\left|\prod_{k=1}^N\beta_k^{N-1}\prod_{i<j}\left(
\frac{\alpha_i}{\beta_i}-\frac{\alpha_j}{\beta_j}\right)\right|^2.
\eq
The second product in the right hand side of this equation is a van
der Monde determinant. Expanding it and inserting in
Eq. (\ref{sphere-z}) we find
\bq
Z_N(2)=e^{N^2/2}(2\pi L)^Na^NN!\prod_{k=1}^N\frac{(k-1)!(N-k)!}{N!}.
\eq
This result is similar to the result (\ref{plane-z}) on the plane
apart from the fact that now only complete gamma functions are
involved. The excess free energy per particle is identical to the
result (\ref{plane-a2}) for the plane.

For the distribution functions we find \cite{Caillol81}
\bq
g(1,2,\ldots,n;N)=\det[(\alpha_i\bar{\alpha}_j+\beta_i\bar{\beta}_j)^{N-1}],
\eq
where $\bar{\alpha}$ is the complex conjugate of $\alpha$. In
particular 
\bq
g(1;N)&=&1,\\
g(1,2;N)&=&1-\left(\frac{1+\cos\theta_{12}}{2}\right)^{N-1}.
\eq
The system appears to be homogeneous for all $N$ and the distribution
functions are invariant under a rotation of the sphere. 

The thermodynamic limit is obtained defining $\rho_i=R\theta_i$ and
taking the limit $N\to\infty$ and $R\to\infty$ at $n$ constant,
keeping $\rho_i$ and $\varphi_i$ constant for each particle $i$. For
an infinitely large sphere the particles will be situated in the
tangent plane at the North pole and there positions will be
characterized by the polar coordinates $(\rho_i,\varphi_i)$. The
solution for the planar geometry of section \ref{sec:ocp-plane} is
thereby recovered.

\section{The pseudosphere} 
\label{sec:ocp-pseudosphere}

The pseudosphere is non-embeddable in the three dimensional Euclidean
space and it is a non-compact Riemannian surface of constant negative
curvature. Unlike the sphere it has an infinite area and this fact
makes it interesting from the point of view of statistical physics
because one can take the thermodynamic limit on it. 

Riemannian surfaces of negative curvature play a special role in the
theory of dynamical systems \cite{Steiner}. Hadamard study of the geodesic 
flow of a point particle on a such
surface \cite{Hadamard} has been of great importance for the future
development of ergodic theory and of modern chaos
theory. In 1924 the mathematician Emil Artin \cite{Artin} studied 
the dynamics of a free point particle of mass $m$ on a pseudosphere
closed at infinity by a reflective boundary (a billiard). Artin' s
billiard belongs to the class of the so called Anosov systems. All
Anosov systems are ergodic and posses the mixing property 
\cite{Arnold/Avez}. Sinai \cite{Sinai63} translated the problem of the 
Boltzmann-Gibbs gas into a study of the by now famous ``Sinai' s
billiard'', which in turn could relate to Hadamard' s model of 1898. 
Recently, smooth experimental versions of Sinai' s billiard have been 
fabricated at semiconductor interfaces as arrays of nanometer
potential wells and have opened the new field of
mesoscopic physics \cite{Beenakker}.

The following important theorem holds for Anosov systems 
\cite{Arnold/Kozlov},\cite{Anosov}:
\begin{theo}
Let $M$ be a connected, compact, orientable analytic surface
which serves as the configurational manifold of a dynamical system
whose Hamiltonian is $H=K+U$. Let the dynamical system be closed and
its total energy be $h$. Consider the manifold ${\cal M}$ defined by
the Maupertuis Riemannian metric $d\sss^2=2(h-U)K\,dt^2$ on $M$, where
$t$ is time. If the curvature of ${\cal M}$ is negative everywhere
then the dynamical system is an Anosov system and in particular is
ergodic on $M_h=\{{h=H}\}$.  

If the dynamical system is composed of $N$ particles, the same
conclusions hold, we need only require that the curvature be negative
when we keep the coordinates of all the particles but anyone constant. 
\end{theo}

The metric tensor of the {\sl pseudosphere} in the coordinates
$\qq=(\theta,\varphi)$ with $\theta\in [0,\infty[$ is,
\bq
\gggg=\left(
\begin{array}{cc}
a^2 & 0\\
0 & a^2\sinh^2\theta
\end{array}\right),
\eq
where $a$ is the ``radius'' of the pseudosphere. 

Introducing the alternative coordinates $\qq=(r,\varphi)$ with
$r/2a=\tanh(\theta/2)$ we find
\bq
\gggg=\left(
\begin{array}{cc}
  [1-(r/2a)^2]^{-2} & 0\\
0 & r^2[1-(r/2a)^2]^{-2}
\end{array}\right).
\eq
These are the polar coordinates $\omega=(r/2a,\varphi)$ of a disk of
the unitary disk, $\cdd=\{\omega\in\complexn \mid |\omega|<1\}$,
which with such a metric is called the {\sl Poincar\'e disk}.

A third set of coordinates used is $\qq=(x,y)$ obtained from
$(r/2a,\varphi)$ through the Cayley transformation,
\bq
z= x+iy=\frac{\omega+i}{1+i\omega}.
\eq 
which establishes a bijective transformation between the unitary disk
and the complex half plane,
\bq
{\cal H}=\{z=x+iy\mid x\in\realn,y>0\}.
\eq
The center of the unitary disk corresponds to the point $z_o=i$,
``the center of the plane''. The metric becomes,
\bq
\gggg=\left(
\begin{array}{cc}
a^2/y^2 & 0\\
0 & a^2/y^2
\end{array}\right).
\eq
The complex half plane with such a metric is called the {\sl hyperbolic
plane}, and the metric the {\sl Poincar\'{e}' s metric}.

Cayley transformation is a particular M\"{o}bius transformation. 
Poincar\'{e} metric is invariant under M\"{o}bius transformations. 
And any transformation that preserves Poincar\'{e} metric is a 
M\"{o}bius transformation.

The geodesic distance $d_{01}$ between any two points
$\qq_0=(\tau_0,\varphi_0)$ and $\qq_1=(\tau_1,\varphi_1)$ on the
pseudosphere $\cal S$ is given by, 
\bq \label{2dOCP:geodesic-distance}
\cosh(d_{01}/a)=\cosh\tau_1\cosh\tau_0-\sinh\tau_1\sinh\tau_0
\cos(\varphi_1-\varphi_0).
\eq

Given the set of points $\Omega_d$ at a geodesic distance from the
origin less or equal to $d$,
\bq
\Omega_d=\{(\tau,\varphi)\in\css\mid \tau a \le d, \varphi\in[0,2\pi)\},
\eq
that we shall call a disk of radius $d$, we can determine its 
circumference,
\bq\nonumber
{\cal C}&=&\cll(\partial \Omega_d)=a\int_{\tau=d/a}\sqrt{{\dot
\tau}^2+\sinh^2\tau \,\dot{\varphi}^2}\,dt\\
&=&2\pi\, a\, \sinh\left(\frac{d}{a}\right)
\begin{array}[t]{c}
\mbox{{\Huge$\sim$}}\\[-10pt]
\mbox{\scriptsize $d\rightarrow
\infty$}
\end{array}
\pi \, a\, e^{d/a},
\eq
and its area,
\bq\nonumber
{\cal
A}&=&\cvv(\Omega_d)=\int_0^{2\pi}d\varphi\int_0^{d/a}d\tau\,a^2 \sinh\tau\\
&=&4\pi\,a^2\, \sinh^2\left(\frac{d}{2a}\right)
\begin{array}[t]{c}
\mbox{{\Huge$\sim$}}\\[-10pt]
\mbox{\scriptsize $d\rightarrow
\infty$}
\end{array}
\pi\,a^2\, e^{d/a}.
\eq

The Laplace-Beltrami operator on $\cal S$ is,
\bq\nonumber
\Delta&=&\frac{1}{\sqrt{g}}\frac{\partial}{\partial q^\mu}
\left(\sqrt{g}\,g^{\mu\nu}\frac{\partial}{\partial q^\nu}\right)\\
&=&\frac{1}{a^2}\left(\frac{1}{\sinh\tau}\frac{\partial}{\partial
\tau}\sinh\tau \frac{\partial}{\partial\tau}+\frac{1}{\sinh^2\tau}
\frac{\partial^2}{\partial\varphi^2}\right),
\eq
where $g$ is the determinant of the metric tensor 
$g=\det[g_{\mu\nu}]$.

The characteristic component of the Riemann tensor is, 
\bq
R^\tau_{\;\:\varphi\tau\varphi}=-\sinh^2\tau.
\eq
The Gaussian curvature is given by
\bq
{R^{\tau\varphi}}_{\tau\varphi}=g^{\varphi\varphi}{R^\tau}_{\varphi\tau\varphi}
=-\frac{1}{a^2},
\eq
except at its singular cusp, in agreement with Hilbert's theorem.
Contraction gives the components of the Ricci tensor,
\bq
{R^{\tau}}_{\tau}={R^{\varphi}}_{\varphi}=-\frac{1}{a^2}~~,~~~~
{R^{\tau}}_{\varphi}=0,
\eq
and further contraction gives the scalar curvature,
\bq
R=-\frac{2}{a^2}.
\eq

The ensemble of $N$ identical point-wise particles of charge $e$ are
constrained to move in a connected and compact domain $\Omega\subset
{\cal S}$ by an infinite potential barrier on the boundary of the
domain $\partial\Omega$ with a number density $n=N/\cvv(\Omega)$.

\subsection{The Coulomb potential} 

The pair Coulomb potential between two unit charges a geodesic
distance $d$ apart, satisfies Poisson equation on ${\cal S}$, 
\bq
\Delta G(d)=-2\pi\delta^{(2)}(d),
\eq
where $\delta^{(2)}(d_{01})=\delta(\qq_0-\qq_1)/\sqrt{g}$ is the Dirac
delta function on the curved manifold. Poisson equation admits a  
solution vanishing at infinity,
\bq\label{coulomb interaction}
G(d_{ij})=-\ln\left[\tanh\left(\frac{d_{ij}}{2a}\right)\right].
\eq

\subsection{The background} 

If we choose $\Omega=\Omega_{a\tau_0}$, the electrostatic potential of the
background inside $\Omega$ can be chosen (see appendix
\ref{app:pseudosphere-epotb}) to be just a function of $\tau$, 
\bq \label{pbinteraction}
v_b(\tau)=en_b2\pi a^2\left\{\ln\left[\frac{1-\tanh^2(\tau_0/2)}
{1-\tanh^2(\tau/2)}\right]+\sinh^2(\tau_0/2)\ln[\tanh^2(\tau_0/2)]
\right\}.
\eq

\subsection{Ergodicity}
\label{ergodicity}

Consider a closed one component Coulomb plasma of $N$ charges and
total energy $h$, confined in the domain $\Omega_{a\tau_0}\subset{\cal
S}$. Let the coordinates of particle $i$ be $\qq_i=\qqq{i}{\alpha}
\vec{e}_{\alpha}=(\qqq{i}{1},\qqq{i}{2})\in \Omega_{a\tau_0}$, where
$\vec{e}_\alpha=\partial/\partial q^\alpha$ ($\alpha=1,2$) is a
coordinate basis for ${\cal S}$. The trajectory of the dynamical
system,
\bq
{\cal T}_{t_0}=\{q^N(t)\equiv(\qq_1,\ldots,\qq_N)\mid
t\in[0,t_0]\},
\eq
is a geodesic on the $2N$ dimensional manifold $\cal M$ defined by
the metric,
\bq
{\cal G}_{\alpha\beta}=(h-V_N)g_{\mu\nu}(\qq_i)
\otimes\cdots\otimes g_{\mu\nu}(\qq_N),
\eq
on ${\cal S}^N$. We now assume $n_b=n$ and rewrite
$V_N^{pb}=v_1+v_{pb}$ where 
\bq
v_1=N\,2\pi a^2 \,e^2n\,\{\ln[1-\tanh^2(\tau_0/2)]+\sinh^2(\tau_0/2)
\ln[\tanh^2(\tau_0/2)]\},
\eq
is a constant. Since the interaction between the particles is
repulsive we conclude that, up to an additive constant ($V_N^0+v_1$),
the potential $V_N$ is a positive function of the coordinates of the
particles. Since $v_{pb}$ and $V_N^{pp}$ are positive on
$\Omega_{a\tau_0}$ we have,
\bq
{\cal G}_{\alpha\beta}<{\cal G}^\prime_{\alpha\beta}=(h-V_N^0-v_1)
g_{\mu\nu}(\qq_i)\otimes\cdots\otimes g_{\mu\nu}(\qq_N),
\eq
where $\cal G^\prime$ has a negative curvature along the coordinates of 
any given particle. In the next subsection we will calculate the 
curvature of $\cal G$ along the coordinates of one particle. According
to the theorem stated in the introduction we will require the 
curvature to be negative everywhere on ${\cal S}^N$. This will determine
a condition on the kinetic and potential energy of the system, sufficient 
for its ergodicity to hold on $M_h$.

Let $\tilde{\pp}_i=\ppp{i}{\alpha}\tilde{\omega}^\alpha$ be the 
momentum of charge $i$, where $\tilde{\omega}^\alpha=\dt q^\alpha$ are
the 1-forms of the dual coordinate basis, and define 
$p^N(t)\equiv(\tilde{p}_1,\ldots,\tilde{p}_N),
q^N(t)\equiv(\qq_1,\ldots,\qq_N)$. The ergodicity of the 
system tells us that given any dynamical quantity $A(q^N,p^N)$, its
time average,
\bq
\langle A\rangle_t=\lim_{T\rightarrow\infty}\frac{1}{T}\int_0^TA(q^N,p^N)
\,dt,
\eq
coincides with its microcanonical phase space average,
\bq
\langle A\rangle_h=\frac{\int_{M_{ps}}A(q^N,p^N)\,\delta(h-H)\,d^{4N}\mu_{ps}}
{\int_{M_{ps}}\delta(h-H)\,d^{4N}\mu_{ps}},
\eq
where the phase space of the system is,
\bq\nonumber
M_{ps}=\{(q^N,p^N)&\mid& \qq_i\in{\cal S}~~i=1,\ldots,N;\\
&& \ppp{i}{\alpha}\in[-\infty,\infty]~~i=1,\ldots,N, \alpha=1,2\},
\eq
the phase space measure is,
\bq
d^{4N}\mu_{ps}=\prod_{\alpha=1}^2d\qqq{1}{\alpha}\cdots 
d\qqq{N}{\alpha}d\ppp{1}{\alpha}\cdots d\ppp{N}{\alpha}, 
\eq
and $\delta$ is the Dirac delta function.

\subsection{Calculation of the curvature of \texorpdfstring{${\cal M}$}{M}} 

We calculate the curvature of ${\cal M}$ along particle $1$ using 
Cartan structure equations. Let $K=h-U(\tau,\varphi)$ be the kinetic 
energy of the $N$ particle system of total energy $h$, as a function
of the coordinates of particle $1$ (all the other particles having 
fixed coordinates). We choose an orthonormal basis,
\bq
\left\{
\begin{array}{l}
\om{\tau}=a\sqrt{K}\dt \tau\\
\om{\varphi}=a\,\sinh(\tau)\sqrt{K}\dt \varphi
\end{array}
\right.
\eq

By Cartan second theorem we know that the connection 1-form satisfies 
$\omdd{\alpha}{\beta}+\omdd{\beta}{\alpha}=0$. Then we must have,
\bq
\left\{
\begin{array}{l}
\omud{\tau}{\tau}=\omud{\varphi}{\varphi}=0\\
\omud{\tau}{\varphi}=-\omdu{\varphi}{\tau}=-\omud{\varphi}{\tau}
\end{array}
\right.
\eq

We use Cartan first theorem to calculate $\omud{\tau}{\varphi}$,
\bq 
\dt\om{\tau}&=&-\omud{\tau}{\varphi}\wedge\om{\varphi}\\\nonumber
&=&\dt(a\sqrt{K}\dt\tau)\\\nonumber
&=&a\,{K^{\oh}}_{,\varphi}\,\dt\varphi\wedge\dt\tau=0,
\eq
where in the last equality we used the fact that the pair interaction 
is a function of $\varphi_i-\varphi_j$ and that the interaction with the 
background is a function of $\tau$ only (being the system confined in 
a domain which is symmetric under translations of $\varphi$).
We must then conclude that $\omud{\tau}{\varphi}$ is either zero or 
proportional to $\om{\varphi}$. We proceed then calculating,
\bq
\dt\om{\varphi}&=&-\omud{\varphi}{\tau}\wedge\om{\tau}\\\nonumber
&=&\dt(a\,\sinh(\tau)\sqrt{T}\dt\varphi)\\\nonumber
&=&a(\sinh(\tau)K^{\oh})_{,\tau}\,\dt\tau\wedge\dt\varphi,
\eq
which tells us that indeed,
\bq
\omud{\varphi}{\tau}=\frac{(\sinh(\tau)K^{\oh})_{,\tau}}{a\,\sinh(\tau)K}
\om{\varphi}.
\eq

Next we calculate the characteristic component of the curvature 2-form 
$\rc{\alpha}{\beta}=\dt\omud{\alpha}{\beta}+\omud{\alpha}{\gamma}
\wedge\omud{\gamma}{\beta}$,
\bq\nonumber
\rc{\tau}{\varphi}&=&\dt\omud{\tau}{\varphi}\\\nonumber
&=&\dt[-(\sinh(\tau)K^{\oh})_{,\tau}K^{-\oh}\dt\varphi]\\
&=&-\frac{[(\sinh(\tau)K^{\oh})_{,\tau}K^{-\oh}]_{,\tau}}
{a^2 \sinh(\tau)K}\om{\tau}\wedge\om{\varphi}.
\eq
and use Cartan third theorem to read off the characteristic component of 
the Riemann tensor,
\bq
\ruddd{\tau}{\varphi}{\tau}{\varphi}=-\frac{[(\sinh(\tau)K^{\oh})_{,\tau}
K^{-\oh}]_{,\tau}}{a^2 \sinh(\tau)K}.
\eq 

We find then for the scalar curvature,
\bq\nonumber
R=\ruudd{\alpha}{\beta}{\alpha}{\beta}&=&2\ruudd{\tau}{\varphi}{\tau}
{\varphi}\\
&=&-\frac{2}{a^2}\left\{\frac{[(\sinh(\tau)\,K^{\oh})_{,\tau}K^{-\oh}]_{,\tau}}
{\sinh(\tau)\,K}\right\},
\eq
which can be rewritten in terms of the Laplacian as follows,
\bq
R=-\frac{2}{a^2K}\left\{1+\frac{1}{2K}\left[-a^2\Delta U+
\frac{U_{,\varphi\varphi}}{\sinh^2\tau}-\frac{(U_{,\tau})^2}{K}\right]
\right\}.
\eq

For finite values of $h$, the condition for $R$ to be negative on
all the accessible region of ${\cal S}^N$ is then,
\bq \label{ergo}
2\pi a^2\,q^2n-\frac{U_{,\varphi\varphi}}{\sinh^2\tau}+\frac{(U_{,\tau})^2}{K}
<2K.
\eq

\subsection{Ergodicity of the semi-ideal Coulomb plasma}

Consider a one component Coulomb plasma where we switch off the mutual 
interactions between the particles, leaving unchanged the interaction 
between the particles and the neutralizing background
($U=V_N^0+V_N^{pb}$). We will call it the ``semi-ideal'' system. Define,
\bq
\Omega(h,\tau_0)=\{q^N|\qq_i\in\Omega_{a\tau_0}~~\forall i,
h-U(q^N)\ge 0\},
\eq 
and call $h'=h-V_N^0-v_1$ and
\bq \nonumber
f(N)&=&-N\ln[1-\tanh^2(\tau_0/2)]=N\ln[1+\sinh^2(\tau_0/2)]\\
&=&N\ln\left(1+\frac{N}{4\pi a^2 n}\right).
\eq
We will have ($\alpha=2\pi a^2ne^2$)
\bq
r=\inf_{q^N\in\Omega(h,\tau_0)}2K^2=\left\{
\begin{array}[c]{ll}
2[h'-\alpha f(N)]^2 &h'>\alpha f(N)\\
0    &h'\le \alpha f(N)
\end{array}\right.,
\eq

Notice that for large $N$, at constant $n$, we have (see appendix
\ref{app:pseudosphere-epotb}), 
\bq
-V_N^0/\alpha&=&\frac{\alpha}{e^2}\left[-2\frac{N}{4\pi a^2n}+
\ln\left(1+\frac{N}{4\pi a^2n}\right)+\frac{1}{2}\right]+O(1/N),\\
-v_1/\alpha&=&f(N)+N-\frac{\alpha}{e^2}+O(1/N).
\eq
Using the extensive property of the energy we may assume that
$h=Nh_0$, where $h_0$ is the total energy per particle. Then for large
$N$ we will have
\bq \label{ergo hbar}
h'=Nh_0+\alpha f(N)+\left(\frac{\alpha}{e}\right)^2\left[
\ln\left(1+\frac{N}{4\pi a^2n}\right)-\frac{1}{2}\right]+O(1/N)>
\alpha f(N),
\eq
if $h_0\ge 0$.

On the other hand for $h'>\alpha f(N)$ we have
\bq \nonumber
l&=&\sup_{q^N\in\Omega(h,\tau_0)}[\alpha K+(U_{,\tau})^2]\le
\sup_{q^N\in\Omega(h,\tau_0)}[\alpha K]+
\sup_{q^N\in\Omega(h,\tau_0)}[(U_{,\tau})^2]\\
&=&l_+=\alpha h'+\alpha^2\tanh^2(\tau_0/2),
\eq
Condition (\ref{ergo}) is always satisfied if $l<r$. Then the
semi-ideal system is ergodic if, 
\bq\label{ergo ni}
h'>h'_{+}=\alpha f(N)+
\frac{\alpha}{4}\left[1+\sqrt{1+8f(N)+8\tanh^2(\tau_0/2)}\right],
\eq 
where $h'_{+}$ is the largest root of the equation
$l_+=r$. Recalling that $\tanh^2(\tau_0/2)\to 1$ at lare $N$, one can 
verify that, given Eq. (\ref{ergo hbar}), Eq. (\ref{ergo
ni}) must be satisfied at large $N$ if $h_0>0$.

We conclude that the semi ideal system is certainly ergodic if the
total enery is extensive and the total energy per particle is
positive. 

\subsection{Partition function and densities at a special temperature} 
\label{sec:pseudosphere-g2}

Working with the set of coordinates $(r,\varphi)$ on the pseudosphere
(the Poincar\'e disk representation), the particle $i$-particle $j$
interaction term in the Hamiltonian can be written as~\cite{Jancovici1998}
\begin{equation} \label{4.1}
G(d_{ij})=-\ln\tanh(d_{ij}/2a)=
-\ln\left|
\frac{(z_i-z_j)/2a}{1-(z_i\bar{z}_j/4a^2)}
\right|,
\end{equation}
where $z_j=r_j e^{i\varphi_j}$ and $\bar{z}_j$ is the complex conjugate
of $z_j$. This interaction (\ref{4.1}) happens to be the Coulomb 
interaction in a flat disc of radius $2a$ with ideal conductor walls.
Therefore, it is possible to use the techniques which have been 
developed~\cite{Forrester91,Jancovici96} for dealing with ideal conductor 
walls, in the grand canonical ensemble.

The grand canonical partition function of the OCP at fugacity $\zeta$
with a fixed background density $n_b$, when $\Gamma=\beta e^2=2$, is 
\begin{equation}
\Xi(2)=C_0 \left[1+\sum_{N=1}^{\infty} \frac{1}{N!}
\int \prod_{i=1}^N \frac{r_i dr_i d\varphi_i}{[1-(r_i^2/4a^2)]}
\prod_{i<j}\left|\frac{(z_i-z_j)/2a}{1-(z_i
\bar{z}_j/4a^2)}\right|^2
\prod_{i=1}^N \zeta(r_i)\right]
\end{equation}
where for $N=1$ the product $\prod_{i<j}$ must be replaced by 1. 
We have defined a position-dependent fugacity $\zeta(r)=\zeta
[1-r^2/4a^2]^{4\pi n_b a^2-1} e^C$ which includes the
particle-background interaction~(\ref{pbinteraction}) and only one factor
$[1-r^2/4a^2]^{-1}$ from the integration measure
$dS=[1-r^2/4a^2]^{-2}\,d\r$. This should prove to be convenient
later. The $e^C$ factor is
\begin{equation}
\label{eq:cste-exp-c}
e^C=\exp\left[4\pi n_b a^2\left(\ln\cosh^2\frac{\tau_0}{2}
-\sinh^2\frac{\tau_0}{2}\ln\tanh^2\frac{\tau_0}{2}\right)\right]
\end{equation}
which is a constant term coming from the particle-background
interaction term~(\ref{pbinteraction}) and
\begin{equation}
\label{eq:logC0}
\ln C_0=\frac{(4\pi n_b a^2)^2}{2}\left[
\ln\cosh^2\frac{\tau_0}{2}+\sinh^2\frac{\tau_0}{2}\,\left(
\sinh^2\frac{\tau_0}{2}
\ln\tanh^2\frac{\tau_0}{2}-1
\right)
\right]
\end{equation}
which comes from the background-background
interaction. Notice that for large domains, when
$\tau_0\to\infty$, we have
\begin{equation}
\label{eq:cste-exp-c-asymptot}
e^C \sim \left[ \frac{e^{\tau_0+1}}{4} \right]^{4\pi n_b a^2}
\end{equation}
and
\begin{equation}
\label{eq:logC0-asymptot}
\ln C_0 \sim -\frac{(4\pi n_b a^2)^2 e^{\tau_0}}{4}
\end{equation}
Let us define a set of reduced complex coordinates $u_i=(z_i/2a)$
inside the Poincar\'e disk and its corresponding images
$u_i^*=(2a/\bar{z}_i)$ outside the disk. By using the following Cauchy
identity \cite{Aitken} 
\begin{equation} \label{pseudosphere-Cauchy}
\det 
\left(
\frac{1}{u_i-u_j^*}
\right)_{(i,j)\in\{1,\cdots,N\}^2}
=
(-1)^{N(N-1)/2}\:
\frac{\prod_{i<j} (u_i-u_j)(u^*_i-u^*_j)}{\prod_{i,j} (u_i-u_j^*)}
\end{equation}
the particle-particle interaction term together with the
$[1-(r_i^2/4a^2)]^{-1}$ other term from the integration measure can be
cast into the form
\begin{equation}
\prod_{i<j}\left|
\frac{(z_i-z_j)/2a}{1-(z_i \bar{z}_j/4a^2)}
\right|^2
\prod_{i=1}^N [1-(r_i^2/4a^2)]^{-1}
=
\det
\left(
\frac{1}{1-u_i \bar{u}_j}
\right)_{(i,j)\in\{1,\cdots,N\}^2}
\end{equation}
The grand canonical partition function then is
\begin{equation}
\label{eq:GrandPart-prelim}
\Xi(2)= \left[1+\sum_{N=1}^{\infty} \frac{1}{N!}
\int \prod_{i=1}^N d^2\r_i \prod_{i=1}^N \zeta(r_i)
\det
\left(
\frac{1}{1-u_i \bar{u}_j}
\right)
\right]
C_0
\end{equation}

We shall now show that this expression can be reduced to an infinite
continuous determinant, by using a functional integral representation
similar to the one which has been developed for the two-component
Coulomb gas~\cite{ZinnJustin}. Let us consider the Gaussian partition 
function 
\begin{equation}
\label{eq:part-fun-Grass-libre}
Z_0=\int {\cal D}\psi {\cal D}\bar{\psi} \,\exp\left[\int \bar{\psi}(\r)
M^{-1}(z,\bar{z}') \psi(\r')\, d^2\r\, d^2\r'\right]
\end{equation}
The fields $\psi$ and $\bar{\psi}$ are anticommuting Grassmann
variables.  The Gaussian measure in~(\ref{eq:part-fun-Grass-libre}) is
chosen such that its covariance is equal to\footnote{Actually the
operator $M$ should be restricted to act only on analytical functions
for its inverse $M^{-1}$ to exist.}
\begin{equation}
\left<\bar{\psi}(\r_i)\psi(\r_j)\right>
=
M(z_i,\bar{z}_j)=\frac{1}{1-u_i \bar{u}_j}
\end{equation}
where $\langle\ldots\rangle$ denotes an average taken with the Gaussian
weight of (\ref{eq:part-fun-Grass-libre}). By construction we have
\begin{equation} \label{Z_0}
Z_0=\det(M^{-1})
\end{equation}
Let us now consider the following partition function
\begin{equation}
Z=\int {\cal D}\psi {\cal D}\bar{\psi} \exp\left[\int \bar{\psi}(\r)
M^{-1}(z,\bar{z}') \psi(\r') d^2\r d^2\r' +\int \zeta(r)
\bar{\psi}(\r)\psi(\r) \,d^2\r \right]
\end{equation}
which is equal to
\begin{equation}
Z=\det(M^{-1}+\zeta)
\end{equation}
and then
\begin{equation} \label{Z/Z_0}
\frac{Z}{Z_0}=\det[M(M^{-1}+\zeta)]=\det[1+K]
\end{equation}
where 
\begin{equation} \label{K}
K(\r,\r')=M(z,\bar{z}')\,\zeta(r')=
\frac{\zeta(r')}{1-u\bar{u}'}
\end{equation}
The results which follow can also be obtained by exchanging the order of
the factors $M$ and $M^{-1}+\zeta$ in (\ref{Z/Z_0}), i.e. by replacing 
$\zeta(r')$ by $\zeta(r)$ in (\ref{K}), however using the definition 
(\ref{K}) of $K$ is more convenient. Expanding the ratio $Z/Z_0$ in 
powers of $\zeta$ we have
\begin{equation}
\label{eq:expans-ZZ0}
\frac{Z}{Z_0}=
1+
\sum_{N=1}^{\infty}
\frac{1}{N!}
\int \prod_{i=1}^N d^2\r_i
\prod_{i=1}^N
\zeta(r_i)
\left<\bar{\psi}(\r_1)\psi(\r_1)\cdots
\bar{\psi}(\r_N)\psi(\r_N)\right>
\end{equation}
Now, using Wick theorem for anticommuting variables~\cite{ZinnJustin},
we find that
\begin{equation}
\label{eq:WickFerms}
\left<\bar{\psi}(\r_1)\psi(\r_1)\cdots
\bar{\psi}(\r_N)\psi(\r_N)\right>
=\det M(z_i,\bar{z}_j)=\det\left(\frac{1}{1-u_i \bar{u}_j}\right)
\end{equation}
Comparing equations~(\ref{eq:expans-ZZ0})
and~(\ref{eq:GrandPart-prelim}) with the help of
equation~(\ref{eq:WickFerms}) we conclude that
\begin{equation}
\label{eq:Xi-det}
\Xi(2)=C_0\,\frac{Z}{Z_0}=C_0\det(1+K)
\end{equation}

The problem of computing the grand canonical partition function has
been reduced to finding the eigenvalues of the operator $K$. The
eigenvalue problem for $K$ reads
\begin{equation}
\label{eq:vpK}
\int \zeta e^C 
\frac{\left(\displaystyle 1-\frac{r'^2}{4a^2}\right)^{4\pi n_b a^2-1}}
{\displaystyle 1-\frac{z\bar{z}'}{4a^2}}\,
\Phi(\r')\,r'\,dr'd\varphi'
=
\lambda \Phi(\r)
\end{equation}
For $\lambda\neq 0$ we notice from equation~(\ref{eq:vpK}) that
$\Phi(\r)=\Phi(z)$ is an analytical function of $z$. Because of
the circular symmetry it is natural to try
$\Phi(z)=\Phi_{\ell}(z)=z^{\ell}=r^{\ell}e^{i\ell\varphi}$ with $\ell$
a positive integer. Expanding
\begin{equation}
\frac{1}{\displaystyle1-\frac{z\bar{z}'}{4a^2}}=
\sum_{n=0}^{\infty}
\left(\frac{z\bar{z}'}{4a^2}\right)^{n}
\end{equation}
and replacing $\Phi_{\ell}(z)=z^{\ell}$ in equation~(\ref{eq:vpK}) one
can show that $\Phi_{\ell}$ is actually an eigenfunction of $K$ with
eigenvalue
\begin{equation}
\label{eq:lambda-vp-de-K}
\lambda_{\ell}=
4\pi a^2 \zeta e^C
B_{t_0}(\ell+1,4\pi n_b a^2)
\end{equation}
with $t_0=r_0^2/4a^2=\tanh^2 (\tau_0/2)$ and 
\begin{equation}
B_{t_0}(\ell+1,4\pi n_b a^2)=
\int_0^{t_0} (1-t)^{4\pi n_b a^2-1} t^{\ell} \, dt
\end{equation}
the incomplete beta function. So we finally arrive to the result
for the grand potential 
\begin{equation}
\label{eq:grand-potential-somme}
\beta\Omega = -\ln\Xi(2) 
=
-\ln C_0
-
\sum_{\ell=0}^{\infty}
\ln\left(
1+4\pi a^2 \zeta e^C B_{t_0}(\ell+1,4\pi n_b a^2)
\right)
\end{equation}
with $e^C$ and $\ln C_0$ given by equations~(\ref{eq:cste-exp-c})
and~(\ref{eq:logC0}). This result is valid for any disk domain of
radius $a\tau_0$. A more explicit expression of the grand potential
for large domains $\tau_0\to\infty$ can also be obtained
\cite{Fantoni03jsp}.

As usual one can compute the density by doing a functional derivative
of the grand potential with respect to the position-dependent fugacity:
\begin{equation}
\label{eq:n-funct-deriv}
n^{(1)}(\r)=
\left(1-\frac{r^2}{4a^2}\right)^{2}
\zeta(r)\frac{\delta\ln\Xi(2)}{\delta \zeta(r)}
\end{equation}
The factor $[1-(r^2/4a^2)]^2$ is due to the
curvature~\cite{Jancovici1998}, so that $n^{(1)}(\r)\, dS$ is the average
number of particles in the surface element $dS=[1-(r^2/4a^2)]^{-2}\,d\r$.
Using a Dirac-like notation, one can formally write
\begin{equation}
\ln\Xi(2)=\Tr \ln(1+K)+\ln C_0=
\int \left<\r\left|
\ln(1+\zeta(r)M)\right|\r\right>
\,d\r
+\ln C_0
\end{equation}
Then, doing the functional derivative~(\ref{eq:n-funct-deriv}), one
obtains
\begin{equation}
n^{(1)}(\r)= \left(1-\frac{r^2}{4a^2}\right)^{2}
\zeta(r)\left<\r\left| (1+K)^{-1}M \right|\r\right> =
4\pi a\left(1-\frac{r^2}{4a^2}\right)^2\zeta(r)\tilde{G}(\r,\r)
\end{equation}
where we have defined $\tilde{G}(\r,\r')$ by\footnote{The factor $4\pi
a$ is there just to keep the same notations as in 
Ref.~\cite{Jancovici1998}.} $\tilde{G}=(1+K)^{-1}M/4\pi a$. More
explicitly, $\tilde{G}$ is the solution of $(1+K)\tilde{G}=M/4\pi a$, 
that is
\begin{equation}
\label{eq:eq-Green-function}
\tilde{G}(\r,\r') + \zeta e^C \int
\tilde{G}(\r'',\r')\,\frac{\left(\displaystyle
1-\frac{r''^2}{4a^2}\right)^{4\pi n_b a^2-1}}{\displaystyle
1-\frac{z\bar{z}''}{4a^2}}\, d\r'' = \frac{1}{\displaystyle 4\pi a
\left[1-\frac{z\bar{z}'}{4a^2}\right]}
\end{equation}
and the density is given by
\begin{equation}
n^{(1)}(\r)=4\pi a \zeta e^C \left(1-\frac{r^2}{4a^2}\right)^{4\pi n_b
a^2+1}
\tilde{G}(\r,\r) 
\end{equation}
 From the integral equation~(\ref{eq:eq-Green-function}) one can see
that $\tilde{G}(\r,\r')$ is an analytical function of $z$. Trying a solution
of the form
\begin{equation}
\tilde{G}(\r,\r')=\sum_{\ell=0}^{\infty} a_{\ell}(\r') z^{\ell}
\end{equation}
into equation~(\ref{eq:eq-Green-function}) yields
\begin{equation}
\label{eq:solution-G}
\tilde{G}(\r,\r')=
\frac{1}{4\pi a}
\sum_{\ell=0}^{\infty}
\left(\frac{z\bar{z}'}{4a^2}\right)^{\ell}
\frac{1}{1+4\pi a^2 \zeta e^{C} B_{t_0} (\ell+1,4\pi n_b a^2)}
\end{equation}
Then the density is given by
\begin{equation}
\label{eq:densite-somme}
n^{(1)}(r)=\zeta e^C \left(1-\frac{r^2}{4a^2}\right)^{4\pi n_b a^2+1}
\sum_{\ell=0}^{\infty}
\left(\frac{r^2}{4a^2}\right)^{\ell}
\frac{1}{1+4\pi a^2 \zeta e^{C} B_{t_0} (\ell+1,4\pi n_b a^2)}
\end{equation}
After some calculation (see appendix \ref{app:pseudosphere-flat}), it
can be shown that, in the limit $a\rightarrow\infty$, the result for
the flat disk in the canonical ensemble \cite{Jancovici81} 
\begin{equation} \label{surf}
\frac{n^{(1)}(r)}{n_b}=\exp(-\pi n_b r^2)
\sum_{\ell =0}^{N_b-1}\frac{(\pi n_b r^2)^{\ell}}{\gamma
(\ell +1,\,N_b)}
\end{equation}
is recovered. up to a correction due to the non-equivalence of ensembles
in finite systems. In (\ref{surf}), $\gamma$ is the incomplete gamma
function
\begin{equation} \label{gamma}
\gamma(\ell +1,\,x)=\int_0^x t^{\ell}e^{-t}dt
\end{equation} 
In that flat-disk case, in the thermodynamic limit (half-space), 
$n^{(1)}(r_0)=n_{\mathrm{contact}}\rightarrow n_b\ln\,2$.

In a flat space, the neighborhood of the boundary of a large domain has
a volume which is a negligible fraction of the whole volume. This is
why, for the statistical mechanics of ordinary fluids, usually there is
a thermodynamic limit: when the volume becomes infinite, quantities such
as the free energy per unit volume or the pressure have a unique limit,
independent of the domain shape and of the boundary conditions. However,
even in a flat space, the one-component plasma is special. For the OCP, 
it is possible to define several non-equivalent pressures, some of
which, for instance the kinetic pressure \cite{Fantoni03jsp},
obviously are surface-dependent even in the infinite-system limit.

Even for ordinary fluids, statistical mechanics on a pseudosphere is
expected to have special features, which are essentially related to the 
property that, for a large domain, the area of the neighborhood of the 
boundary is of the same order of magnitude as the whole area. Although 
some bulk properties, such as correlation functions far away from the 
boundary, will exist, extensive quantities such as the free energy 
or the grand potential are strongly dependent on the boundary 
neighborhood and surface effects. For instance, in the large-domain
limit, no unique limit is expected for the free energy per unit area
$F/{\cal A}$ or the pressure 
$-(\partial F/\partial {\cal A})_{\beta,N}$.

In the present section, we have studied the 2D OCP on a pseudosphere, for
which surface effects are expected to be important for both reasons:
because we are dealing with a one-component plasma and because the space
is a pseudosphere. Therefore, although the
correlation functions far away from the boundary have unique
thermodynamic limits~\cite{Jancovici1998}, many other properties are 
expected to depend on the domain shape and on the boundary conditions.
This is why we have considered a special well-defined geometry: the
domain is a disk bounded by a plain hard wall, and we have studied the
corresponding large-disk limit. Our results have been derived only for 
that geometry.

\section{The Flamm paraboloid} 
\label{sec:ocp-flamm}

\begin{figure}[htbp]
\begin{center}
\includegraphics[width=10cm]{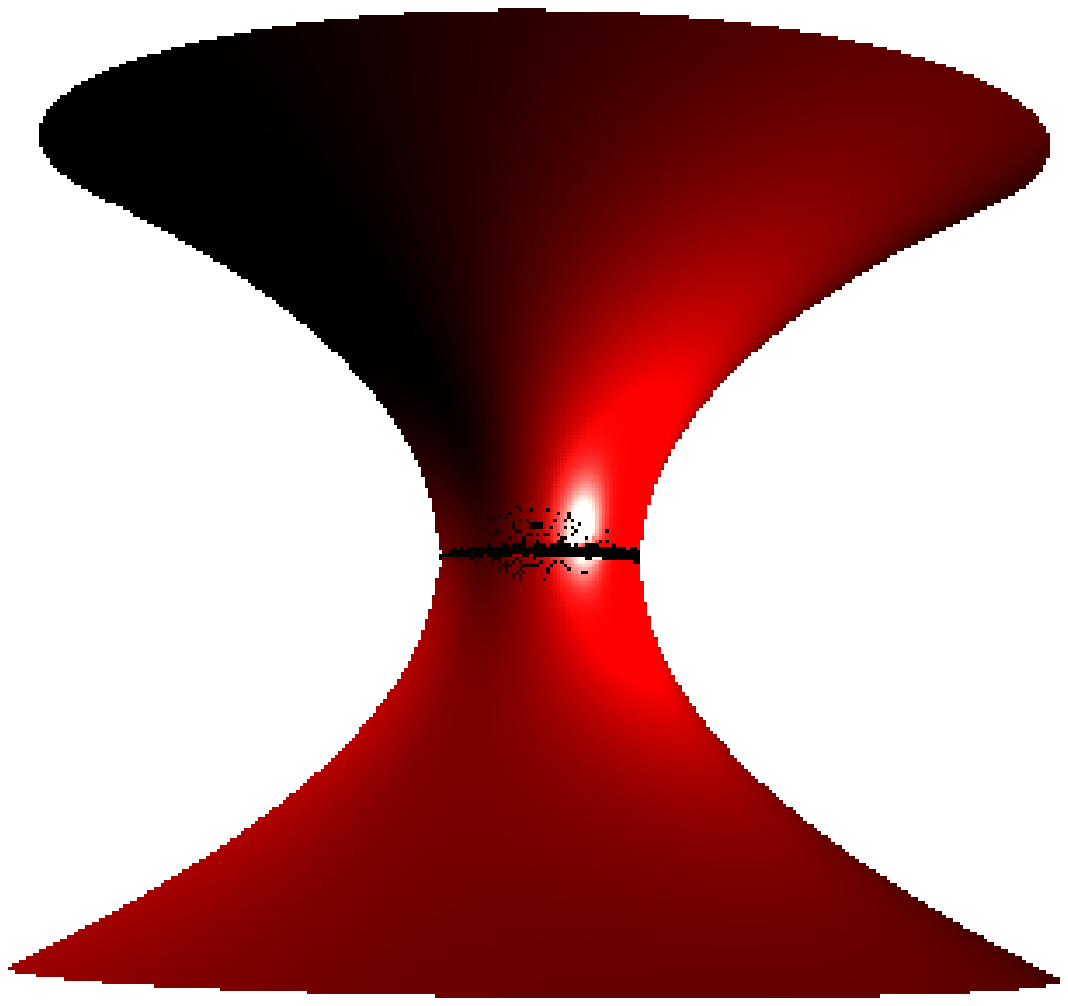}
\end{center}
\caption{The Riemannian surface ${\cal S}$ of Eq. (\ref{flamm-surf}).}
\label{fig:surf}
\end{figure}

The metric tensor of {\sl Flamm's paraboloid} in the coordinates
$\qq=(r,\varphi)$ is now, 
\bq \label{flamm-g}
\gggg=\left(
\begin{array}{cc}
(1-2M/r)^{-1} & 0\\
0 & r^2
\end{array}\right),
\eq
where $M$ is a constant. This is an embeddable surface in the
three-dimensional Euclidean space with cylindrical coordinates
$(r,\varphi,Z)$ with $d\mathbf{s}^2=dZ^2+dr^2+r^2d\varphi^2$, whose
equation is
\begin{eqnarray} \label{flamm-surf} 
Z(r)=\pm 2\sqrt{2M(r-2M)}.  
\end{eqnarray} 
This surface is illustrated in Fig. \ref{fig:surf}. It has a hole of
radius $2M$. As the hole shrinks to a point 
(limit $M\to 0$) the surface becomes flat. We will from now on call the
$r=2M$ region of the surface its ``horizon''. The Schwarzschild
geometry in general relativity is a vacuum solution to the Einstein
field equation which is spherically symmetric and in a two dimensional
world its spatial part is a Flamm paraboloid $\css$. In general
relativity, $M$ (in appropriate units) is the mass of the source of the
gravitational field. 

The ``Schwarzschild wormhole'' provides a path from the upper
``universe'' $\css_+$ ($Z>0$) to the lower one $\css_-$ ($Z<0$). These
are both multiply connected surfaces. We will study the OCP on a
single universe, on the whole surface, and on a single universe with
the ``horizon'' (the region $r=2M$) grounded. 

Since the curvature of the surface is not a constant but varies from
point to point, the plasma will not be uniform even in the
thermodynamic limit.

The system of coordinates $(r,\varphi)$ with the
metric~(\ref{flamm-g}) has the disadvantage that it requires two
charts to cover the whole surface ${\cal S}$. It can be more
convenient to use the variable 
\begin{eqnarray} \label{u}
u=\frac{Z}{4M}=
\pm\sqrt{\frac{r}{2M}-1}
\end{eqnarray}
instead of $r$. Replacing $r$ as a function of $Z$ using
equation~(\ref{flamm-surf}) gives the following metric when using the
system of coordinates $\qq=(u,\varphi)$,
\bq
\gggg=\left(
\begin{array}{cc}
(4M)^2(1+u^2) & 0\\
0 & 4M^2(1+u^2)^2
\end{array}\right),
\eq
The region $u>0$ corresponds to ${\cal S}_+$ and the region $u<0$
to ${\cal S}_-$.

Let us consider that the OCP is confined in a disk defined as
\begin{eqnarray}
\Omega_R^{+}=\{\qq=(r,\varphi)\in {\cal S}_+ |0\le\varphi\le 2\pi, 
2M\le r\le R\}~.
\end{eqnarray}
The area of this disk is given by
\begin{eqnarray} \label{volume}
\mathcal{A}_R=\int_{\Omega_R} dS=\pi\left[\sqrt{R(R-2M)}(3M+R)+
6M^2\ln\left(\frac{\sqrt{R}+\sqrt{R-2M}}{\sqrt{2M}}\right)\right]~,
\end{eqnarray}
where $dS=\sqrt{g}\,dr\,d\varphi$ and $g=\det[g_{\mu\nu}]$. The
perimeter is $\mathcal{C}_R=2\pi R$.

The Riemann tensor characteristic component is 
\begin{eqnarray} 
  {R^r}_{\varphi r\varphi}=-\frac{M}{r}~.
\end{eqnarray} 
The scalar curvature is then given by the following indexes
contractions
\begin{eqnarray}
\mathcal{R}={R^\mu}_\mu={R^{\mu\nu}}_{\mu\nu}=2{R^{r\varphi}}_{r\varphi}
=2g^{\varphi\varphi}{R^r}_{\varphi r\varphi}= -\frac{2M}{r^3}~, 
\end{eqnarray}
and the (intrinsic) Gaussian curvature is
$K=\mathcal{R}/2=-M/r^3$. The (extrinsic) mean curvature of the
manifold turns out to be $H=-\sqrt{M/8r^3}$.

The Euler characteristic (\ref{ec}) of the disk $\Omega_R^{+}$ turns
out to be $\chi=0$, in agreement with the Gauss-Bonnet theorem
$\chi=2-2h-b$ where $h=0$ is the number of handles and $b=2$ the
number of boundaries. 

We can also consider the case where the system is confined in a
``double'' disk
\begin{equation}
\Omega_R=\Omega_R^{+}\cup\Omega_R^{-}\,,  
\end{equation}
with $\Omega_R^{-}=\{\qq=(r,\varphi)\in {\cal S}_{-} |0\le\varphi\le
2\pi, 2M\le r\le R\}$, the disk image of $\Omega_{R}^{+}$ on the lower
universe ${\cal S}_{-}$ portion of ${\cal S}$. The Euler
characteristic of $\Omega_R$ is also $\chi=0$.

The fact that the Euler characteristic is zero implies that the
asymptotic expansion in the thermodynamic limit of the free energy  
does not exhibit the logarithmic corrections predicted by
Ref. \cite{Jancovici94}. 

The Laplacian for a function $f$ is
\begin{eqnarray} \nonumber
\Delta f&=&
\frac{1}{\sqrt{g}}\frac{\partial}{\partial q^\mu}
\left(\sqrt{g}\,g^{\mu\nu}\frac{\partial}{\partial q^\nu}\right)f\\
&=&\left[\left(1-\frac{2M}{r}\right)\frac{\partial^2}{\partial r^2}
+\frac{1}{r^2}\frac{\partial^2}{\partial \varphi^2}
+\left(\frac{1}{r}-\frac{M}{r^2}\right)\frac{\partial}{\partial r}\right]
f~,
\end{eqnarray}
where $\qq\equiv(r,\varphi)$. In 
appendix~\ref{app:flamm-green}, we show how, finding the
Green function of the Laplacian, naturally leads to consider the
system of coordinates $(x,\varphi)$, with
\begin{equation}
  x=(\sqrt{u^2+1}+u)^{2}
  \,.
\end{equation}
The range for the variable $x$ is $\left]0,+\infty\right[$. The lower
    paraboloid ${\cal S}_{-}$ corresponds to the region $0<x<1$ and
    the upper one ${\cal S}_{+}$ to the region $x>1$. A point in the
    upper paraboloid with coordinate $(x,\varphi)$ has a mirror image
    by reflection ($u\to -u$) in the lower paraboloid, with
    coordinates $(1/x,\varphi)$, since if
\begin{equation}
  x=(\sqrt{u^2+1}+u)^{2}
\end{equation}
then
\begin{equation}
  \frac{1}{x}=(\sqrt{u^2+1}-u)^{2}
  \,.
\end{equation}
In the upper paraboloid ${\cal S}_{+}$, the new coordinate $x$ can be
expressed in terms of the original one, $r$, as
\begin{equation} \label{flamm-x}
x=\frac{(\sqrt{r}+\sqrt{r-2M})^2}{2M}  
\,.
\end{equation}

Using this system of coordinates, the metric takes the form of a
flat metric multiplied by a conformal factor
\begin{equation}
  \label{eq:metric-in-x}
\gggg=\left(
\begin{array}{cc}
(M/2)^2(1+1/x)^4 & 0\\
0 & (M/2)^2(1+1/x)^4x^2
\end{array}\right),
\end{equation}
The Laplacian also takes a simple form
\begin{equation}
  \Delta f =\frac{4}{M^2\left(1+\frac{1}{x}\right)^4}
  \,\Delta_{\mathrm{flat}}f
\end{equation}
where 
\begin{equation}
  \Delta_{\mathrm{flat}}f=
  \frac{\partial^2 f}{\partial x^2}
  +\frac{1}{x}\frac{\partial f}{\partial x}
  +\frac{1}{x^2}\frac{\partial^2 f}{\partial \varphi^2}
\end{equation}
is the Laplacian of the flat Euclidean space $\mathbb{R}^2$.  The
determinant of the metric is now given by $g=[M^2 x (1+x^{-1})^4
/4]^2$.

With this system of coordinates $(x,\varphi)$, the area of a disk
$\Omega_{R}^{+}$ of radius $R$, in the original system $(r,\varphi)$,
is given by
\begin{equation}
  \label{eq:arpx}
  \mathcal{A}_R=\frac{\pi M^2}{4}\,p(x_m)
\end{equation}
with
\begin{equation}
  \label{eq:p(x)}
  p(x)=x^2+8 x-\frac{8}{x}-\frac{1}{x^2}+12\ln x
\end{equation}
and $x_m=(\sqrt{R}+\sqrt{R-2M})^2/(2M)$.

The Coulomb potential $G(x,\varphi;x_0,\varphi_0)$ created at
$(x,\varphi)$ by a unit charge at $(x_0,\varphi_0)$ is given by
the Green function of the Laplacian
\begin{equation}
  \label{eq:LaplaceGreen}
  \Delta G(x,\varphi;x_0,\varphi_0)
  =-2\pi \delta^{(2)}(x,\varphi;x_0,\varphi_0)
\end{equation}
with appropriate boundary conditions. The Dirac distribution on $\css$
is given by
\begin{equation}
  \delta^{(2)}(x,\varphi;x_0,\varphi_0)
  =\frac{4}{M^2 x (1+x^{-1})^4}\,\delta(x-x_0)\delta(\varphi-\varphi_0)
\end{equation}

Notice that using the system of coordinates $(x,\varphi)$ the
Laplacian Green function equation takes the simple form
\begin{equation}
  \label{eq:GreenLaplace-flat}
  \Delta_{\mathrm{flat}}  G(x,\varphi;x_0,\varphi_0)
    =-2\pi\frac{1}{x}\,\delta(x-x_0)\delta(\varphi-\varphi_0)
\end{equation}
which is formally the same Laplacian Green function equation for
flat space.

We shall consider three different situations: when the
particles can be in the whole surface ${\cal S}$, or when the
particles are confined to the upper paraboloid universe ${\cal
  S}_{+}$, confined by a hard wall or by a grounded perfect conductor.

The geodesic distance on the Flamm paraboloid is determined in
appendix \ref{app:flamm-geodesic}.  

\subsection{Coulomb potential in the whole surface (ws)}  

To complement the Laplacian Green function
equation~(\ref{eq:LaplaceGreen}), we impose the usual boundary
condition that the electric field $-\nabla G$ vanishes at infinity
($x\to\infty$ or $x\to0$). Also, we require the usual interchange
symmetry $G(x,\varphi;x_0,\varphi_0)=G(x_0,\varphi_0;x,\varphi)$ to be
satisfied. Additionally, due to the symmetry between each universe
${\cal S}_{+}$ and ${\cal S}_{-}$, we require that the Green function
satisfies the symmetry relation
\begin{equation}
  \label{eq:symmetry-S+S-}
  G^{\mathrm{ws}}(x,\varphi;x_0,\varphi_0)=
  G^{\mathrm{ws}}(1/x,\varphi;1/x_0,\varphi_0)
\end{equation}

The Laplacian Green function equation~(\ref{eq:LaplaceGreen}) can be
solved, as usual, by using the decomposition as a Fourier
series, as shown in appendix \ref{app:flamm-green}. Since
equation~(\ref{eq:LaplaceGreen}) reduces to the flat 
Laplacian Green function equation~(\ref{eq:GreenLaplace-flat}), the
solution is the standard one
\begin{equation}
  \label{eq:Fourier}
  G(x,\varphi;x_0,\varphi_0)=
  \sum_{n=1}^{\infty}
  \frac{1}{n}\left(\frac{x_{<}}{x_{>}}\right)^{n}
  \cos\left[ n(\varphi-\varphi_0)\right]
  +g_0(x,x_0)
\end{equation}
where $x_{>}=\max(x,x_0)$ and $x_{<}=\min(x,x_0)$. The Fourier
coefficient for $n=0$, has the form
\begin{equation}
  g_0(x,x_0)=
  \begin{cases}
  a_0^{+}\ln x+b_0^{+}\,,&x>x_0\\    
  a_0^{-}\ln x+b_0^{-}\,,&x<x_0\,.
  \end{cases}
\end{equation}
The coefficients $a_0^{\pm},b_0^{\pm}$ are determined by the boundary
conditions that $g_0$ should be continuous at $x=x_0$, its derivative
discontinuous $\partial_x g_0|_{x=x_0^{+}}-\partial_x
g_0|_{x=x_0^{-}}=-1/x_0$, and the boundary condition at infinity
$\nabla g_0|_{x\to\infty}=0$ and $\nabla g_0|_{x\to 0}=0$.
Unfortunately, the boundary condition at infinity is trivially
satisfied for $g_0$, therefore $g_0$ cannot be determined only with
this condition. In flat space, this is the reason why the Coulomb
potential can have an arbitrary additive constant added to
it. However, in our present case, we have the additional symmetry
relation~(\ref{eq:symmetry-S+S-}) which should be satisfied. This
fixes the Coulomb potential up to an additive constant $b_0$. We find
\begin{equation}
  \label{eq:g0-ws}
  g_0(x,x_0)=-\frac{1}{2}\ln\frac{x_{>}}{x_{<}} + b_0\,,
\end{equation}
and summing explicitly the Fourier series~(\ref{eq:Fourier}), we obtain
\begin{equation}
\label{eq:Gws}
G^{\mathrm{ws}}(x,\varphi;x_0,\varphi_0)=
-\ln\frac{\left|z-z_0\right|}{\sqrt{\left|z z_0  \right|}}
+b_0
~,
\end{equation}
where we defined $z=xe^{i\varphi}$ and $z_0=x_0e^{i\varphi_0}$. Notice
that this potential does not reduce exactly to the flat one when
$M=0$. This is due to the fact that the whole surface $\mathcal{S}$ in
the limit $M\to0$ is not exactly a flat plane $\mathbb{R}^2$, but
rather it is two flat planes connected by a hole at the origin, this
hole modifies the Coulomb potential.

\subsection{Coulomb potential in the half surface (hs) confined by
  hard walls}  

We consider now the case when the particles are restricted to live in
the half surface ${\cal S}_{+}$, $x>1$, and they are confined by a
hard wall located at the ``horizon'' $x=1$. The region $x<1$ (${\cal
  S}_{-}$) is empty and has the same dielectric constant as the upper
region occupied by the particles. Since there are no image charges,
the Coulomb potential is the same $G^{\mathrm{ws}}$ as above. However,
we would like to consider here a new model with a slightly different
interaction potential between the particles. Since we are dealing only
with half surface, we can relax the symmetry
condition~(\ref{eq:symmetry-S+S-}). Instead, we would like to consider
a model where the interaction potential reduces to the flat Coulomb
potential in the limit $M\to0$. The solution of the Laplacian Green
function equation is given in Fourier series by
equation~(\ref{eq:Fourier}). The zeroth order Fourier component $g_0$ can
be determined by the requirement that, in the limit $M\to0$, the
solution reduces to the flat Coulomb potential
\begin{equation}
  G^{\mathrm{flat}}(\rr,\rr')=-\ln\frac{|\rr-\rr'|}{L} 
\end{equation}
where $L$ is an arbitrary constant length. Recalling that $x\sim 2r/M$,
when $M\to0$, we find
\begin{equation}
  \label{eq:g0-hs}
  g_0(x,x_0)=-\ln x_{>}-\ln\frac{M}{2L}
\end{equation}
and 
\begin{equation}
  \label{flamm-cgreen}
  G^{\mathrm{hs}}
  (x,\varphi;x_0,\varphi_0)=-\ln |z-z_0|-\ln\frac{M}{2L}
  \,.
\end{equation}

\subsection{Coulomb potential on half surface with a grounded horizon
  (gh)}

Let us consider now that the particles are confined to ${\cal S}_{+}$
by a grounded perfect conductor at $x=1$ which imposes Dirichlet
boundary condition to the electric potential. The Coulomb potential
can easily (see appendix \ref{app:flamm-green}) be found from the
Coulomb potential $G^{\mathrm{ws}}$~(\ref{eq:Gws}) using the method of
images 
\begin{equation}
  \label{flamm-ghgreen}
  G^{\mathrm{gh}}(x,\varphi;x_0,\varphi_0)=
  -\ln\frac{|z-z_0|}{\sqrt{|z z_0|}}
  +\ln\frac{|z-\bar{z}_0^{-1}|}{\sqrt{|z \bar{z}_0^{-1}|}}
  =
  -\ln\left|\frac{z-z_0}{1-z\bar{z}_0}\right|
\end{equation}
where the bar over a complex number indicates its complex
conjugate. We will call this the grounded horizon Green
function. Notice how its shape is the same of the Coulomb potential on
the pseudosphere~\cite{Fantoni03jsp} or in a flat disk confined by
perfect conductor boundaries~\cite{Jancovici96}.

This potential can also be found using the Fourier
decomposition. Since it will be useful in the following, we note that
the zeroth order Fourier component of $G^{\mathrm{gh}}$ is
\begin{equation}
  \label{eq:g0-gh}
  g_0(x,x_0)=\ln x_{<}\,.
\end{equation}

\subsection{The background}

The Coulomb potential generated by the background, with a constant
surface charge density $\rho_b$ satisfies the Poisson equation, for
$r>2M$, 
\begin{eqnarray}
\Delta v_b =-2\pi \rho_b~,
\end{eqnarray}
Assuming that the system occupies an area $\mathcal{A}_R$, the
background density can be written as
$\rho_b=-qN_b/\mathcal{A}_R=-qn_b$, where we have defined here
$n_b=N_b/\mathcal{A}_R$ the number density associated to the
background. For a neutral system $N_b=N$. The Coulomb potential of the
background can be obtained by solving Poisson equation with the
appropriate boundary conditions for each case. Also, it can be
obtained from the Green function computed in the previous section
\begin{equation}
  v_b(x,\varphi)=\int G(x,\varphi;x',\varphi') \rho_b \,dS'
\end{equation}
This integral can be performed easily by using the Fourier series
decomposition~(\ref{eq:Fourier}) of the Green function $G$. Recalling
that $dS=\frac{1}{4}M^2 x (1+x^{-1})^4\,dx\,d\varphi$, after the angular
integration is done, only the zeroth order term in the Fourier series
survives
\begin{equation}
  v_b(x,\varphi)=\frac{\pi  \rho_b M^2}{2}
  \int_{1}^{x_m} g_0(x,x') \, x' \left(1+\frac{1}{x'}\right)^4 \,dx'
  \,.
\end{equation}
The previous expression is for the half surface case and the grounded
horizon case. For the whole surface case, the lower limit of
integration should be replaced by $1/x_m$, or, equivalently, the
integral multiplied by a factor two.

Using the explicit expressions for $g_0$,~(\ref{eq:g0-ws}),
(\ref{eq:g0-hs}), and (\ref{eq:g0-gh}) for each case, we find, for the
whole surface,
\begin{equation}
  v_b^{\mathrm{ws}}(x,\varphi)=-\frac{\pi \rho_b M^2}{8}
  \left[ h(x)-h(x_m) +2 p(x_m) \ln x_m - 4b_0 p(x_m)\right]
\end{equation}
where $p(x)$ was defined in equation~(\ref{eq:p(x)}), and
\begin{equation}
  h(x)=x^2+16x+\frac{16}{x}+\frac{1}{x^2}+12(\ln x)^2 - 34
  \,.
\end{equation}
Notice the following properties satisfied by the functions $p$ and $h$
\begin{equation}
  \label{eq:p-h-symmetry}
  p(x)=-p(1/x) \,,\qquad h(x)=h(1/x)
\end{equation}
and
\begin{equation}
  \label{eq:p-h-deriv}
  p(x)=x h'(x)/2 \,,\qquad p'(x)=2x\left(1+\frac{1}{x}\right)^4
\end{equation}
where the prime stands for the derivative.

The background potential for the half surface case, with the pair
potential $-\ln(|z-z'|M/2L)$ is
\begin{equation}
  v_b^{\mathrm{hs}}(x,\varphi)=-\frac{\pi \rho_b M^2}{8}
  \left[ h(x)-h(x_m) + 2 p(x_m) \ln\frac{x_m M}{2L} \right]
  \,.
\end{equation}
Also, the background potential in the half surface case, but with the pair
potential $-\ln(|z-z'|/\sqrt{|zz'|})+b_0$ is
\begin{equation}
  v_b^{\overline{\mathrm{hs}}}(x,\varphi)=
  -\frac{\pi \rho_b M^2}{8}
  \left[ h(x)-\frac{h(x_m)}{2}+p(x_m)\left(\ln\frac{x_m}{x}-2
    b_0\right)
    \right]\,.
\end{equation}

Finally, for the grounded horizon case,
\begin{equation}
  v_b^{\mathrm{gh}}(x,\varphi)=-\frac{\pi \rho_b M^2}{8}
  \left[ h(x) - 2 p(x_m) \ln x\right]
  \,.
\end{equation}

\subsection{Partition function and densities at a special temperature} 
\label{sec:flamm-g2}

We will now show how at the special value of the coupling constant
$\Gamma=\beta e^2=2$ the partition function and $n$-body correlation
functions can be calculated exactly. 

\subsubsection{The 2D OCP on half surface with potential
  \texorpdfstring{$-\ln|z-z'|-\ln M/(2L)$}{...}} 
\label{sec:half-surface-1} 

For this case, we work in the canonical ensemble with $N$ particles
and the background neutralizes the charges: $N_b=N$, and
$n=N/\mathcal{A}_R=n_b$. The potential energy of the system takes the
explicit form
\begin{eqnarray}
  V^{\mathrm{hs}} & = &
  -e^2\sum_{1\leq i<j\leq N}\ln|z_i-z_j|
  +\frac{e^2}{2}\alpha \sum_{i=1}^N h(x_i)
  +\frac{e^2}{2} N \ln \frac{M}{2L}
  -\frac{e^2}{4}N\alpha h(x_m)
  \nonumber\\
  &&+\frac{e^2}{2} N^2 \ln x_m
  -\frac{e^2}{4} \alpha^2 \int_{1}^{x_m} h(x) p'(x)\,dx 
  \label{eq:pot1}
\end{eqnarray}
where we have used the fact that $dS=\pi M^2 x (1+x^{-1})^4\,dx/2=\pi
M^2 p'(x)\,dx/4$, and we have defined
\begin{equation}
  \alpha=\frac{\pi n_b M^2}{4}\,.
\end{equation}
Integrating by parts the last term of~(\ref{eq:pot1}) and
using~(\ref{eq:p-h-deriv}), we find
\begin{eqnarray}
  V^{\mathrm{hs}} & = &
  -e^2\sum_{1\leq i<j\leq N}\ln|z_i-z_j|
  +\frac{e^2}{2}\alpha \sum_{i=1}^N h(x_i)
  +\frac{e^2}{2} N \ln \frac{M}{2L}
  +\frac{e^2}{2} N^2 \ln x_m
  \nonumber\\
  &&
  +\frac{e^2}{2}\alpha^2\int_{1}^{x_m} \frac{[p(x)]^2}{x}\,dx
  -\frac{e^2}{2} N \alpha h(x_m)
  \,.
  \label{eq:Vhs}
\end{eqnarray}

When $\beta e^2=2$, the canonical partition function can be written as
\begin{equation}
  Z^{\mathrm{hs}}=\frac{1}{\Lambda^{2N}}\,Z_0^{\mathrm{hs}} 
  \exp(-\beta F_0^{\mathrm{hs}})
\end{equation}
with 
\begin{equation}
  \label{eq:F0}
  -\beta F_0^{\mathrm{hs}}=
   -N \ln \frac{M}{2L}
  - N^2 \ln x_m
  -\alpha^2\int_{1}^{x_m} \frac{[p(x)]^2}{x}\,dx
  + N \alpha h(x_m)
\end{equation}
and
\begin{equation}
  Z_0^{\mathrm{hs}}(2)=\frac{1}{N!}\int
  \prod_{i=1}^N dS_{i}\, e^{-\alpha h(x_i)} 
  \prod_{1\leq i<j \leq N} |z_i-z_j|^2
  \,.
\end{equation}
where $\Lambda=\sqrt{2\pi\beta\hbar^2/m}$ is the de Broglie thermal
wavelength. $Z_0(2)$ can be computed using the original method for the
OCP in flat space~\cite{Jancovici81b,Alastuey81}, which was
originally introduced in the context of random matrices~\cite{Mehta91,
  Ginibre65}, and which was presented in section
\ref{sec:ocp-plane}. By expanding the van der Monde determinant 
$\prod_{i<j}(z_i-z_j)$ and performing the integration over the angles,
the partition function can be written as
\begin{eqnarray} \label{cpf}
Z_0^{\mathrm{hs}}(2)&=&
\prod_{k=0}^{N-1}{\cal B}_N(k)~,
\end{eqnarray}
where
\begin{eqnarray} 
{\cal B}_N(k)&=&
\int x^{2k} e^{-\alpha h(x)}\,dS
\\
&=&
\frac{\alpha}{n_b}\int_{1}^{x_m}
x^{2k} e^{-\alpha h(x)} p'(x)\,dx
\,.
\label{flamm-gamma}
\end{eqnarray}

In the flat limit $M\to0$, we have $x\sim 2r/M$, with $r$ the radial
coordinate of flat space $\mathbb{R}^2$, and $h(x)\sim p(x)\sim
x^2$. Then, $\mathcal{B}_N$ reduces to
\begin{equation}
  \mathcal{B}_N(k)\sim\frac{1}{n_b \alpha^k}\, \gamma(k+1,N)
\end{equation}
where $\gamma(k+1,N)=\int_0^{N} t^k e^{-t}\,dt$ is the incomplete
Gamma function. Replacing into~(\ref{cpf}), we recover the partition
function (\ref{plane-a2}) for the OCP in a flat disk of radius
$R$~\cite{Alastuey81} 
\begin{equation}
  \ln Z^{\mathrm{hs}}(2)=\frac{N}{2}\ln\frac{\pi L^2}{n_b\Lambda^4}
  +\frac{3N^2}{4}-\frac{N^2}{2}\ln N
  +\sum_{k=1}^{N} \ln \gamma(k,N)
  \,.
\end{equation}

\begin{figure}[htbp]
\begin{center}
\includegraphics[width=8cm]{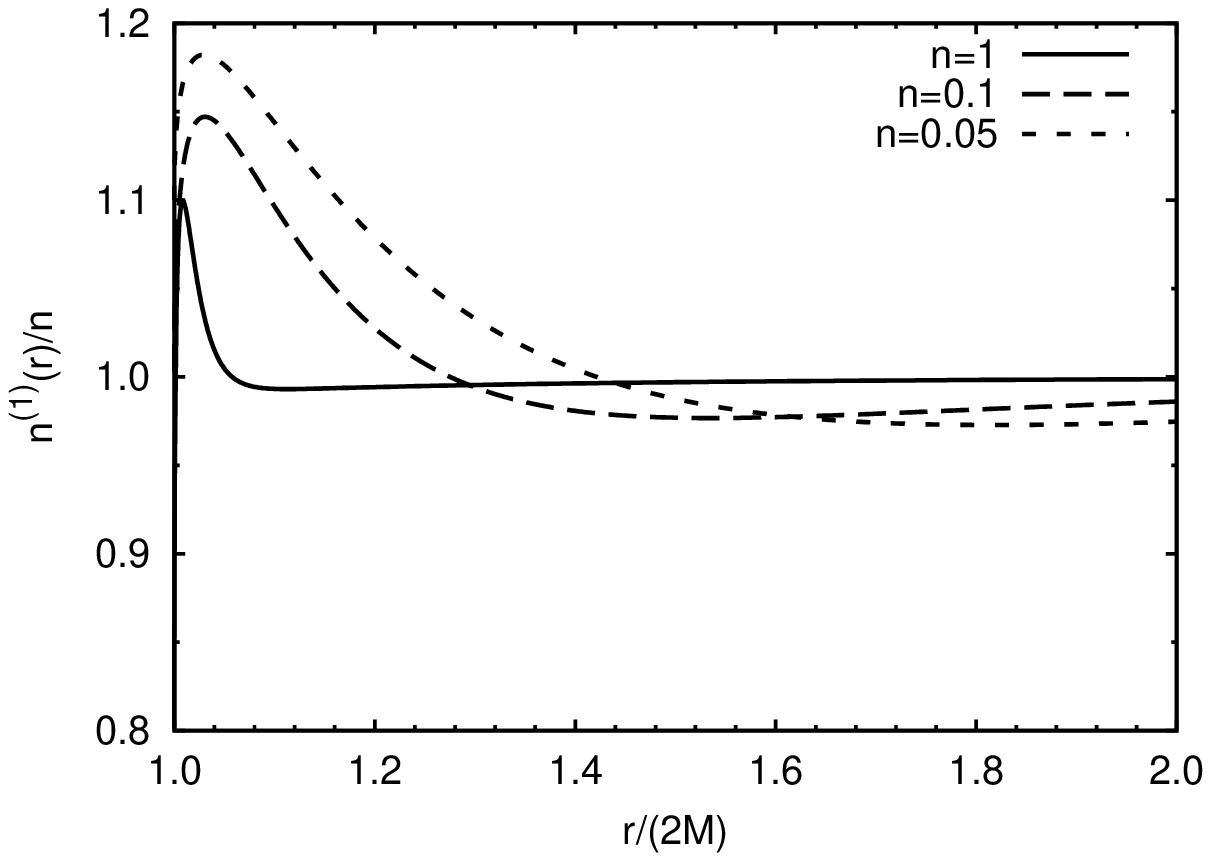}
\includegraphics[width=8cm]{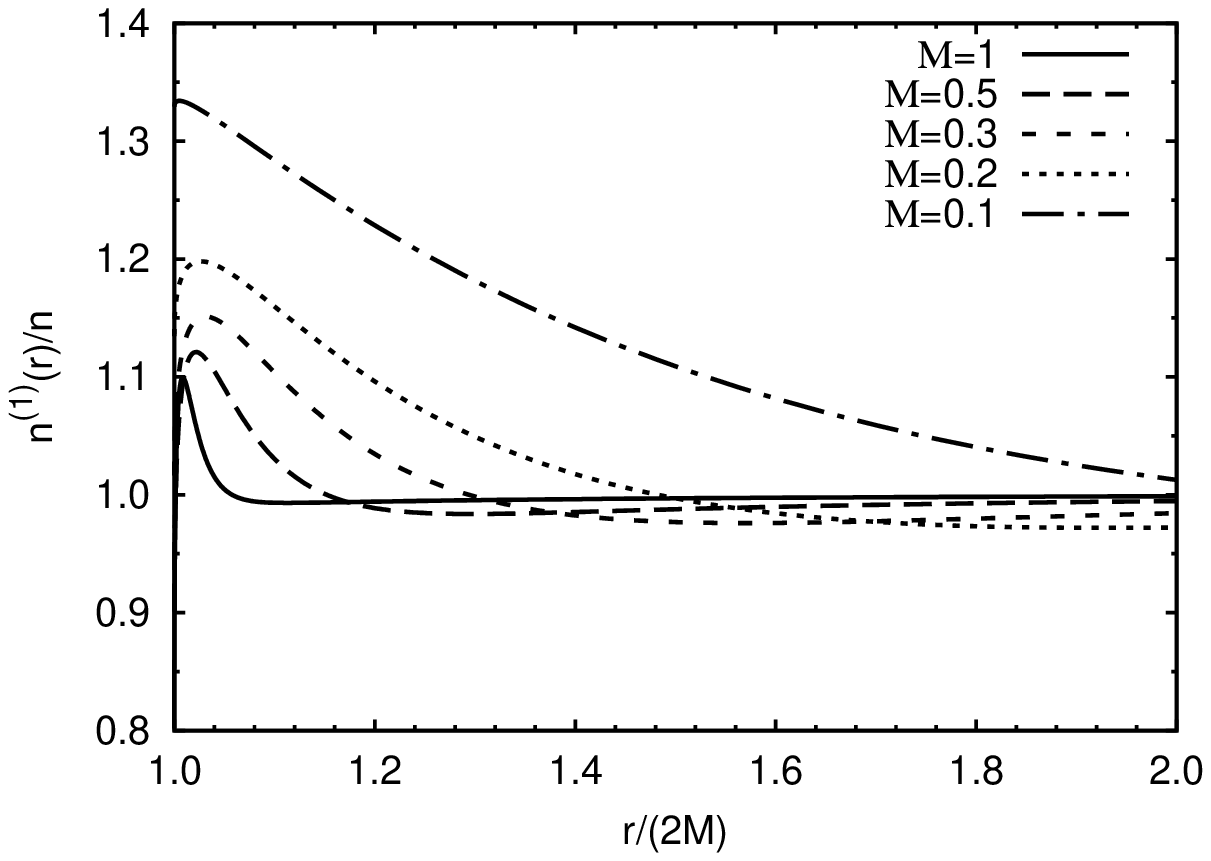}
\end{center}
\caption{The one body density $n^{\mathrm{hs}}(r)/n$ of
  Eq. (\ref{eq:density-Nfinite}), for the 2D OCP on just 
one universe of the surface ${\cal S}$, obtained with $N=300$. 
On the left at fixed $M=1$ and on the right at fixed $n=1$.} 
\label{fig:g1}
\end{figure}

Following~\cite{Jancovici81b}, we can also find the $k$-body distribution
functions
\begin{eqnarray} \label{cf}
n^{(k){\mathrm{hs}}}(\qq_1,\ldots,\qq_k)= 
\det[{\cal K}_N^{\mathrm{hs}}(\qq_i,\qq_j)]_{(i,j)\in\{1,\ldots,k\}^2}~,
\end{eqnarray}
where $\qq_i=(x_i,\varphi_i)$ is the position of the particle $i$, and
\begin{eqnarray} \label{KN}
{\cal K}_N^{\mathrm{hs}}(\qq_i,\qq_j)
=\sum_{k=0}^{N-1}\frac{z_{i}^{k}\bar{z}_j^{k} 
e^{-\alpha[h(|z_i|)+h(|z_j|)]/2}}{{\cal B}_N(k)}~.
\end{eqnarray}
where $z_k=x_k e^{i\varphi_k}$. In particular, the one-body density is
given by
\begin{equation}
  \label{eq:density-Nfinite}
  n^{\mathrm{hs}}(x)=\mathcal{K}_N(\qq,\qq)=
  \sum_{k=0}^{N-1} \frac{x^{2k} e^{-\alpha
      h(x)}}{\mathcal{B}_N(k)}
\,.
\end{equation}
The density shows a peak in the neighborhoods of each boundary,
tends to a finite value at the boundary and to the background density
far from it, in the bulk. This is shown in Fig. \ref{fig:g1}
%

\subsubsection{Internal screening}

Internal screening means that at equilibrium, a particle of the system
is surrounded by a polarization cloud of opposite charge. It is usually
expressed in terms of the simplest of the multipolar sum rules
\cite{Martin88}: the charge or electroneutrality sum rule, which for
the OCP reduces to the relation
\begin{eqnarray} \label{csr}
\int n^{(2){\mathrm{hs}}}(\qq_1,\qq_2)\,dS_2=(N-1)n^{(1){\mathrm{hs}}}(\qq_1)~,
\end{eqnarray}
This relation is trivially satisfied because of the particular
structure~(\ref{cf}) of the correlation function expressed as a
determinant of the kernel $\mathcal{K}_{N}^{\mathrm{hs}}$, and the
fact that $\mathcal{K}_{N}^{\mathrm{hs}}$ is a projector
\begin{equation}
  \int dS_3\, \mathcal{K}_N^{\mathrm{hs}}(\qq_1,\qq_3) 
  \mathcal{K}_N^{\mathrm{hs}}(\qq_3,\qq_2)
  = \mathcal{K}_N^{\mathrm{hs}}(\qq_1,\qq_2)
  \,.
\end{equation}
Indeed,
\begin{eqnarray} \nonumber
\int n^{(2){\mathrm{hs}}}(\qq_1,\qq_2)\,dS_2&=&
\int
[{\cal K}_N^{\mathrm{hs}}(\qq_1,\qq_1)
  {\cal K}_N^{\mathrm{hs}}(\qq_2,\qq_2)-
{\cal K}_N^{\mathrm{hs}}(\qq_1,\qq_2)
{\cal K}_N^{\mathrm{hs}}(\qq_2,\qq_1)]\,dS_2
\\ \nonumber
&=&\int n^{(1){\mathrm{hs}}}(\qq_1)n^{(1){\mathrm{hs}}}(\qq_2)\,dS_2-
{\cal K}_N^{\mathrm{hs}}(\qq_1,\qq_1)
\\ 
&=&
(N-1)n^{(1){\mathrm{hs}}}(\qq_1)
\,.
\end{eqnarray}

\subsubsection{External screening}

External screening means that, at equilibrium, an external charge
introduced into the system is surrounded by a polarization cloud of
opposite charge.  When an external infinitesimal point charge $Q$ is
added to the system, it induces a charge density
$\rho_Q(\qq)$. External screening means that
\begin{eqnarray} \label{es}
  \int \rho_Q(\qq)\, dS=-Q~.
\end{eqnarray}
Using linear response theory we can calculate $\rho_Q$ to first order
in $Q$ as follows. Imagine that the charge $Q$ is at $\qq$. Its
interaction energy with the system is $\hat{H}_{int}=Q\hat{\phi}(\qq)$
where $\hat{\phi}(\qq)$ is the microscopic electric potential created
at $\qq$ by the system. Then, the induced charge density at $\qq'$ is
\begin{eqnarray}
 \rho_Q(\qq')=-\beta\langle\hat{\rho}(\qq')\hat{H}_{int}\rangle_T=
 -\beta Q \langle\hat{\rho}(\qq')\hat{\phi}(\qq)\rangle_T~,
\end{eqnarray}
where $\hat{\rho}(\qq')$ is the microscopic charge density at $\qq'$,
$\langle AB\rangle_T=\langle AB\rangle-\langle A\rangle\langle
B\rangle$, and $\langle\ldots\rangle$ is the thermal average.
Assuming external screening~(\ref{es}) is satisfied, one obtains the
Carnie-Chan sum rule \cite{Martin88}
\begin{eqnarray} \label{cc}
  \beta\int \langle\hat{\rho}(\qq')\hat{\phi}(\qq)\rangle_T\,dS'=1~.
\end{eqnarray}
Now in a uniform system starting from this sum rule one can derive the
second moment Stillinger-Lovett sum rule \cite{Martin88}. This is not
possible here because our system is not homogeneous since the
curvature is not constant throughout the surface but varies from point
to point.  If we apply the Laplacian respect to $\qq$ to this
expression and use Poisson equation
\begin{eqnarray}
  \Delta_{\qq}\langle\hat{\rho}(\qq')\hat{\phi}(\qq)\rangle_T=
  -2\pi\langle\hat{\rho}(\qq')\hat{\rho}(\qq)\rangle_T~,
\end{eqnarray}
we find
\begin{eqnarray} \label{csr'}
  \int \rho_e^{(2)}(\qq',\qq)\,dS'=0~,
\end{eqnarray}
where
$\rho_e^{(2)}(\qq',\qq)=\langle\hat{\rho}(\qq')\hat{\rho}(\qq)\rangle_T$
is the excess pair charge density function. Eq.~(\ref{csr'}) is
another way of writing the charge sum rule Eq.~(\ref{csr}) in the
thermodynamic limit.

\subsubsection{The 2D OCP on the whole surface with potential 
\texorpdfstring{$-\ln(|z-z'|/\sqrt{|zz'|})$}{...}}

Until now we studied the 2D OCP on just one universe. Let us find the
thermodynamic properties of the 2D OCP on the whole surface ${\cal
  S}$. In this case, we also work in the canonical ensemble with a
global neutral system. The position $z_k=x_k e^{i\varphi_k}$ of each
particle can be in the range $1/x_m<x_k<x_m$. The total number
particles $N$ is now expressed in terms of the function $p$ as
$N=2\alpha p(x_m)$. Similar calculations to the ones of the previous
section lead to the following expression for the partition function,
when $\beta e^2=2$,
\begin{equation}
  Z^{\mathrm{ws}}=\frac{1}{\Lambda^{2N}}Z_0^{\mathrm{ws}}
  \exp(-\beta F_0^{\mathrm{ws}})
\end{equation}
now, with
\begin{equation}
  -\beta F_0^{\mathrm{ws}}
  = N b_0+N\alpha h(x_m)- \frac{N^2}{2} \ln x_m -\alpha^2
  \int_{1/x_m}^{x_m} \frac{\left[p(x)\right]^2}{x}\,dx
\end{equation}
and
\begin{equation}
  Z_0^{\mathrm{ws}}(2)=\frac{1}{N!}\int
  \prod_{i=1}^N dS_{i}\, e^{-\alpha h(x_i)} x_{i}^{-N+1} 
  \prod_{1\leq i<j \leq N} |z_i-z_j|^2
  \,.
\end{equation}
Expanding the van der Monde determinant and performing the angular
integrals we find
\begin{equation}
  Z_0^{\mathrm{ws}}(2)=\prod_{k=0}^{N-1} 
  \tilde{\mathcal{B}}_N(k)
\end{equation}
with
\begin{eqnarray} 
\tilde{{\cal B}}_N(k)&=&
\int x^{2k-N+1} e^{-\alpha h(x)}\,dS
\\
&=&
\frac{\alpha}{n}\int_{1/x_m}^{x_m}
x^{2k-N+1} e^{-\alpha h(x)} p'(x)\,dx
\,.
\label{gamma-tilde}
\end{eqnarray}
The function $\tilde{\mathcal{B}}_N(k)$ is very similar to
$\mathcal{B}_{N}$, and its asymptotic behavior for large values of $N$
can be obtained by Laplace method as explained in
Ref. \cite{Fantoni2008}.  

\begin{figure}[htbp]
\begin{center}
\includegraphics[width=8cm]{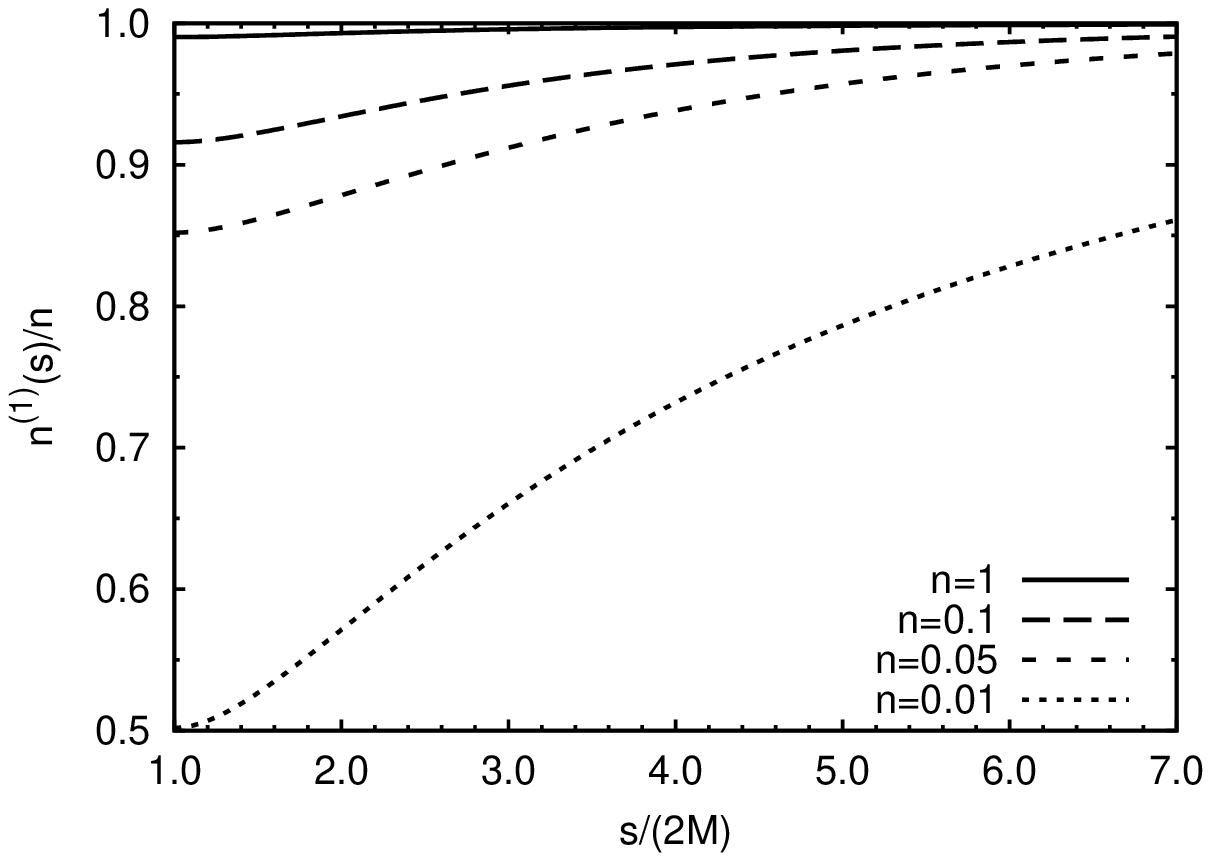}
\includegraphics[width=8cm]{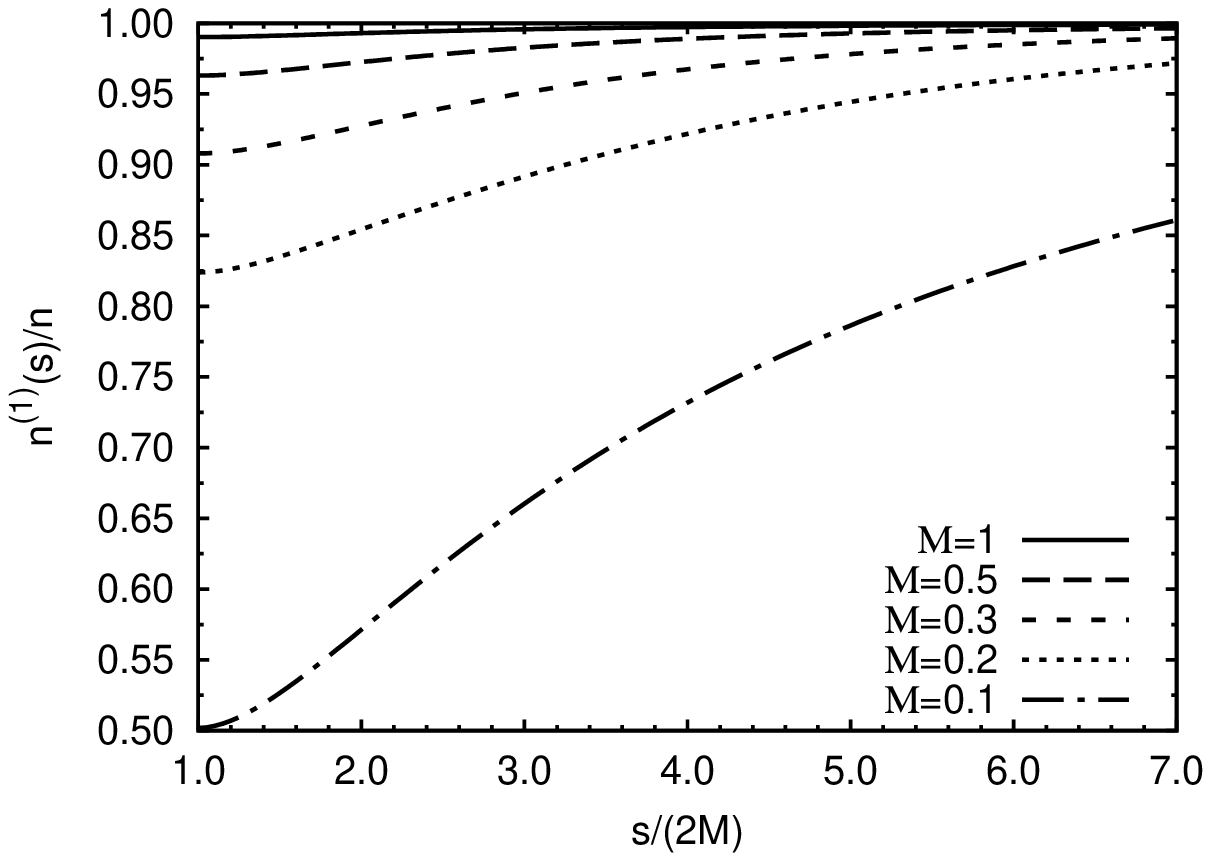}
\end{center}
\caption{The one body density $n^{(1)}(s)/n$, where $s=2Mx$, for the
  2D OCP on the whole manifold, obtained using Eq. (\ref{gamma-tilde}) 
with $N=300$. On the left at fixed $M=1$ and on the right at fixed
$n=1$.}  
\label{fig:g1wm}
\end{figure}

The one body density for the 2D OCP on the whole manifold is drawn in
Fig. \ref{fig:g1wm}. From the figure we can see how the peaks in the
neighborhood of the horizon are now disappeared. The density approaches
the horizon with zero slope.
%
\subsubsection{The 2D OCP on the half surface with potential 
\texorpdfstring{$-\ln(|z-z'|/\sqrt{|zz'|})$}{...}}

In this case, we have $N=\alpha p(x_m)$. In this case the
  partition function at $\beta e^2=2$ is
\begin{equation}
  Z^{\overline{\mathrm{hs}}}=Z_0^{\overline{\mathrm{hs}}}
  e^{-\beta F_0^{\overline{\mathrm{hs}}}}
\end{equation}
with 
\begin{equation}
  -\beta F_0^{\overline{\mathrm{hs}}}
  = \alpha^2 p(x_m) h(x_m) - p(x_m)^2\ln x_m
  +\int_1^{x_m} \frac{\left[p(x)\right]^2}{x}\,dx
  -Nb_0
\end{equation}
and 
\begin{equation}
  Z_0^{\overline{\mathrm{hs}}}(2)=\prod_{k=0}^{N-1} \hat{\mathcal{B}}_N(k)
\end{equation}
with
\begin{equation}
  \hat{\mathcal{B}}_N(k)=\frac{\alpha}{n_b}\int_1^{x_m}
  x^{2k+1}e^{-\alpha h(x)}\,dx
\end{equation}

\begin{figure}[htbp]
\begin{center}
\includegraphics[width=10cm]{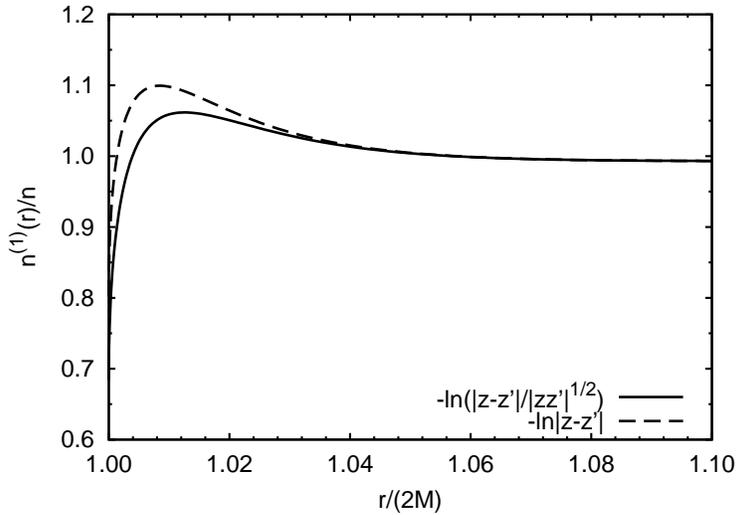}
\end{center}
\caption{The one body density $n^{(1)}(r)/n$, for the 2D OCP on just
one universe of the surface ${\cal S}$, obtained using both the pair
potential $-\ln|z-z'|$ and $-\ln(|z-z'|/\sqrt{|zz'|})$ at fixed $M=n=1$.} 
\label{fig:g1hsc}
\end{figure}

In Fig. \ref{fig:g1hsc} we compare the one body density obtained in
this case with the one of the previous section. 
%
\subsubsection{The grounded horizon case}

In order to find the partition function for the system in the half
space, with a metallic grounded boundary at $x=1$, when the charges
interact through the pair potential of Eq.~(\ref{flamm-ghgreen}) it is
convenient to work in the grand canonical ensemble instead, and use
the techniques developed in Refs.~\cite{Forrester85,Jancovici96}.  We
consider a system with a fixed background density $\rho_b$. The
fugacity $\zeta=e^{\beta \mu}/\Lambda^2$, where $\mu$ is the
chemical potential, controls the average number of particles $\langle
N\rangle$, and in general the system is non-neutral $\langle N\rangle
\neq N_b$, where $N_b=\alpha p(x_m)$. The excess charge is expected to
be found near the boundaries at $x=1$ and $x=x_m$, while in the bulk
the system is expected to be locally neutral. In order to avoid the
collapse of a particle into the metallic boundary, due to its
attraction to the image charges, we confine the particles to be in a
disk domain $\tilde{\Omega}_R$, where $x\in[1+w,x_m]$. We
introduced a small gap $w$ between the metallic boundary and the
domain containing the particles, the geodesic width of this gap is
$W=\sqrt{\alpha p'(1)/(2\pi n_b)}\,w$. On the other hand, for
simplicity, we consider that the fixed background extends up to the
metallic boundary.

In the potential energy of the system~(\ref{eq:tpe}) we
should add the self energy of each particle, that is due to the fact
that each particle polarizes the metallic boundary, creating an
induced surface charge density. This self energy is $\frac{e^2}{2}\ln
[|x^2-1|M/2L]$, where the constant $\ln (M/2L)$ has been added to
recover, in the limit $M\to0$, the self energy of a charged particle
near a plane grounded wall in flat space.

The grand partition function, when $\beta e^2=2$, is
\begin{equation}
\Xi(2)=e^{-\beta F_0^{\text{gh}}} \left[1+\sum_{N=1}^{\infty}
  \frac{\zeta^N}{N!}  \int \prod_{i=1}^N dS_i
  \prod_{i<j}\left|\frac{z_i-z_j}{1-z_i \bar{z}_j}\right|^2
  \prod_{i=1}^{N} \left| |z_i|^2-1\right|^{-1} 
  \prod_{i=1}^{N} e^{-\alpha [ h(x_i)-2N_b \ln x_i ]}
  \right]
\end{equation}
where for $N=1$ the product $\prod_{i<j}$ must be replaced by 1. The
domain of integration for each particle is $\tilde{\Omega}_R$. We
have defined a rescaled fugacity $\zeta=2L\zeta/M$ and
\begin{equation}
  -\beta F_0^{\text{gh}}=\alpha N_b h(x_m) - N_b^2\ln x_m 
  -\alpha^2 \int_{1}^{x_m} \frac{[p(x)]^2}{x}\,dx
\end{equation}
which is very similar to $F_0^{\text{hs}}$, except that here
$N_b=\alpha p(x_m)$ is not equal to $N$ the number of particles.

Let us define a set of reduced complex coordinates $u_i=z_i$
and its corresponding images $u_i^*=1/\bar{z}_i$. By using Cauchy
identity (\ref{pseudosphere-Cauchy}),
\begin{eqnarray} \label{flamm-Cauchy}
\det
\left(
\frac{1}{u_i-u_j^*}
\right)_{(i,j)\in\{1,\cdots,N\}^2}
=
(-1)^{N(N-1)/2}\:
\frac{\prod_{i<j} (u_i-u_j)(u^*_i-u^*_j)}{\prod_{i,j} (u_i-u_j^*)},
\end{eqnarray}
the particle-particle interaction and self energy terms can be
cast into the form
\begin{eqnarray}
\prod_{i<j}\left|
\frac{z_i-z_j}{1-z_i \bar{z}_j}
\right|^2
\prod_{i=1}^N \left(|z_i|^2-1\right)^{-1}
=(-1)^{N}
\det
\left(
\frac{1}{1-z_i \bar{z}_j}
\right)_{(i,j)\in\{1,\cdots,N\}^2}
\,.
\end{eqnarray}
The grand canonical partition function is then
\begin{eqnarray}
\label{eq:flamm-GrandPart-prelim}
\Xi(2)= e^{-\beta F_0^{\text{gh}}}\left[1+\sum_{N=1}^{\infty} \frac{1}{N!}
\int \prod_{i=1}^N dS_i 
\prod_{i=1}^{N} \left[-\zeta(x_i)\right]\,\det
\left(
\frac{1}{1-z_i \bar{z}_j}
\right)
\right]
\,,
\end{eqnarray}
with $\zeta(x)=\zeta  e^{-\alpha[h(x)-2N_b \ln x]}$.
We now notice that we already found an analogous expression
(\ref{eq:GrandPart-prelim}) when studying the
pseudosphere. We therefore proceed as we did for that case. For ease
of reading we repeat here the relevant steps reducing this expression
to a Fredholm determinant~\cite{Forrester85}. Then let us consider the
Gaussian partition function
\begin{equation}
\label{eq:flamm-part-fun-Grass-libre}
Z_0=\int {\cal D}\psi {\cal D}\bar{\psi} \,\exp\left[\int \bar{\psi}(\qq)
A^{-1}(z,\bar{z}') \psi(\qq')\, dS\, dS' \right]
\end{equation}
The fields $\psi$ and $\bar{\psi}$ are anticommuting Grassmann
variables.  The Gaussian measure in~(\ref{eq:flamm-part-fun-Grass-libre}) is
chosen such that its covariance is equal to
\begin{equation}
\left<\bar{\psi}(\qq_i)\psi(\qq_j)\right>
=
A(z_i,\bar{z}_j)=\frac{1}{1-z_i \bar{z}_j}
\end{equation}
where $\langle\ldots\rangle$ denotes an average taken with the Gaussian
weight of (\ref{eq:flamm-part-fun-Grass-libre}). By construction we have
\begin{equation} \label{flamm-Z_0}
Z_0=\det(A^{-1})
\end{equation}
Let us now consider the following partition function
\begin{equation}
Z=\int {\cal D}\psi {\cal D}\bar{\psi} \exp\left[\int \bar{\psi}(\qq)
A^{-1}(z,\bar{z}') \psi(\qq') dS dS' -\int \zeta(x) 
\bar{\psi}(\qq)\psi(\qq) \,dS \right]
\end{equation}
which is equal to
\begin{equation}
Z=\det(A^{-1}-\zeta)
\end{equation}
and then
\begin{equation} \label{flamm-Z/Z_0}
\frac{Z}{Z_0}=\det[A(A^{-1}-\zeta)]=\det(1+K)
\end{equation}
where $K$ is an integral operator (with integration measure $dS$)
with kernel
\begin{equation} \label{flamm-K}
K(\qq,\qq')=-\zeta(x')\, A(z,\bar{z}')=
-\frac{\zeta(x')}{1-z\bar{z}'}
\,.
\end{equation}

Expanding the ratio $Z/Z_0$ in 
powers of $\zeta$ we have
\begin{equation}
\label{eq:flamm-expans-ZZ0}
\frac{Z}{Z_0}=
1+
\sum_{N=1}^{\infty}
\frac{1}{N!}
\int \prod_{i=1}^N dS_i
(-1)^{N}\prod_{i=1}^N
\zeta(x_i)
\left<\bar{\psi}(\qq_1)\psi(\qq_1)\cdots
\bar{\psi}(\qq_N)\psi(\qq_N)\right>
\end{equation}
Now, using Wick theorem for anticommuting variables~\cite{ZinnJustin},
we find that
\begin{equation}
\label{eq:flamm-WickFerms}
\left<\bar{\psi}(\qq_1)\psi(\qq_1)\cdots
\bar{\psi}(\qq_N)\psi(\qq_N)\right>
=\det A(z_i,\bar{z}_j)=\det\left(\frac{1}{1-z_i \bar{z}_j}\right)
\end{equation}
Comparing equations~(\ref{eq:flamm-expans-ZZ0})
and~(\ref{eq:flamm-GrandPart-prelim}) with the help of
equation~(\ref{eq:flamm-WickFerms}) we conclude that
\begin{equation}
\label{eq:flamm-Xi-det}
\Xi(2)=e^{-\beta F_0^{\text{gh}}}\,\frac{Z(2)}{Z_0(2)}=
e^{-\beta F_0^{\text{gh}}}\det(1+K)
\end{equation}

The problem of computing the grand canonical partition function has
been reduced to finding the eigenvalues $\lambda$ of the operator
$K$. The eigenvalue problem for $K$ reads
\begin{equation}
\label{eq:flamm-vpK}
-\int_{\tilde{\Omega}_R} 
\frac{\zeta(x')}
{ 1-z\bar{z}'}\,
\Phi(x',\varphi') dS'
=
\lambda \Phi(x,\varphi)
\end{equation}
For $\lambda\neq 0$ we notice from equation~(\ref{eq:flamm-vpK}) that
$\Phi(x,\varphi)=\Phi(z)$ is an analytical function of
$z=xe^{i\varphi}$ in the region $|z|>1$. Because of the circular
symmetry, it is natural to try $\Phi(z)=\Phi_{\ell}(z)=z^{-\ell}$ with
$\ell\ge 1$ a positive integer. Expanding
\begin{equation}
\frac{1}{1-z\bar{z}'}=
-\sum_{n=1}^{\infty}\left(z\bar{z}'\right)^{-n}
\end{equation}
and replacing $\Phi_{\ell}(z)=z^{-\ell}$ in equation~(\ref{eq:flamm-vpK}) we
show that $\Phi_{\ell}$ is indeed an eigenfunction of $K$ with
eigenvalue
\begin{equation}
\label{eq:flamm-lambda-vp-de-K}
\lambda_{\ell}=
\zeta \mathcal{B}_{N_b}^{\text{gh}}(N_b-\ell)
\end{equation}
where 
\begin{equation}
  \label{eq:BNgh}
  \mathcal{B}_{N_b}^{\text{gh}}(k)=\frac{\alpha}{n_b}\int_{1+w}^{x_m}
  x^{2k} e^{- \alpha h(x)}\,p'(x)\,dx
\end{equation}
which is very similar to $\mathcal{B}_N$ defined in Eq.~(\ref{flamm-gamma}),
except for the small gap $w$ in the lower limit of integration. So, we
arrive to the result for the grand potential
\begin{equation}
\label{eq:flamm-grand-potential-somme}
\beta\Omega = -\ln\Xi(2) 
=
\beta F_0
-
\sum_{\ell=1}^{\infty}
\ln\left[
1+\zeta {\cal B}_{N_b}^{\text{gh}}(N_b-\ell)
\right]\,.
\end{equation}

As usual one can compute the density by doing a functional derivative
of the grand potential with respect to a position-dependent fugacity
$\zeta(\qq)$
\begin{equation}
\label{eq:flamm-n-funct-deriv}
n^{\text{gh}}(\qq)=
\zeta(\qq)\frac{\delta\ln\Xi(2)}{\delta \zeta(\qq)}
\,.
\end{equation}
For the present case of a curved space, we shall understand the
functional derivative with the rule $\delta \zeta(\qq')/\delta
  \zeta(\qq)=\delta^{(2)}(\qq;\qq')$ where
$\delta^{(2)}(\qq;\qq')=\delta(x-x')\delta(\varphi-\varphi')/\sqrt{g}$
  is the Dirac distribution on the curved surface.

\begin{figure}[htbp]
\begin{center}
\includegraphics[width=8cm]{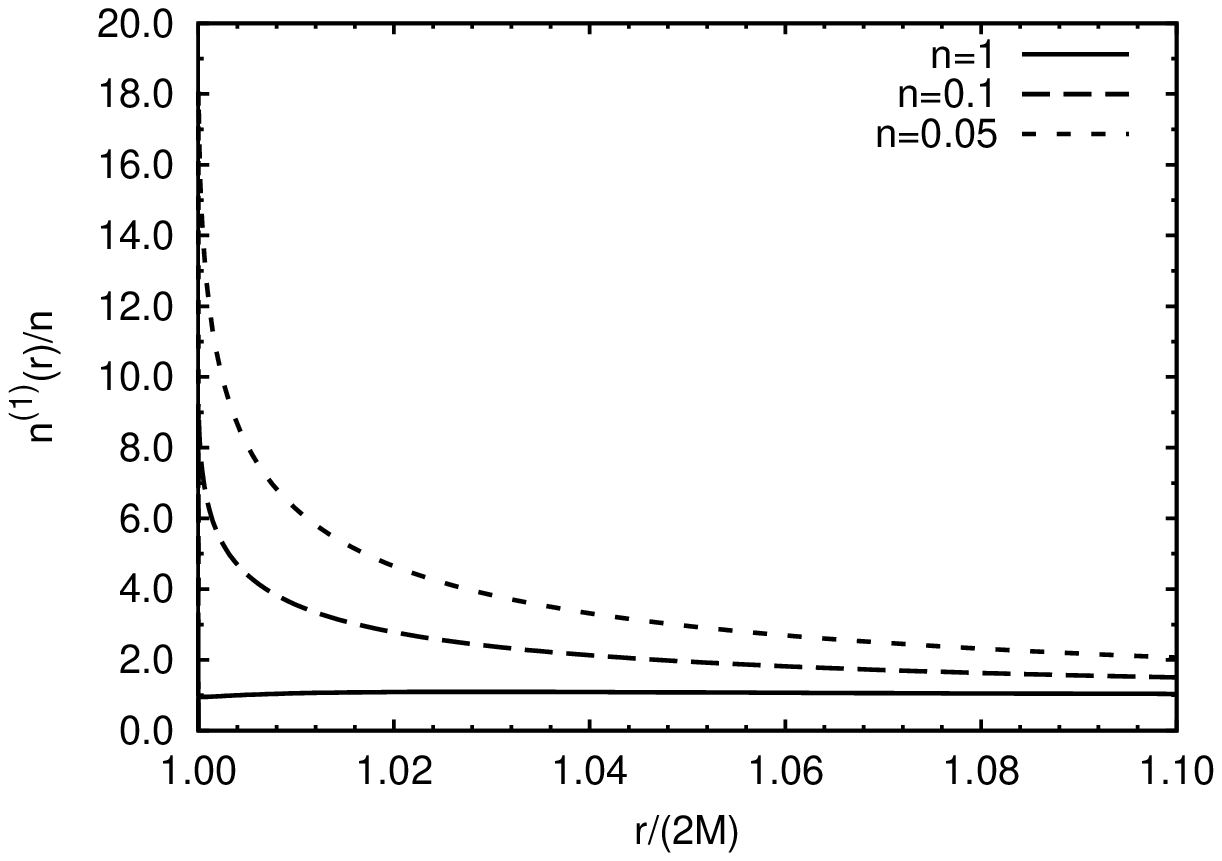}
\includegraphics[width=8cm]{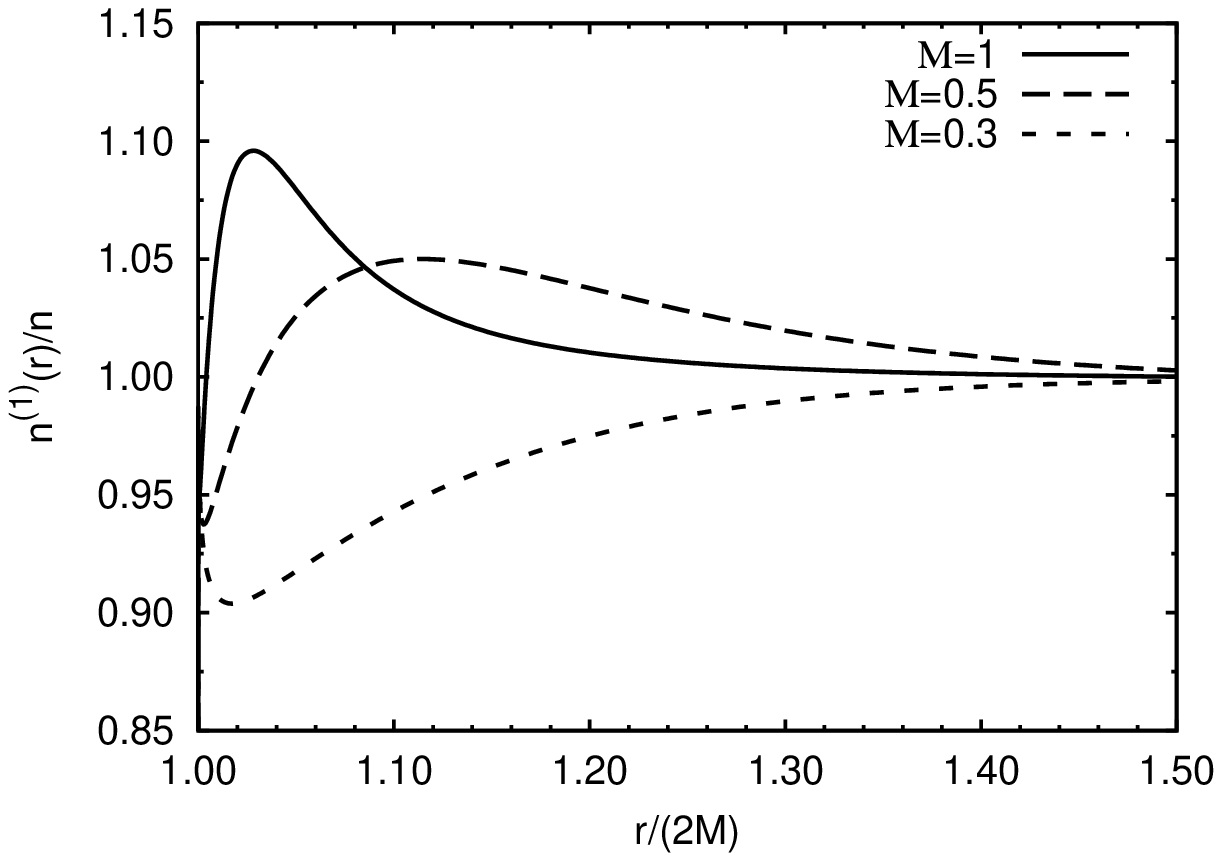}
\includegraphics[width=8cm]{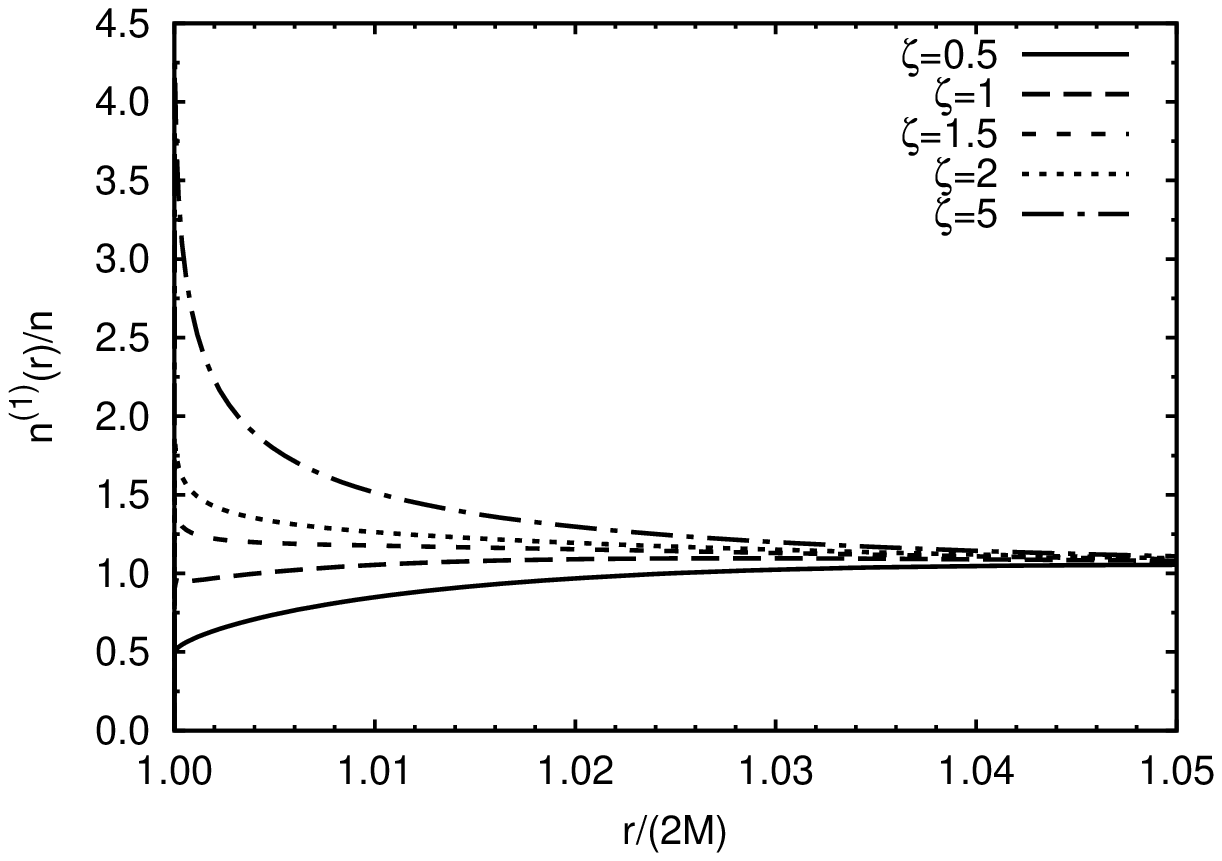}
\end{center}
\caption{The one body density $n^{\text{gh}}(r)/n$ obtained truncating
the sum of Eq. (\ref{eq:flamm-densite-somme}) after the first $300$
terms and choosing $(\sqrt{R}+\sqrt{R-2M})^2/2M=10$. On top on the
left at fixed $M=\zeta=1$ and on the right at fixed $n=\zeta=1$. On
the bottom at fixed $M=n=1$.}   
\label{fig:g1gh}
\end{figure}

Using a Dirac-like notation, one can formally write
\begin{equation}
\ln\Xi(2)=\Tr \ln(1+K)-\beta F_0^{\text{gh}}=
\int \left<\qq\left|
\ln(1-\zeta(\qq)A)\right|\qq\right>
\,dS
-\beta F_0^{\text{gh}}
\end{equation}
Then, doing the functional derivative~(\ref{eq:flamm-n-funct-deriv}), one
obtains
\begin{equation}
n^{\text{gh}}(\qq)=
\zeta\left<\qq\left| (1+K)^{-1}(-A) \right|\qq\right> =
\zeta \tilde{G}(\qq,\qq)
\end{equation}
where we have defined $\tilde{G}(\qq,\qq')$ by
$\tilde{G}=(1+K)^{-1}(-A)$. More explicitly, $G$ is the solution
of $(1+K)\tilde{G}=-A$, that is
\begin{equation}
\label{eq:flamm-eq-Green-function}
\tilde{G}(\qq,\qq') - \int_{\tilde{\Omega}_R}
\zeta(x'')\frac{\tilde{G}(\qq'',\qq')}{1-z\bar{z}''}
\, dS'' = -\frac{1}{1-z\bar{z}'}
\,.
\end{equation}
From this integral equation, one can see that $\tilde{G}(\qq,\qq')$ is
an analytical function of $z$ in the region $|z|>1$. Then, we look for
a solution in the form of a Laurent series
\begin{equation}
\tilde{G}(\qq,\qq')=\sum_{\ell=1}^{\infty} a_{\ell}(\rr') z^{-\ell}
\end{equation}
into equation~(\ref{eq:flamm-eq-Green-function}) yields
\begin{equation}
\label{eq:flamm-solution-G}
\tilde{G}(\qq,\qq')=
\sum_{\ell=1}^{\infty}
\frac{\left(z\bar{z}'\right)^{-\ell}}{1+\lambda_\ell}
\,.
\end{equation}
Recalling that $\lambda_{\ell}=\zeta\mathcal{B}_{N}^{\text{gh}}(N_b-\ell)$,
the density is given by
\begin{equation}
\label{eq:flamm-densite-somme}
n^{\text{gh}}(x)=
\zeta
\sum_{k=-\infty}^{N_b-1}
\frac{x^{2k}e^{-\alpha h(x)}}{1+\zeta \mathcal{B}_{N}^{\text{gh}}(k)}
\end{equation}
The density reaches the background density far from the boundaries. In
this case, the fugacity and the background density control the density
profile close to the metallic boundary (horizon). In the bulk and
close to the outer hard wall boundary, the density profile is
independent of the fugacity. 
In Fig. \ref{fig:g1gh} we show the density for various choices of the
parameters $M,n$, and $\zeta$. The figure shows how the density tends
to the background density far from the horizon. The value of the
density at the horizon depends on $n$ and $\zeta$.
%
\part{The Two-Component Plasma} 
\label{sec:tcp}

A {\sl two-component plasma} is a neutral mixture of two species of
$2N$ point charges of opposite charge $\pm e$. 

\section{The plane} 
\label{sec:tcp-plane}

We represent the Cartesian components of the position $\qq=(x,y)$ of a
particle by the complex number $z=x+iy$. For a system of $N$ positive
charges with complex coordinates $u_i$ and $N$ negative charges with
complex coordinates $v_i$ the Boltzmann factor at $\Gamma=\beta e^2=2$
is,
\bq \nonumber
e^{2\sum_{i<j}\left[\ln\frac{|u_i-u_j|}{L}+\ln\frac{|v_i-v_j|}{L}\right]-
2\sum_{i,j}\ln\frac{|u_i-v_j|}{L}}&=&L^{2N}\left|
\frac{\prod_{i<j}(u_i-u_j)(v_i-v_j)}{\prod_{i,j}(u_i-v_j)}\right|^2\\ 
\label{tcp-plane-boltzmann}
&=&L^{2N}\left|\det
\left(
\frac{1}{u_i-v_j}
\right)_{(i,j)\in\{1,\cdots,N\}^2}\right|^2,
\eq
where the last equality stems from the Cauchy identity
(\ref{pseudosphere-Cauchy}). Following Ref. \cite{Cornu87}, it is
convenient to start with a 
discretized model for which there are no divergencies. Two interwoven
sublattices $U$ and $V$ are introduced. The positive (negative)
particles sit on the sublattice $U(V)$. Each lattice site is occupied
no or one particle. A possible external potential is described by
position dependent fugacities $\zeta_+(u_i)$ and $\zeta_-(v_i)$. The
the grand partition function reorganized as a sum including only
neutral systems is
\bq \label{pf0}
\Xi(2)&=&1+\sum_{N=1}^\infty L^{2N}\prod_{i=1}^N\zeta_+(u_i)\zeta_-(v_i)
\mathop{\sum_{u_1,\ldots,u_N\in U}}_{v_1,\ldots,v_N\in V}
\left|\det\left(\frac{1}{u_i-v_j}\right)_{(i,j)\in\{1,\ldots,N\}^2}
\right|^2~,
\eq
where the sums are defined with the prescription that configurations
which differ only by a permutation of identical particles are counted
only once. This grand partition function is the determinant of
an anti-Hermitian matrix $\mathbf{M}$ explicitly shown in
Ref. \cite{Cornu89}.

When passing to the continuum limit in the element $\mathbf{M}_{ij}$
one should replace $u_i$ or $v_i$ by $z$ and $u_j$ or $v_j$ by
$z^\prime$, {\sl i.e.} $i\to z$ and $j\to z^\prime$. 
Each lattice site is characterized by its complex coordinate $z$ and
an isospinor which is \mbox{\tiny
  $\left(\begin{array}{c}1\\0\end{array}\right)$}  
if the site belongs to the positive sublattice $U$ and
\mbox{\tiny $\left(\begin{array}{c}0\\1\end{array}\right)$} if it
belongs to the negative sublattice $V$. We then define a matrix
$\pmb{\mathcal M}$ by
\bq \label{M}
\langle z|\pmb{\mathcal M}|z^\prime\rangle=
\frac{\pmb{\sigma_x}+i\pmb{\sigma_y}}{2}\frac{L}{z-z^\prime}+
\frac{\pmb{\sigma_x}-i\pmb{\sigma_y}}{2}
\frac{L}{\overline{z}-\overline{z}^\prime}~,
\eq
where the $\pmb{\sigma}$ are the $2\times 2$ Pauli matrices
operating in the isospinor space.

The matrix $\pmb{\mathcal M}$ can be expressed in terms of a simple
Dirac operator
$\slashed{\partial}=\pmb{\sigma_x}\partial_x+\pmb{\sigma_y}\partial_y$
as follows,  
\bq \label{tcp-plane-do}
\langle z|\pmb{\mathcal M}|z^\prime\rangle=L
(\pmb{\sigma_x}\partial_x+\pmb{\sigma_y}\partial_y)
\ln|z-z^\prime|~,
\eq
and the grand partition function can be rewritten as 
\bq \nonumber
\Xi(2)&=&\det\left\{\mathbf{1}\delta^{(2)}(z;z^\prime)
+\left[\zeta_+(z)\frac{\mathbf{1}+\pmb{\sigma_z}}{2}+
\zeta_-(z)\frac{\mathbf{1}-\pmb{\sigma_z}}{2}\right]
\langle z|\pmb{\mathcal M}|z^\prime\rangle\right\}\\ \label{gpf}
&=&\det[\mathbf{1}+\pmb{\mathcal K}^{-1}]~,
\eq
where $\mathbf{1}$ is the $2\times 2$ identity matrix and
\bq
\pmb{\lambda}&=&\zeta_+\frac{\mathbf{1}+\pmb{\sigma_z}}{2}+
\zeta_-\frac{\mathbf{1}-\pmb{\sigma_z}}{2}~,\\
\pmb{\mathcal K}^{-1}&=&\pmb{\lambda}\pmb{\mathcal M}~.
\eq
Then, since $\Delta\ln|z|=2\pi\delta(r)\delta(\varphi)/r=2\pi\delta(z)$,
where $(r=|z|,\varphi=\arg z)$ are the polar coordinates in the plane, the
inverse operator is $\pmb{\mathcal K}=\pmb{\mathcal{O}}\mathbf{m}^{-1}$, where
\bq 
\mathbf{m}(z)&=&m_+(z)\frac{\mathbf{1}+\pmb{\sigma_z}}{2}+
m_-(z)\frac{\mathbf{1}-\pmb{\sigma_z}}{2}~,\\ \label{tcp-plane-o}
\pmb{\mathcal{O}}&=&\pmb{\sigma_x}\partial_x+
\pmb{\sigma_y}\partial_y=\slashed{\partial}~.
\eq
Here $m_{\pm}(z)=2\pi L\zeta_\pm(z)/S$ are rescaled position dependent
fugacities and $S$ is the area per lattice site which appears when the
discrete sums are replaced by integrals. 

We then find 
\bq \nonumber
\ln\Xi(2)&=&\Tr\left\{\ln\left[\mathbf{1}+\pmb{\mathcal K}^{-1}
\right]\right\}~,
\eq
which expresses the well known equivalence between the 2D OCP at
$\Gamma=2$ and a free Fermi field \cite{Samuel1978}.

The one-body densities and $n$-body truncated densities
\cite{Martin88} can be obtained in the usual way by taking functional
derivatives of the logarithm of the grand partition function with
respect to the fugacities $\zeta_{\pm}$. Marking the sign of the
particle charge at $z_i$ by an index $p_i=\pm$, and defining the
matrix  
\bq \label{Green}
R_{p_1p_2}(z_1,z_2)=\langle z_1p_1|
\pmb{\mathcal K}^{-1}(1+\pmb{\mathcal K}^{-1})^{-1}
|z_2p_2\rangle~,
\eq
it can then be shown \cite{Cornu87,Cornu89} that they are given by 
\bq \label{rho1}
\rho_{p_1}^{(1)}(z_1)&=&
R_{p_1p_1}(z_1,z_1)~,\\ \label{rho2}
\rho_{p_1p_2}^{(2)T}(z_1,z_2)&=&
-R_{p_1p_2}(z_1,z_2)R_{p_2p_1}(z_2,z_1)~,\\ \label{rhon}
\rho_{p_1p_2,\ldots,p_n}^{(n)T}
(z_1,z_2,\ldots,z_n)&=&
(-)^{n+1}\sum_{(i_1,i_2,\ldots,i_n)}
R_{p_{i_1}p_{i_2}}(z_{i_1},z_{i_2})\cdots
R_{p_{i_n}p_{i_1}}(z_{i_n},z_{i_1})~,
\eq
where the summation runs over all cycles $(i_1,i_2,\ldots,i_n)$ built
with $\{1,2,\ldots,n\}$. 

\subsection{Symmetries of Green's function \texorpdfstring{$R$}{R}}

Since $\mathbf{m}^\dagger=\mathbf{m}$ and $\pmb{\mathcal{O}}^\dagger
=-\pmb{\mathcal{O}}$ we find 
\bq
\overline{R_{p_1p_2}(z_1,z_2)}=\langle z_2p_2|\mathbf{m}(z)
(\mathbf{m}(z)-\pmb{\mathcal{O}})^{-1}|z_1p_1\rangle~.
\eq
Expanding in $\pmb{\mathcal{O}}$ and comparing with the definition
$R_{p_1p_2}(z_1,z_2)=\langle z_1p_1|\mathbf{m}(z)
(\mathbf{m}(z)+\pmb{\mathcal{O}})^{-1}|z_2p_2\rangle$ we find
\bq
\overline{R_{pp}(z_1,z_2)}=R_{pp}(z_2,z_1)~,\\
\overline{R_{p-p}(z_1,z_2)}=-R_{-pp}(z_2,z_1)~.
\eq
From which also follows that $R_{pp}(z_1,z_1)$ has to be real. If
$\zeta_+=\zeta_-$ then we additionally must have
\bq
R_{pp}(z_1,z_2)=R_{-p-p}(z_1,z_2)~.
\eq

\subsection{Two-body truncated correlation functions and perfect
  screening sum rule}

For the two-body truncated correlation functions of Eq. (\ref{rho2})
we then find
\bq \label{rhopp}
\rho_{++}^{(2)T}(z_1,z_2)&=&-|R_{++}(z_1,z_2)|^2~,\\ \label{rhopm}
\rho_{+-}^{(2)T}(z_1,z_2)&=&|R_{+-}(z_1,z_2)|^2~.
\eq

Notice that the total correlation function for the like particles
$h_{++}(z_1,z_2)$ $=$ $\rho_{++}^{(2)T}(z_1,z_2)$ $/
\rho_{+}^{(1)}(z_1)\rho_{+}^{(1)}(z_2)$ goes to $-1$ when the
particles coincide $z_1\to z_2$ as follows from the structure of
Eqs. (\ref{rho1}) and (\ref{rho2}). Moreover the truncated densities of 
any order has to decay to zero as two groups of particles are
infinitely separated. In particular
$|R_{++}(z_1,z_2)|=|R_{++}(r_1,r_2,\varphi_2-\varphi_1)|$ 
has to decay to zero as $|\qq_1-\qq_2|\to \infty$. 

The perfect screening sum rule has to be satisfied for the symmetric
mixture 
\bq \label{perfect-screening}
\int[\rho_{+-}^{(2)T}(z_1,z_2)-\rho^{(2)T}_{++}(z_1,z_2)]
\sqrt{g_1}dr_1d\varphi_1=\rho_\pm(z_2)~,
\eq
where $g_1$ is $g$ calculated on particle 1.

\subsection{Determination of Green's function \texorpdfstring{$R$}{R}}

The Green function matrix $\mathbf{R}$ is the solution of a system of
four coupled partial differential equations, namely
\bq
(\mathbf{1}+\pmb{\mathcal K}^{-1})\pmb{\mathcal K} \mathbf{R}(z_1,z_2)=
(\mathbf{1}+\pmb{\mathcal K}) \mathbf{R}(z_1,z_2)=
\mathbf{1}\delta^{(2)}(z_1;z_2)
\eq
where $\delta^{(2)}(z_1;z_2)$ $=$ $(\sqrt g)^{-1} 
\delta(r-r_0)\delta(\varphi-\varphi_0)$, with $\sqrt g=r$ is the Dirac
delta function on the plane which we will call $\delta(z_1-z_2)$ the
{\sl flat} Dirac delta and $\mathbf{1}$ is the $2\times 2$ identity
matrix. These coupled equations can be rewritten as follows  
\bq \nonumber
\left[\pmb{\mathcal{O}}+\mathbf{m}(z_1)\right]
\mathbf{R}(z_1,z_2)=\mathbf{m}(z_1)\delta^{(2)}(z_1;z_2)~.
\eq

If instead of $\mathbf{R}$ one uses
$\mathbf{R}=\mathbf{G}\mathbf{m}$, $\mathbf{G}$ satisfies the
equation
\bq \label{Geq}
[\pmb{\mathcal{O}}+\mathbf{m}(z_1)]\mathbf{G}(z_1,z_2) =\mathbf{1}
\delta^{(2)}(z_1;z_2)~.  
\eq 
By combining the components of this equation one obtains decoupled
equations for $G_{++}$ and $G_{--}$ as follows
\bq \label{plane:G++}
\left\{m_+(z_1)+A^\dagger [m_-(z_1)]^{-1}A
\right\}G_{++}(z_1,z_2)&=&
\delta^{(2)}(r_1,\varphi_1;r_2,\varphi_2)~,\\ \label{plane:G--}
\left\{m_-(z_1)+A[m_+(z_1)]^{-1}A^\dagger
\right\}G_{--}(z_1,z_2)&=&
\delta^{(2)}(r_1,\varphi_1;r_2,\varphi_2)~,
\eq
where $A=\partial_x+i\partial_y$, while
\bq \label{plane:G-+}
G_{-+}(z_1,z_2)&=&-\left[m_-(z_1)\right]^{-1}
AG_{++}(z_1,z_2)~,\\
G_{+-}(z_1,z_2)&=&+\left[m_+(z_1)\right]^{-1}
A^\dagger G_{--}(z_1,z_2)~,
\eq
Then Eq. (\ref{G++}) can be rewritten in Cartesian coordinates as
\bq \label{plane:G++flateq}
\left[m_+m_--
\frac{1}{r_1}\partial_{r_1}(r_1\partial_{r_1})-
\frac{1}{r_1^2}\partial^2_{\varphi_1}\right]G_{++}(z_1,z_2)=
\frac{m_-}{r_1}\delta(r_1-r_2)\delta(\varphi_1-\varphi_2)~.
\eq
which, when $m_+(z)=m_-(z)=m$, has the
following solution \cite{Cornu89,Cornu87} 
\bq
G_{++}(z_1,z_2)&=&\frac{m}{2\pi}K_0(m|\qq_1-\qq_2|)~,\\
G_{-+}(z_1,z_2)&=&\frac{m}{2\pi}\frac{(x_1-x_2)+i(y_1-y_2)}{|\qq_1-\qq_2|}
K_1(m|\qq_1-\qq_2|)~,
\eq
where $K_0$ and $K_1$ are modified Bessel functions. These functions
decay at large distances on a characteristic length scale
$m^{-1}$. The $n$-body truncated densities (\ref{rhon}) are well
defined quantities for the point particle system. The two-body
truncated densities, for example, have the simple forms
\bq
\rho_{++}^{(2)T}(r)&=&-\left(\frac{m^2}{2\pi}\right)^2K_0^2(mr),\\
\rho_{+-}^{(2)T}(r)&=&-\left(\frac{m^2}{2\pi}\right)^2K_1^2(mr).
\eq
The one-body densities, however, as given by Eq. (\ref{rho1}), are
infinite since $K_0(mr)$ diverges logarithmically as $r\to 0$. This
divergence can be suppressed by a short distance cutoff $R$. We
replace the point particles by small hard discs of diameter $R$ and
use a regularized form of Eq. (\ref{rho1}),
\bq \label{rpm}
\rho_{\pm}=\frac{m^2}{2\pi}K_0(mR)\sim\frac{m^2}{2\pi}
\left[\ln\frac{2}{mR}-\gamma\right],
\eq
where $\gamma=0.5772$ is Euler's constant. Keeping the point charge
expression for the correlation functions for separations larger than
$R$ the perfect screening rule (\ref{perfect-screening}) is
satisfied. 

Integrating $\rho_++\rho_-=m\partial(\beta p)/\partial m$, from
Eq. (\ref{rpm}) one obtains for the pressure $p$, 
\bq
\beta p=\frac{1}{2}(\rho_++\rho_-)+\frac{m^2}{4\pi}.
\eq
The same result can be obtained by using the regularized form of
Eq. (\ref{gpf}). In the limit $mR\to 0$ one finds the expected result
for an ideal gas of collapsed neutral pairs.

\section{The sphere} 
\label{sec:tcp-sphere}

We consider the stereographic projection \cite{Forrester1992} of the 
sphere of radius $a$ on the plane tangent to its south pole. The
coordinates of the point $\pp=(x,y)$ stereographic projection of a
point $\qq=(\theta,\varphi)$ of the sphere from the north pole is
given in terms of the complex coordinate $z=x+iy$ by
$z=2ae^{i\varphi}\cotan(\theta/2)$. This projection is a conformal
transformation. The conformal metric in the new coordinates $(x,y)$ is
then
\bq \label{eq:sphere-metric-stereo}
\gggg=\left(
\begin{array}{cc}
e^\omega & 0\\
0 & e^\omega
\end{array}\right),
\eq
with the conformal factor given by
\bq
e^\omega=\sin^2\frac{\theta}{2}=\frac{1}{1+(|z|/2a)^2}.
\eq
The length $r_{ij}$ (\ref{sphere-rij}) of the chord joining two
particles $i$ and $j$ has a simple relation with its projection
$|z_i-z_j|$, 
\bq
r_{ij}=e^{\omega_i/2}|z_i-z_j|e^{\omega_j/2}=\sin\frac{\theta_i}{2}
|z_i-z_j|\sin\frac{\theta_j}{2}.
\eq 

We can then follow the same steps as in section \ref{sec:tcp-plane}
with $z-z'$ replaced by $e^{\omega/2}(z-z^\prime)e^{\omega'/2}$. In
particular the matrix $\pmb{\mathcal M}$ will now become,
\bq
\langle z|\pmb{\mathcal M}|z^\prime\rangle=
\frac{\pmb{\sigma_x}+i\pmb{\sigma_y}}{2}
\frac{L}{e^{\omega/2}(z-z^\prime)e^{\omega'/2}}+
\frac{\pmb{\sigma_x}-i\pmb{\sigma_y}}{2}
\frac{L}{e^{\omega/2}(\overline{z}-\overline{z}^\prime)e^{\omega'/2}}~,
\eq
In the inverse operator $\pmb{\mathcal K}$ we now have
\bq \label{sphere-Dirac}
\pmb{\mathcal O}=e^{-3\omega/2}\slashed{\partial}e^{\omega/2}=\slashed{D},
\eq
since the Dirac delta function on the sphere
$\delta^{(2)}(z;z')=e^{-2\omega}\delta(z-z')$ where $\delta$ is the
flat Dirac delta function. 

Thus, the Dirac operator $\slashed{\partial}$ in the plane has to be
replaced by $\slashed{D}$ defined by (\ref{sphere-Dirac}). It turns
out that $\slashed{D}$ is the Dirac operator on the sphere. 
The Dirac operators in curved spaces have been investigated by many
authors. 

\subsection{Thermodynamic properties}

If we define $m=2\pi L\zeta/S$ in terms of the fugacity $\zeta$
and the area per lattice site $S$ (a local property of the surface),
we have 
\bq
\ln\Xi(2)=\Tr\ln[1+m\slashed{D}^{-1}].
\eq
The eigenvalues of $\slashed{D}$ are \cite{Jayewardena88} $\pm in/a$
where $n$ is any positive integer, with multiplicity $2n$. Thus the
pressure is given by
\bq
\beta p=\frac{\ln\Xi(2)}{4\pi a^2}=
\frac{1}{8\pi a^2}\Tr\ln[1-m^2\slashed{D}^{-2}]=
\frac{1}{2\pi a^2}\sum_{n=1}^\infty n\ln\left[1+\frac{m^2a^2}{n^2}\right],
\eq
and the densities are
\bq
\rho_++\rho_-=m\frac{\partial}{\partial m}(\beta p)=
\frac{m^2}{4\pi a^2}\Tr\frac{1}{m^2-\slashed{D}^2}=
\frac{m^2}{\pi}\sum_{n=1}^\infty\frac{n}{m^2a^2+n^2}.
\eq
These pressure and densities are divergent quantities, unless they are 
regularized by a short distance cutoff, as in the planar case. In the
limit $a\to\infty$, setting $k=n/a$, one retrieves the non-regularized
planar results.  

\subsection{Determination of Green's function \texorpdfstring{$G$}{G}}

Eq. (\ref{Geq}) now becomes
\bq
(\slashed{D}+\mathbf{m})\mathbf{G}(\pp,\pp')=e^{-2\omega}
\mathbf{1}\delta(\pp-\pp'),
\eq
which in terms of 
\bq \label{Gt}
\mathbf{\widetilde{G}}(\pp,\pp')=e^{\omega/2}\mathbf{G}(\pp,\pp')
e^{\omega'/2},
\eq
can be rewritten as 
\bq \label{Gteq}
(\slashed{\partial}+\mathbf{m}e^{\omega})\mathbf{\widetilde{G}}(\pp,\pp')
=\mathbf{1}\delta(\pp-\pp').
\eq
This equation has a remarkably simple
interpretation. $\mathbf{\widetilde{G}}(\pp,\pp')$ is the Green
function of the planar problem with a position dependent fugacity
$me^\omega=m/[1+(r/2a)^2]$. This equation correctly reduces to the
flat analogue (\ref{Geq}) in the $a\to\infty$ limit. Moreover, it
admits solutions in term of some hypergeometric functions
\cite{Forrester1992}.  


\section{The pseudosphere} 
\label{sec:tcp-pseudosphere}

The pseudosphere has already been discussed in section
\ref{sec:ocp-pseudosphere}. 

We then observe that the curved system can be mapped onto a flat
system in the Poincar\'e disk. The Boltzmann factor gain a
multiplicative contribution $[1-(r_i/2a)^2]$ for each particle and in
the computation of the partition function the area element
$dS_i=[1-(r_i/2a)^2]^{-2}\,d\rr_i$. Thus, the original system with a
constant fugacity $\zeta$ maps onto a flat system with a position
dependent fugacity $\zeta[1-(r_i/2a)^2]^{-1}$.   

The Dirac operator on the pseudosphere is then,
\bq
\slashed{D}=\left(1-\frac{r^2}{4a^2}\right)^{3/2}\slashed{\partial}
\left(1-\frac{r^2}{4a^2}\right)^{-1/2}.
\eq

\subsection{Determination of Green's function \texorpdfstring{$G$}{G}}

Eq. (\ref{Gt}) now becomes,
\bq
\mathbf{\widetilde{G}}(z_1,z_2)=\left(1-\frac{r_1^2}{4a^2}\right)^{-1/2}
\mathbf{G}(z_1,z_2)\left(1-\frac{r_2^2}{4a^2}\right)^{-1/2},
\eq
and Eq. (\ref{Gteq}) becomes,
\bq
\left[\slashed{\partial}+\frac{\mathbf{m}}{1-(r/2a)^2}\right]
\mathbf{\widetilde{G}}(z,z')=\mathbf{1}\delta(z-z').
\eq
where $\delta$ is the flat Dirac delta.

Thus $\mathbf{G}$ is the Green function of $\slashed{D}+\mathbf{m}$ on
the pseudosphere. The solution of these coupled partial differential
equations can be found in terms of hypergeometric functions
\cite{Tellez98}. Again the flat limit results by taking $a\to\infty$
at a fixed value of $m$. 

\subsection{Thermodynamic properties}

If we define $m=4\pi a\zeta/S$ in terms of the fugacity $\zeta$
and the area per lattice site $S$ (a local property of the surface),
we have,
\bq
\Xi(2)=\det [1+m\slashed{D}^{-1}].
\eq

Then the equation of state can be otained integrating
$n=m\partial(\beta p)/\partial m$ where $n=2\rho_+$. The
one-body density $\rho_+$ can be obtained from Eq. (\ref{rho1}) where
$\mathbf{R}=\mathbf{G}\mathbf{m}$. However, the integration cannot be
performed in terms of known functions for arbitrary $m$. 

\section{The Flamm paraboloid} 
\label{sec:tcp-flamm}

Flamm's paraboloid has already been discussed in section
\ref{sec:ocp-flamm}. 

\subsection{Half surface with an insulating horizon} 
\label{sec:tcp-flamm-hs}

When the TCP lives in the half surface with an insulating horizon the
Coulomb potential is given by Eq. (\ref{flamm-cgreen}). We will use
$u_i=s_ie^{i\varphi_i}$ and $v_j=s_je^{i\varphi_j}$ to denote the
complex coordinates of the positively and negatively charged particles
respectively, where, according to (\ref{flamm-x}), we set
$s=(\sqrt{r}+\sqrt{r-2M})^2/2M > 1$. Note that the 
following small $M$ behaviors holds: $
s = 2r/M-2-M/2r+O(M^2)$ and $\sqrt g = rM/2+O(M^2)$.

The Boltzmann factor at $\Gamma=\beta e^2=2$ now becomes
\bq
\left(\frac{2L}{M}\right)^{2N}\left|\det
\left(
\frac{1}{u_i-v_j}
\right)_{(i,j)\in\{1,\cdots,N\}^2}\right|^2,
\eq
where $L$ is a length scale.

We can then repeat the analysis of
Eqs. (\ref{tcp-plane-boltzmann})-(\ref{perfect-screening}) noticing
that now $\delta^{(2)}(z_1;z_2)$ $=$ $(\sqrt g)^{-1}
\delta(s-s_0)\delta(\varphi-\varphi_0)$ is the Dirac delta function on 
the curved surface and $\delta(s-s_0)\delta(\varphi-\varphi_0)/s$ $=$
$\delta(z-z_0)$ is the flat Dirac delta. Which gives the following,
\bq
m_{\pm}(z)=(2\pi L\zeta_\pm\sqrt g/sS)(2/M)^2,
\eq
rescaled position dependent fugacities which tends to
$\widetilde{m}_\pm=2\pi L\zeta_\pm/S$, the ones of the flat system, in the
$M\to 0$ limit. Here $S$ is a local property of the surface
independent of its curvature. Moreover Eqs. (\ref{tcp-plane-do}) and
(\ref{tcp-plane-o}) read
\bq 
\langle z|\pmb{\mathcal M}_\text{hs}|z^\prime\rangle&=&\frac{2L}{M}
(\pmb{\sigma_x}\partial_x+\pmb{\sigma_y}\partial_y)
\ln|z-z^\prime|~,\\
\pmb{\mathcal{O}}_\text{hs}&=&\frac{2}{M}(\pmb{\sigma_x}\partial_x+
\pmb{\sigma_y}\partial_y)=\frac{2}{M}\slashed{\partial}~.
\eq

\subsection{Determination of Green's function \texorpdfstring{$R$}{R}}

Upon defining $\mathbf{R}=\mathbf{G}\mathbf{\widetilde{m}}$,
$\mathbf{G}$ satisfies the equation
\bq
[\pmb{\mathcal{O}}+\mathbf{m}(z_1)]\mathbf{G}(z_1,z_2) =\mathbf{1}
(4/M^2) \delta(z_1;z_2)~.  
\eq 
which in the flat limit $M\to 0$ reduces to
Eq. (\ref{Geq}). Unfortunately this equation does not admit an
analytical solution for $\mathbf{G}$.
By combining the components of this equation one obtains decoupled
equations for $G_{++}$ and $G_{--}$ as follows
\bq \label{G++}
\left\{m_+(z_1)+A^\dagger [m_-(z_1)]^{-1}A
\right\}G_{++}(z_1,z_2)&=&
\frac{4}{M^2}\delta(s_1,\varphi_1;s_2,\varphi_2)~,\\ \label{G--}
\left\{m_-(z_1)+A[m_+(z_1)]^{-1}A^\dagger
\right\}G_{--}(z_1,z_2)&=&
\frac{4}{M^2}\delta(s_1,\varphi_1;s_2,\varphi_2)~,
\eq
while
\bq \label{G-+}
G_{-+}(z_1,z_2)&=&-\left[m_-(z_1)\right]^{-1}
AG_{++}(z_1,z_2)~,\\
G_{+-}(z_1,z_2)&=&+\left[m_+(z_1)\right]^{-1}
A^\dagger G_{--}(z_1,z_2)~,
\eq

Then Eq. (\ref{G++}) can be rewritten in Cartesian coordinates as
\bq \nonumber
&&\left\{m_+(z_1)m_-(z_1)-\left(\frac{2}{M}\right)^2
\left[(\partial_{x_1}^2+\partial_{y_1}^2)-\right.\right.\\ \nonumber
&&\left.\left.\frac{4(-x_1+iy_1)}{s_1^2(1+s_1)}
(\partial_{x_1}+i\partial_{y_1})\right]\right\}G_{++}(z_1,z_2)= 
\\ \nonumber
&&\left(\frac{2}{M}\right)^4
\frac{\widetilde{m}_-\sqrt{g_1}}{s_1^2}\delta(s_1-s_2)
\delta(\varphi_1-\varphi_2)=\\ \label{G++cart}
&&\left(\frac{2}{M}\right)^4\frac{\widetilde{m}_-\sqrt{g_1}}
{\sqrt{x_1^2+y_1^2}}
\delta(x_1-x_2)\delta(y_1-y_2)~,
\eq
where $s=\sqrt{x^2+y^2}$. From the
expression of the gradient in polar coordinates follows  
\bq
\left\{\begin{array}{c} \displaystyle
\partial_x=\cos\varphi\partial_s
-\frac{\sin\varphi}{s}\partial_\varphi~,\\ \displaystyle
\partial_y=\sin\varphi\partial_s+\frac{\cos\varphi}{s}\partial_\varphi~.
\end{array}\right.
\eq
Which allows us to rewrite Eq. (\ref{G++cart}) in polar coordinates as  
\bq \nonumber
&&\left[\widetilde{m}_+\widetilde{m}_-\left(1+\frac{1}{s_1}\right)^8-
\left(\frac{2}{M}\right)^2\left(\frac{1}{s_1}\partial_{s_1}(s_1\partial_{s_1})+
\frac{1}{s_1^2}\partial^2_{\varphi_1}+\right.\right. \\ \nonumber
&&\left.\left.\frac{4}{s_1(1+s_1)}\partial_{s_1}+
\frac{4i}{s_1^2(1+s_1)}\partial_{\varphi_1}\right)\right] 
G_{++}(z_1,z_2)=\\ \label{G++polar}
&&\left(\frac{2}{M}\right)^4\frac{\widetilde{m}_-\sqrt{g_1}}{s_1^2}
\delta(s_1-s_2)\delta(\varphi_1-\varphi_2)~.
\eq
From this equation we immediately see that $G_{++}(z_1,z_2)$ cannot be real.
Notice that in the flat limit $M\to 0$ we have $s\sim 2r/M$ and
Eq. (\ref{G++polar}) reduces to  
\bq \nonumber
&&\left[\widetilde{m}_+\widetilde{m}_--
\frac{1}{r_1}\partial_{r_1}(r_1\partial_{r_1})-
\frac{1}{r_1^2}\partial^2_{\varphi_1}\right]G_{++}(z_1,z_2)=
\\ \label{G++flateq}
&&\frac{\widetilde{m}_-}{r_1}\delta(r_1-r_2)\delta(\varphi_1-\varphi_2)~.
\eq
which, when $\widetilde{m}_+=\widetilde{m}_-=\widetilde{m}$, has the following
well known solution \cite{Cornu89,Cornu87} 
\bq
G_{++}(z_1,z_2)=\frac{\widetilde{m}}{2\pi}K_0(\widetilde{m}|\rr_1-\rr_2|)~,
\eq
where $K_0$ is a modified Bessel function.

Let us from now on restrict to the case of equal fugacities of the two
species. Then $\zeta_-=\zeta_+=\zeta$ with  
\bq
\widetilde{m}=\frac{2\pi L}{S}\zeta
=\frac{2\pi Le^{\beta\mu}}{\Lambda^2}
=\left(2\pi L\frac{me^2}{4\pi\hbar^2}\right)e^{2\mu/e^2}~,
\eq
where $\hbar$ is Planck's constant, $m$ is the mass of the particles,
and $\mu$ the chemical potential. So $\widetilde{m}$ has the dimensions of
an inverse length. From the symmetry of the problem we can say that
$G_{++}=G_{++}(s_1,s_2;\varphi_1-\varphi_2)$. We can then express the
Green function as the following Fourier series expansion 
\bq \label{fourier-expansion}
G_{++}(s_1,s_2;\varphi)=\frac{1}{2\pi}
\sum_{k=-\infty}^\infty g_{++}(s_1,s_2;k)e^{ik\varphi}~.
\eq 
Then, using the expansion of the Dirac delta function, $\sum_k
e^{ik\varphi}=2\pi\delta(\varphi)$, we find that $g_{++}$, a
continuous real function symmetric under exchange of $s_1$ and $s_2$,
has to satisfy the following equation    
\bq \nonumber
\left[Q_0\left(k,s_1\right)
+Q_1\left(s_1\right) \partial_{s_1}
+ Q_2\left(s_1\right) \partial_{s_1}^2\right]
g_{++}\left(s_1,s_2;k\right)=\\ \label{Eq0} 
\left(\frac{2}{M}\right)^2\widetilde{m}s_1^3(1+s_1)^5
\delta\left(s_1-s_2\right)~,  
\eq
where
\bq \nonumber
Q_0\left(k,s\right)&=&{\widetilde{m}}^2 \left(1+s\right)^9+
\left(\frac{2}{M}\right)^2
ks^6\left(4+k\left(1+s\right)\right)~,\\ \nonumber 
Q_1\left(s \right)&=&-\left(\frac{2}{M}\right)^2s^7\left(5+s\right)~,
\\ \nonumber
Q_2\left(s \right)&=&-\left(\frac{2}{M}\right)^2s^8\left(1+s\right)~.
\eq
And the coefficients $Q_i$ are polynomials of up to degree 9.

\subsection{Method of solution}

We start from the homogeneous form of Eq. (\ref{Eq0}). We note that,
for a given $k$, the two linearly independent solutions 
$f_\alpha(s;k)$ and $f_\beta(s;k)$ of this linear homogeneous second
order ordinary differential equation are not 
available in the mathematical literature to the best of our knowledge.
Assuming we knew those solutions we would then find the Green function,
$g_{++}(s_1,s_2;k)$, writing \cite{Jackson}
\bq
f(t_1,t_2;k)=c_kf_\alpha(s_<;k)f_\beta(s_>;k)~,
\eq
where $s_<=\min(s_1,s_2)$, $s_>=\max(s_1,s_2)$, and $f_\beta$ has the
correct behavior at large $s$. Then
we determine $c_k$ by imposing the kink in $f$ due to the Dirac delta
function at $s_1=s_2$ as follows 
\bq
\partial_{s_1} f(s_1,s_2;k)|_{s_1=s_2+\epsilon}-\partial_{s_1}
f(s_1,s_2;k)|_{s_1=s_2-\epsilon}=-\widetilde{m}\frac{(1+s_2)^4}{s_2^5}~,
\eq 
where $\epsilon$ is small and positive.

The Green function, symmetric under exchange of $s_1$ and $s_2$, is
reconstructed as follows  
\bq 
G_{++}(z_1,z_2)&=&G_{++}(s_1,s_2;\varphi)=
\frac{1}{2\pi}\sum_{k=-\infty}^\infty
c_kf_\alpha(s_<;k)f_\beta(s_>;k)e^{ik\varphi} \label{Gppms}
\eq

\subsection{Whole surface} 
\label{sec:tcp-flamm-ws}

On the whole surface, using Eq. (\ref{eq:Gws}) with $b_0=-\ln(L_0/L)$,
we can now write the Boltzmann factor at a coupling constant
$\Gamma=\beta e^2=2$ as follows, 
\bq
\left|\det\left(\frac{L}{L_0}
\frac{\sqrt{|u_j v_j|}}{u_i-v_j}\right)
_{(i,j)\in\{1,\ldots,N\}^2}\right|^2~,
\eq
where $L_0$ is another length scale.

The grand partition function will then be,
\bq
\Xi(2)=\det\left[\mathbf{1}+\pmb{\mathcal K}_\text{ws}^{-1}\right]~,
\eq
where now Eqs. (\ref{tcp-plane-do}) and the ones following read,
\bq 
\langle z|\pmb{\mathcal M}_\text{ws}|z^\prime\rangle&=&\frac{L}{L_0}
(\pmb{\sigma_x}\partial_x+\pmb{\sigma_y}\partial_y)
\ln|z-z^\prime|~,\\
\pmb{\mathcal K}^{-1}_\text{ws} &=&
\pmb{\lambda}_\text{ws}\pmb{\mathcal{M}}_\text{ws}~,\\ 
\pmb{\lambda}_\text{ws}&=&\zeta_+|z|
\frac{\mathbf{1}+\pmb{\sigma_z}}{2}+
\zeta_-|z|
\frac{\mathbf{1}-\pmb{\sigma_z}}{2}~,\\
\mathbf{K_\text{ws}}&=&
\pmb{\mathcal{M}}^{-1}_\text{ws}\pmb{\lambda}_\text{ws}^{-1}~,\\
\pmb{\lambda}_\text{ws}^{-1}&=&\frac{1}{\zeta_+|z|}
\frac{\mathbf{1}+\pmb{\sigma_z}}{2}+
\frac{1}{\zeta_-|z|}
\frac{\mathbf{1}-\pmb{\sigma_z}}{2}~.
\eq

Introducing position dependent fugacities  
\bq
m_\pm(z)=\frac{2\pi (L/L_0)\zeta_\pm\sqrt g}{Ss}
=\widetilde{m}_\pm\frac{\sqrt{g}}{s}~,
\eq
where now $\widetilde{m}_\pm/L_0\to\widetilde{m}_\pm$, we can rewrite
\bq
\pmb{\mathcal K}_{\text{ws}}&=&
\frac{\pmb{\sigma_x}+i\pmb{\sigma_y}}{2}a_-+
\frac{\pmb{\sigma_x}-i\pmb{\sigma_y}}{2}a_+~,
\eq
with the operators
\bq
a_-&=&-\frac{\overline{z}}{m_-(z)|z|^3}+
\frac{1}{m_-(z)|z|}(\partial_x-i\partial_y)~,\\
a_+&=&-\frac{z}{m_+(z)|z|^3}+
\frac{1}{m_+(z)|z|}(\partial_x+i\partial_y)~.
\eq
Then the equation for the Green functions are
\bq
(1-a_-a_+)R_{++}(z_1,z_2)=\delta^{(2)}(z_1;z_2)~,\\
(1-a_+a_-)R_{--}(z_1,z_2)=\delta^{(2)}(z_1;z_2)~,\\
R_{+-}=-a_-R_{--}~,\\
R_{-+}=-a_+R_{++}~.
\eq
The equation for $R_{++}$ in the symmetric mixture case is 
\bq \nonumber
\left[m^2(z_1)-\frac{2}{s_1^4}+\frac{2\partial_{s_1}}{s_1^3}
-\frac{\partial_{s_1}^2}{s_1^2}
-\frac{-i\partial_{\varphi_1}+\partial_{\varphi_1}^2}{s_1^4}
\right]R_{++}(z_1,z_2)=\\ \label{Eq1}
\frac{m^2(z_1)}{\sqrt{g_1}}
\delta(s_1-s_2)\delta(\varphi_1-\varphi_2)=
\frac{\widetilde{m}^2\sqrt{g_1}}{s_1^2}
\delta(s_1-s_2)\delta(\varphi_1-\varphi_2)~,
\eq
From this equation we see that $R_{++}(z_1,z_2)$ will now be real. 

By expanding Eq. (\ref{Eq1}) in a Fourier series in the azimuthal
angle we now find  
\bq \nonumber
\left[Q_0\left(k,s_1\right)+
Q_1(s_1)\partial_{s_1}+ Q_2\left(s_1\right)\partial_{s_1}^2\right]
g_{++}\left(s_1,s_2;k\right)=\\ \label{Eq2} 
\left(\frac{M}{2}\right)^2\widetilde{m}s_1^3(1+s_1)^4
\delta\left(s_1-s_2\right)~,  
\eq
where
\bq \nonumber
Q_0\left(k,s\right)&=&\left(\frac{M}{2}\right)^4{\widetilde{m}}^2 
\left(1+s\right)^8+s^4(k^2-k-2)~,\\ \nonumber
Q_1(s)&=&2s^5~,\\ \nonumber
Q_2\left(s \right)&=&-s^6~.
\eq
And the coefficients $Q_i$ are now polynomials of up to degree 8.

In the flat limit we find, for $G_{++}=R_{++}/\widetilde{m}$,
the following equation 
\bq \nonumber
\left[\widetilde{m}^2-\frac{2}{r_1^4}+\frac{2\partial_{r_1}}{r_1^3}-
\frac{\partial^2_{r_1}}{r_1^2}-
\frac{-i\partial_{\varphi_1}+\partial^2_{\varphi_1}}{r_1^4}
\right]G_{++}(z_1,z_2)=\\
\frac{\widetilde{m}}{r_1}\delta(r_1-r_2)\delta(\varphi_1-\varphi_2)~.
\eq
We then see that we now do not recover the TCP in the
plane \cite{Cornu89,Cornu87}. This has to be expected because in the
flat limit, Flamm's paraboloid reduces to two planes connected by
the origin. 

After the Fourier expansion of Eq. (\ref{fourier-expansion}) we now
get  
\bq \label{Eq3}
[P_0(k,r_1)+P_1(r_1)\partial_{r_1}+P_2(r_1)\partial_{r_1}^2]
g_{++}(r_1,r_2;k)=\widetilde{m}\delta(r_1-r_2)~,
\eq
where
\bq \nonumber
P_0(k,r)&=&\widetilde{m}^2r+\frac{k^2-k-2}{r^3}~,\\ \nonumber
P_1(r)&=&\frac{2}{r^2}~,\\ \nonumber
P_2(r)&=&-\frac{1}{r}~.
\eq
The homogeneous form of this equation admits the following two
linearly independent solutions 
\bq \nonumber
&&\left.\begin{array}{ll}
f_1(r;-1)=&[D_{-1/2}(i\sqrt{2\widetilde{m}}r)
+\overline{D_{-1/2}(i\sqrt{2\widetilde{m}}r)}]/2\\
f_2(r;-1)=&D_{-1/2}(\sqrt{2\widetilde{m}}r)\\
\end{array}\right\}~~~k=-1~,\\ \nonumber
&&\left.\begin{array}{ll}
f_1(r;2)=&[D_{-1/2}((-2)^{1/4}\sqrt{\widetilde{m}}r)+\\
&\overline{D_{-1/2}((-2)^{1/4}\sqrt{\widetilde{m}}r)}]/2\\
f_2(r;2)=&[D_{-1/2}(i(-2)^{1/4}\sqrt{\widetilde{m}}r)+\\
&\overline{D_{-1/2}(i(-2)^{1/4}\sqrt{\widetilde{m}}r)}]/2
\end{array}\right\}~~~k=2~,\\ \nonumber
&&\left.\begin{array}{ll}
f_1(t;k)=&\sqrt rI_{-\sqrt{7-4k+4k^2}/4}(\widetilde{m}r^2/2)\\
f_2(t;k)=&\sqrt rI_{\sqrt{7-4k+4k^2}/4}(\widetilde{m}r^2/2)
\end{array}\right\}~~~\mbox{else}~,
\eq 
where $D_\nu(x)$ are parabolic cylinder functions and $I_\mu(x)$ are the
modified Bessel functions of the first kind which diverge as
$e^x/\sqrt{2\pi x}$ for large $x\gg|\mu^2-1/4|$. 

Again we write $g_{++}(r_1,r_2;k)=c_kf_\alpha(r_<;k)f_\beta(r_>;k)$
and impose the kink condition,
\bq
\partial_{r_1}g_{++}(r_1,r_2;k)|_{r_1=r_2+\epsilon}-
\partial_{r_1}g_{++}(r_1,r_2;k)|_{r_1=r_2-\epsilon}=
-\widetilde{m}r_2~,
\eq 
to find the $c_k$. The Green function is then reconstructed using
Eq. (\ref{Gppms}). But we immediately see that curiously 
$|G_{++}|$ diverges. Even
the structure of the plasma is not well defined in this situation. The
collapse of opposite charges at the horizon shrinking to the origin
makes the structure of the plasma physically meaningless. 

\part{Conclusions}  
\label{part:conclusions}

We presented a review of the analytical exact solutions of the
one-component and two-component plasma at the special value of the
coupling constant $\Gamma=2$ in various Riemannian surfaces. Starting
from the pioneering work \cite{Jancovici81b} of Bernard Jancovici in
1981 showing the analytic exact solution for the Jellium on the plane,
many other curved surfaces with a conformal metric has been
considered. Namely: the cylinder, the sphere, the pseudosphere, and
the Flamm paraboloid. From a physical point of view we can see the
curvature of the surface as an additional external field acting on the
system of charges moving in the corresponding flat space
\cite{Fantoni2012b}. Even if this point of view does not take into
account the fact that the Coulomb pair potential always reflect its
harmonicity inside the given surface. For this reason we did not try a
unifying treatment but rather a detailed presentation of each case
individually as characteristic of the diverse scenarios which stem
out of the various surfaces so far studied in the literature. 

In our review we put light on the description of the surface, of the
Coulomb potential (and the background potential for the OCP) in the
surface, and of the exact solution for the partition function and for
the correlation functions. The surfaces considered exhaust to the best
of our knowledge all the cases considered in the literature until
now. We hope that the review could be a valuable instrument for the
reader who needs to have a broad overview on this fascinating exactly
solvable fluid model giving the opportunity of finding in one place a
self contained summary of various results appeared in the literature
at different times and in different journals. We did our best to fill
in all the conceptual gaps between the lines so that the reader can
follow the various derivations without needing to refer to the
original papers which would require an interruption of the
reading. This choice required a certain degree of detail which we
thought necessary in place of a more conversational presentation.

We decided to leave out the results of taking the thermodynamic limit
of the various finite OCP expressions. If the reader desires he can
always go back to the original references to find this lacking piece of
information. It is well known that Coulomb systems have to exhibit
critical finite-size effects \cite{Jancovici94}. The last surface
considered, Flamm's paraboloid, is the only surface of
non-constant curvature considered. Nonetheless the one-body density of
the plasma is a constant even in this surface in the thermodynamic
limit \cite{Fantoni2012b}. On the Flamm paraboloid two different
thermodynamic limits can be considered \cite{Fantoni2008}: the one where
the radius $R$ of the disk confining the plasma is allowed to become
very big while keeping the surface hole radius $M$ constant, and the
one where both $R\to\infty$ and $M\to\infty$ with the ratio $R/M$ kept
constant (fixed shape limit). When the horizon shrinks to a point the
upper half surface reduces to a plane and one recovers the well known
result valid for the one-component plasma on the plane. In the same 
limit the whole surface reduces to two flat planes connected by a hole
at the origin. When only one-half of the surface is occupied by the
plasma the density shows a peak in the neighborhoods of each boundary,
tends to a finite value at the boundary and to the background density
far from it, in the bulk. In the thermodynamic limit at fixed shape,
we find that the density profile is the same as in flat space near a
hard wall. In the grounded horizon case the density reaches the
background density far from the boundaries. In this case, the fugacity
and the background density control the density profile close to the
metallic boundary (horizon). In the bulk and close to the outer hard
wall boundary, the density profile is independent of the fugacity. In
the thermodynamic limit at fixed shape, the density profile is the
same as for a flat space. 

The importance of having an exactly soluble many-body systems at least
at one special temperature relies in the fact that it can serve as a
guide for numerical experiments or for approximate solutions of the
same system at other temperatures or for different more realistic
systems. For example the 2D OCP thermodynamics and structure can now
be efficiently expanded in Jack polynomials for even values of the
coupling constant $\Gamma$ \cite{Samaj2004,Tellez99,Tellez2012}. And
the TCP can be solved in the whole stability range of temperatures
\cite{Samaj2003}.   

The original 1981 work of Jancovici \cite{Jancovici81b} has been
important for the understanding of the fractional quantum Hall effect
in the Laughlin development \cite{Laughlin1983} of a Jastrow
correlation factor of the variational wave function of the Landau
problem \cite{LandauQM} for an Hall system in its ground state.
We expect the results on the curved surface to be relevant in the
developments towards a general relativistic statistical mechanics
\cite{Rovelli2013} which is still missing. The main difficulty being
the lack of a canonical Hamiltonian in a generally covariant theory
where the dynamics is only given relationally rather than in terms of
evolution in physical time. And without a Hamiltonian it is difficult
to even start doing statistical physics \cite{Rovelli}. 

The quantum 2D OCP does not admit an analytic exact solution but it
has been studied through a computer experiment either in its ground
state \cite{Tanatar1989,Kwon1996} or at finite temperature
\cite{Militzer2003,Ceperley2004,Brown2013,Brown2013b}.

\appendix
\section{Electrostatic potential of the background for the OCP in the
  pseudosphere}  
\label{app:pseudosphere-epotb}

In this appendix we give the expression for the electrostatic potential 
of the background,
\bq\label{backpot}
v_b(\qq_1)=\int\rho_b\,G(d_{10})\,dS_0=-n_be\int_\Omega G(d_{10})\,dS_0.
\eq

The electric potential of the background satisfies equation (\ref{bp}).
Using the coordinates $(r,\varphi)$ we have,
\bq
v_b^{\prime\prime}(r)+\frac{1}{r}v_b^\prime(r)=\alpha_b\frac{4a^2}{(1-r^2)^2},
\eq 
where $\alpha_b=-2\pi\rho_b$ and we denote with a prime a derivative with 
respect to $r$.
This differential equation admits the following solution for $v_b^\prime$,
\bq\nonumber
v_b^\prime(r)&=&e^{-\int_{r_0}^r\frac{1}{r'}\,dr'}\left[v_b^\prime(r_0)+4a^2
\int_{r_0}^r\frac{\alpha_b}{(1-{r'}^2)^2}\,e^{\int_{r_0}^{r'}\frac{1}{s}\,ds}\,dr'
\right]\\
&=&\frac{r_0v_b^\prime(r_0)}{r}+\frac{4a^2}{r}\int_{r_0}^r\alpha_b
\frac{r'}{(1-{r'}^2)^2}\,dr'.
\eq
Since the potential has to be chosen continuous at $r_0$ we set
$v_b'(r_0)=2a^2\alpha_br_0/(1-r_0^2)$ to find, 
\bq\nonumber
v_b^\prime(r)
=2a^2\alpha_b\left\{
\begin{array}[c]{ll}
\displaystyle \frac{r}{1-r^2}&r\le r_0\\
\displaystyle \frac{r_0^2}{1-r_0^2}\frac{1}{r}&r>r_0
\end{array}
\right.,
\eq
where $r_0=\tanh(\tau_0/2)$. For the potential inside
$\Omega_{a\tau_0}$ we then have,
\bq
v_b(r)=-\alpha_b a^2\ln(1-r^2)+{\rm constant},
\eq
or using the coordinates $(\tau,\varphi)$,
\bq
v_b(\tau)=-\alpha_b a^2\ln[1-\tanh^2(\tau/2)]+{\rm constant}.
\eq
We need to adjust the additive constant in such a way that this potential
at $\tau=\tau_0$ has the correct value corresponding to the total
background charge. We then have,
\bq \nonumber
{\rm constant}&=&v_b(0)=-en\int_{\Omega_{a\tau_0}}G(\tau a)\,dS\\\nonumber
&=&2\pi a^2qn\int_0^{\tau_0}\ln[\tanh(\tau/2)]\sinh\tau\,d\tau\\
&=&\alpha_b a^2[\ln[1-\tanh^2(\tau_0/2)]+\sinh^2(\tau_0/2)
\ln[\tanh^2(\tau_0/2)]].
\eq
We reach then the following expression for the potential inside
$\Omega_{a\tau_0}$,
\bq\label{back pot}
v_b(\tau)=\alpha_b a^2\left\{\ln\left[\frac{1-\tanh^2(\tau_0/2)}
{1-\tanh^2(\tau/2)}\right]+\sinh^2(\tau_0/2)\ln[\tanh^2(\tau_0/2)]
\right\}.
\eq

The self energy of the background is,
\bq\label{self e}
V_N^0&=&\frac{1}{2}\int_{\cal S}\rho_bv_b\,dS\\\nonumber
&=&\frac{1}{2}\rho_b\alpha_b a^2 2\pi a^2\left\{
\int_0^{\tau_0}\ln\left[\frac{1-\tanh^2(\tau_0/2)}
{1-\tanh^2(\tau/2)}\right]\sinh\tau\,d\tau+\right.\\\nonumber
&&\hspace{3cm}\left.\sinh^2(\tau_0/2)\ln[\tanh^2(\tau_0/2)]\int_0^{\tau_0}
\sinh\tau\,d\tau\right\}\\\nonumber
&=&-2a^4(\pi\rho_b)^2\{1-\cosh\tau_0+4\,\ln[\cosh(\tau_0/2)]+
2\sinh^4(\tau_0/2)\ln[\tanh^2(\tau_0/2)]\}.
\eq
Notice that if we drop the last term on the right hand side of this 
equation, i.e. if we adjust the additive constant so that the 
potential of the background vanishes on the boundary $\partial 
\Omega_{a\tau_0}$, then in the limit $a\rightarrow \infty$ we recover 
the self energy of the flat system $N^2e^2/8$.

\section{The flat limit for the OCP in the pseudosphere} 
\label{app:pseudosphere-flat}

In this Appendix we study the flat limit $a\to\infty$ of the
expressions found for the density in
section~\ref{sec:pseudosphere-g2}. We shall study the limit
$a\to\infty$ for a finite system and then take the thermodynamic
limit. Since for a large system on the pseudosphere boundary effects
are of the same order as bulk effects it is not clear a priori whether
computing these two limits in different order would give the same
results. In Ref. \cite{Fantoni03jsp} we show that it does.

For a finite disk of radius $d=a\tau_0$, we have in the flat limit
$a\to\infty$, $d\sim r_0$. In equation~(\ref{eq:flamm-densite-somme}), in the
limit $a\to\infty$, the term $e^C$ given by (\ref{eq:cste-exp-c})
becomes
\begin{equation}
e^C\sim \left(\frac{r_0^2}{4a^2}\right)^{-N_b}
e^{N_b}
\end{equation}
where $N_b=\pi n_b r_0^2$ is the number of particles in the background
in the flat limit. Since for large $a$, $t_0=r_0^2/4a^2$ is small, the
incomplete beta function in equation~(\ref{eq:flamm-densite-somme}) is
\begin{equation}
B_{t_0}(\ell+1,\alpha)
=
\int_0^{t_0}
e^{(\alpha-1)\ln(1-t)}
\, t^{\ell}\,dt
\sim
\int_0^{t_0}
e^{-(\alpha-1)t}
\, t^{\ell}\,dt
\sim
\frac{\gamma(\ell+1,N_b)}{\alpha^{\ell+1}}
\end{equation}
Expanding $(1-(r^2/4a^2))^{4\pi n_b a^2}\sim \exp(-\pi n_b r^2)$ in
equation~(\ref{eq:flamm-densite-somme}) we finally find the density as a
function of the distance $r$ from the center
\begin{equation}
n^{(1)}(r)=
n_b e^{-\pi n_b r^2}
\sum_{\ell=0}^{\infty}
\frac{(\pi n_b r^2)^{\ell}}{\alpha^{\ell-N_b} N_b^{N_b} e^{-N_b}
(n_b/\zeta)+
\gamma(\ell+1,N_b)}
\end{equation}
When $\alpha\to\infty$ the terms for $\ell>N_b$ in the sum vanish
because $\alpha^{\ell-N_b}\to\infty$. Then
\begin{equation}
\label{eq:densite-flat}
n^{(1)}(r)=
n_b e^{-\pi n_b r^2}
\sum_{\ell=0}^{E(N_b)-1}
\frac{(\pi n_b r^2)^{\ell}}{\gamma(\ell+1,N_b)}
+\Delta n^{(1)}(r)
\end{equation}
The first term is the density for a flat OCP in the canonical ensemble
with a background with $E(N_b)$ elementary charges ($E(N_b)$ is the
integer part of $N_b$). The second term is a correction due to the
inequivalence of the ensembles for finite systems and it depends on
whether $N_b$ is an integer or not. If $N_b$ is not an integer
\begin{equation}
\Delta n^{(1)}(r)=n_b
\frac{(\pi n_b r^{2})^{E(N_b)} e^{-\pi n_b r^2}}{\gamma(E(N_b)+1,N_b)}
\end{equation}
and if $N_b$ is an integer
\begin{equation}
\Delta n^{(1)}(r)=n_b
\frac{(\pi n_b r^{2})^{N_b} e^{-\pi n_b r^2}}{N_b^{N_b} e^{-N_b}
(n_b/\zeta)+\gamma(N_b+1,N_b)}
\end{equation}
In any case in the thermodynamic limit $r_0\to\infty$,
$N_b\to\infty$, this term $\Delta n^{(1)}(r)$ vanishes giving the
known results for the OCP in a flat space in the canonical
ensemble~\cite{Jancovici81b,Jancovici81}. Integrating the profile
density~(\ref{eq:densite-flat}) one finds the average number of
particles. For a finite system it is interesting to notice that the
average total number of particles $N$ is
\begin{equation}
N=E(N_b)+1
\end{equation}
for $N_b$ not an integer and
\begin{equation}
N=N_b+\frac{1}{\displaystyle 1+
\frac{N_b^{N_b}e^{-N_b} n_b}{\zeta \gamma(N_b+1,N_b)}}
\end{equation}
for $N_b$ an integer. In both cases the departure from the neutral
case $N=N_b$ is at most of one elementary charge as it was noticed
before~\cite{Jancovici86,Jancovici03}.

\section{Green's function of Laplace equation in Flamm's paraboloid} 
\label{app:flamm-green}

In this appendix, we illustrate the calculation of the Green function,
for the various situations considered, using the original system of
coordinates $(r,\varphi)$. 

\subsection{Laplace equation}

We first find a solution $v(\qq)$, not circularly symmetric, to
Laplace equation
\bq \label{laplace}
\Delta v = 0~,
\eq
through the separation of variables technique. We then write 
\bq
v(r,\varphi)=R(r)\phi(\varphi)~,
\eq
so that Laplace equation splits into the two ordinary differential
equations 
\bq
&&\phi^{\prime\prime}=-k^2\phi~,\\
&&(r^2-2Mr)R^{\prime\prime}+(r-M)R^\prime=k^2R~.
\eq
Taking care of the boundary condition
$\phi(\varphi+2\pi)=\phi(\varphi)$ we find that the first
equation admits solution only when $k$ is an integer. The
solutions being 
\bq
\phi_n=C_+e^{in\varphi}+C_-e^{-in\varphi}~~~n=0,1,2,3,\ldots
\eq
The solutions of the second equation are
\bq \label{laplacer1}
R_n=\left\{
\begin{array}{ll}
C_1\cosh(na)+C_2\sinh(na) & r>2M\\ 
C_1\cos(na)+C_2\sin(na) &r<2M
\end{array}\right.
\eq
where
\bq
a=\left\{
\begin{array}{ll} \displaystyle
2\arctan\sqrt{\frac{r}{2M-r}} & r<2M \\\displaystyle
2\ln\frac{\sqrt{r}+\sqrt{r-2M}}{\sqrt{2M}}    & r>2M
\end{array}\right.
\eq
Here $C_-,C_+,C_1,$ and $C_2$ are the integration constants.

Then the general solution is real for $C_+=C_-=C_0$
\bq
v(r,\varphi)=\sum_{n=0}^\infty R_n(r)\phi_n(\varphi)=\left\{
\begin{array}{ll} \displaystyle
C_0\left(C_1+C_2\frac{\sin a}{\cos\varphi-\cos a}\right) & r<2M \\
\displaystyle
C_0\left(C_1+C_2\frac{\sinh a}{\cos\varphi-\cosh a}\right) & r>2M
\end{array}\right.
\eq

If we require the Coulomb potential to go to zero at $r=\infty$ we
must choose $C_1-C_2=0$ so that (for $C_0=1$)
\bq
v(r,\varphi)=\left\{
\begin{array}{ll} \displaystyle
1+\frac{\sin a}{\cos\varphi-\cos a}   & r<2M \\\displaystyle
1+\frac{\sinh a}{\cos\varphi-\cosh a} & r>2M
\end{array}\right.
\eq
Moreover $v(2M,\varphi)=1$.

\subsection{Green's function of Laplace equation}
\label{sec:green}

We now want to find the Coulomb potential generated at $\qq=(r,\varphi)$
by a charge at $\qq_0=(r_0,\varphi_0)$ with $r_0>2M$. We then have to solve
the Poisson equation
\bq
\Delta G(r,\varphi;r_0,\varphi_0) = 
-2\pi\delta(r-r_0)\delta(\varphi-\varphi_0)/\sqrt{g}~,
\eq
where $g=\det (g_{\mu\nu})=r^2/(1-2M/r)$. To this end we expand the
Green function $G$ and the second delta function in a Fourier series
as follows
\bq \label{gexp}
G(r,\varphi;r_0,\varphi_0)&=&
\sum_{n=-\infty}^\infty e^{in(\varphi-\varphi_0)} g_n(r,r_0)~,\\
\delta(\varphi-\varphi_0)&=&\frac{1}{2\pi}
\sum_{n=-\infty}^\infty e^{in(\varphi-\varphi_0)}~,
\eq
to get an ordinary differential equation for $g_n$
\bq
\left[\left(1-\frac{2M}{r}\right)\frac{\partial^2}{\partial r^2}
+\left(\frac{1}{r}-\frac{M}{r^2}\right)\frac{\partial}{\partial
r}-\frac{n^2}{r^2}\right]g_n(r,r_0)=-\delta(r-r_0)/\sqrt{g}~.
\eq
To solve this equation we first solve the homogeneous one for $r<r_0$:
$g_{n,-}(r,r_0)$ and $r>r_0$: $g_{n,+}(r,r_0)$. This equation was already
solved in (\ref{laplacer1}) for $n\neq 0$
\bq
g_{n,\pm}=A_{n,\pm}(\sqrt{r}+\sqrt{r-2M})^{2n}+
B_{n,\pm}(\sqrt{r}+\sqrt{r-2M})^{-2n}~
\eq 
and for $n=0$ one finds
\bq
g_{0,\pm}=A_{0,\pm}+B_{0,\pm}\ln(\sqrt{r}+\sqrt{r-2M})~.
\eq
The form of the solution immediately suggest that it is more
convenient to work with the variable
$x=(\sqrt{r}+\sqrt{r-2M})^{2}/(2M)$. For this reason, we introduced 
this new system of coordinates $(x,\varphi)$ which is used in the main
text.

We then impose the following boundary conditions: (i) the solution at
$r=r_0$ should be continuous, (ii) the first derivative at $r=r_0$
should have a jump due to the delta function, (iii) at $r=2M$ the
solution should tend to the solution of the flat system ($M\to 0$),
and (iv) the solution should vanish at $r=\infty$, namely,
\bq
g_{n,-}(r_0,r_0)&=&g_{n,+}(r_0,r_0)~,\\
g_{n,-}^\prime(r_0,r_0)&=&g_{n,+}^\prime(r_0,r_0)
+\frac{1}{\sqrt{r_0(r_0-2M)}}~,\\
B_{n,-}=0~~~\mbox{for $n>0$}&,&A_{n,-}=0~~~\mbox{for $n<0$}~,\\
A_{n,+}=0~~~\mbox{for $n>0$}&,&B_{n,+}=0~~~\mbox{for $n<0$}~.
\eq
Performing the Fourier series of Eq. (\ref{gexp}) then leads to the
following result,
\bq \label{cgreen}
G^{\rm{hs}}(r,\varphi;r_0,\varphi_0)=
-\ln\left|z-z_0\right|~,
\eq
where the complex coordinates $z=(\sqrt{r}+\sqrt{r-2M})^2e^{i\varphi}$ and 
$z_0=(\sqrt{r_0}+\sqrt{r_0-2M})^2e^{i\varphi_0}$ have been introduced. 
This solution reduces to the correct Coulomb green function on a plane
as $M\to 0$ and it is the Coulomb potential on one universe of the 
surface $\cal S$.  

In order to find the Coulomb potential on the whole surface we can
then start from the definition (\ref{u}) and go back to the
$s=(\sqrt{r}+\sqrt{r-2M})^2$ variable. If we do this we find as
solutions, 
\bq \label{spm}
s_\pm=2M(\sqrt{u^2+1}\pm u)^2~,
\eq
So that for the Coulomb potential one can choose one of the two
definitions depending on which charge is in the upper or lower 
universe. Neglecting an additive constant we could then set
\bq 
G^{\rm{ws}}(u,\varphi;u_0,\varphi_0)=-\ln|z-z_0|~,
\eq
where $z=(\sqrt{u^2+1}+u)^2 e^{i\varphi}$ and
$z_0=(\sqrt{u_0^2+1}+u_0)^2 e^{i\varphi_0}$. Actually this potential
as it stands does not have the correct symmetry properties under the
exchange of the charges from one universe to the other. It can
easily be shown that if $z$ is a point in the upper universe then
$1/z$ is its symmetric in the lower universe. Then we should expect
that if we take $z_0=1$ (in the horizon) the potential created at
$z$ should be the same as the one created at $1/z$, by symmetry. More
generally, one should have $G^{\rm{ws}}(z,z_0)=G^{\rm{ws}}(1/z,1/z_0)$. 

We then need to revise the calculations of the Coulomb potential. We
define the Coulomb potential as the solution of Poisson equation with
the boundary condition that the electric field vanishes at infinity
(this also happens for a flat space). However it turns out that with
this boundary condition one still have several different solutions,
and contrary to the flat 
case, there are some that differ in more than a constant term. One can
see this by solving Poisson equation using the Fourier transform, the
constants of integration for the term which does not depend on the
angular variable cannot be determined. 

However one can impose some additional conditions. For instance we
expect the Coulomb potential to be symmetric in the exchange of $z$
and $z_0$. The previous solution $-\ln |z-z_0|$ does satisfy this, but
it is not the unique solution with this property. Additionally, we can
impose the symmetry relation
$G^{\rm{ws}}(z,z_0)=G^{\rm{ws}}(1/z,1/z_0)$. Then one finds the solution
\bq \label{tcgreen}
G^{\rm{ws}}(z,z_0)=-\ln (|z-z_0| / \sqrt{|z z_0|})~. 
\eq
We have not
verified if this is the only solution (up to a constant) satisfying
this symmetry, but we think so. For the whole surface we think that we
should use this Coulomb potential instead of the original one , which
does not treat on the same foot the upper and lower parts of the
surface. However we have noticed that this potential does not reduce
to the flat one when $M=0$, but this is normal: if we work with the
whole surface the limit $M=0$ is not exactly the flat one, it is two
flat planes connected by a hole at the origin, this hole modifies the
Coulomb potential. 

\subsection{The grounded horizon case}

Imagine now that the horizon at $r=2M$ is a perfect conductor. We
then start from
\bq
g_{n,\pm}=A_{n,\pm}\cosh\left[2n\ln(\sqrt{r}+\sqrt{r-2M})\right]+
B_{n,\pm}\sinh\left[2n\ln(\sqrt{r}+\sqrt{r-2M})\right]~.
\eq  
We fix the four integration constants, for each $n$, requiring that:
(i) the solution at $r=r_0$ should be 
continuous, (ii) the first derivative at $r=r_0$ should have a jump
due to the delta function, (iii) at $r=2M$ the solution should vanish,
and (iv) the solution has the correct behavior at $r=\infty$, namely,
\bq
g_{n,-}(r_0,r_0)&=&g_{n,+}(r_0,r_0)~,\\
g_{n,-}^\prime(r_0,r_0)&=&g_{n,+}^\prime(r_0,r_0)
+\frac{1}{\sqrt{r_0(r_0-2M)}}~,\\
g_{n,-}(2M,r_0)&=&0~,\\
A_{n,+}=B_{n,+}~~~\mbox{for $n\ge 0$}&,&A_{n,+}=-B_{n,+}~~~\mbox{for $n<0$}~.
\eq
Performing the Fourier series of Eq. (\ref{gexp}) then leads to the
following result for $r>r_0$
\bq
G(r,\varphi;r_0,\varphi_0)&=&
-\ln\sqrt{\frac{1+c^2-2c\cos(\varphi-\varphi_0)}
{1+b^2-2b\cos(\varphi-\varphi_0)}}+
2\ln\frac{\sqrt{r_0}+\sqrt{r_0-2M}}{\sqrt{2M}}~,\\
b&=&\left(\frac{\sqrt{r}+\sqrt{r-2M}}
{\sqrt{r_0}+\sqrt{r_0-2M}}\right)^2~,\\
c&=&\left(\frac{(\sqrt{r}+\sqrt{r-2M})
(\sqrt{r_0}+\sqrt{r_0-2M})}{2M}\right)^2~,
\eq
and the solution for $r<r_0$ is obtained by merely exchanging $r$ with
$r_0$. 

In terms of the complex numbers
$z$ and $z_0$ 
this can be rewritten as follows
\bq \label{ghgreen}
G^{\rm{gh}}(r,\varphi;r_0,\varphi_0)=
-\ln\left|\frac{(z-z_0)/2M}{1-z\bar{z}_0/4M^2}\right|
\eq
where the bar over a complex number indicates its complex
conjugate. We will call this the grounded horizon green
function. Notice how its shape is the same of the
Coulomb potential on the pseudosphere \cite{Fantoni03jsp} $M$ playing
the role of the complex radius. This green function could have been
found from the Coulomb one (\ref{cgreen}) by using the images method
from electrostatics.

\section{The geodesic distance on the Flamm paraboloid}  
\label{app:flamm-geodesic}

The geodesics are determined by the following equation
\bq \label{ge1}
\ddot{r}+(\Gamma_{rrr}\dot{r}^2+\Gamma_{r\varphi\varphi}
\dot{\varphi}^2)/g_{rr}&=&0~,\\ \label{ge2}
\ddot{\varphi}+2\Gamma_{\varphi\varphi r}\dot{\varphi}\dot{r}/
g_{\varphi\varphi}&=&0~,
\eq
where the dot stands for a total differentiation with respect to 
time and the Christoffel symbols are as follows
\bq
\Gamma_{rrr}&=&g_{rr,r}/2~,\\
\Gamma_{\varphi\varphi r}&=&-\Gamma_{r\varphi\varphi}=
g_{\varphi\varphi,r}/2~.
\eq
Here the comma means partial differentiation as usual.

The geodesics equation (\ref{ge1})-(\ref{ge2}) is then
\bq \label{geo1}
\ddot{r}-\left[\frac{M}{(r-2M)^2}\dot{r}^2+r
\dot{\varphi}^2\right]\left(1-\frac{2M}{r}\right)&=&0~,\\ \label{geo2}
\ddot{\varphi}+\frac{2}{r}\dot{\varphi}\dot{r}&=&0~,
\eq

The geodesic distance between two points on the surface is
\bq \nonumber
d(\qq_1,\qq_2)&=&\int_{t_1}^{t_2} \frac{d\mathbf{s}}{dt}\,dt=
\int_{r_1}^{r_2} y\,dr 
=\int_{r_1}^{r_2} \sqrt{\frac{1}{1-\frac{2M}{r}}+r^2x^2}\,dr
\eq
where $x(r)=d\varphi/dr$ and $y(r)=d\mathbf{s}/dr$.

Using $\dot{\varphi}=x\dot{r}$ in Eqs. (\ref{geo1}) and
(\ref{geo2}) we find  
\bq
x^\prime&=&\left(\frac{2}{r}+\frac{M}{r^2-2Mr}\right)x
+r\left(1-\frac{2M}{r}\right)x^3~,
\eq
where the prime stands for differentiation with respect to $r$.

The solution for $x(r)$ and $y(r)$ are as follows
\bq
x(r)&=&\pm\sqrt{\frac{15r^3(2M-r)}{r^4(30M^2-24Mr+5r^2)-C}}~,\\ 
y(r)&=&\sqrt{r^2x^2+\frac{r}{r-2M}}~,
\eq
with $C$ the integration constant, so that,
\bq
d(\qq,\qq_0)&=&\int_{r_0}^{r} y(r^\prime)\,dr^\prime~,\\
\varphi-\varphi_0&=&\int_{r_0}^{r} x(r^\prime)\,dr^\prime~.
\eq

\begin{acknowledgments}
...
\end{acknowledgments}

%

\end{document}